\documentclass[preprint,5p,times,twocolumn]{elsarticle}
\usepackage{hyperref}
\hypersetup{
  colorlinks=true, 
}
\usepackage{amssymb}
\usepackage{amsmath}
\usepackage{bbm}
\usepackage{nicefrac}

\DeclareMathAlphabet{\mathdutchcal}{U}{dutchcal}{m}{n}
\newcommand{\ii}{\ensuremath\mathbbm{i}}
\newcommand{\II}{\ensuremath\mathbbm{1}}

\newcommand{\kk}{\ensuremath\mathdutchcal{k}}

\newcommand{\sss}{\ensuremath\mathdutchcal{s}}

\newcommand{\ff}{\ensuremath\mathdutchcal{f}}
\newcommand{\pp}{\ensuremath\mathdutchcal{p}}

\newcommand{\op}[1]{\ensuremath\hat{{\mathcal{#1}}} }

\providecommand{\footref}[1]{\textsuperscript{\ref{#1}}}
\newcommand{\mum}{\ensuremath\mu\mathrm{m}}

\usepackage{tikz}

\newcommand{\myarrow}[1][]{%
  \begin{tikzpicture}[#1]%
    \draw (0,0.7ex) -- (0,0) -- (0.75em,0);
    \draw (0.55em,0.2em) -- (0.75em,0) -- (0.55em,-0.2em);
  \end{tikzpicture}%
}

\newcommand{\revise}[1]{{#1}}

\newcommand{\revisee}[1]{{#1}}

\usepackage{lineno}
\journal{Computer Physics Communications}
\begin{document}

\begin{frontmatter}


\title{High-order exponential solver method for particle-in-cell simulations }

\author[label1]{Szil\'ard Majorosi\corref{cor1} }
\ead{szilard.majorosi@eli-alps.hu}
\cortext[cor1]{Corresponding author.}

\author[label1,label2]{Nasr A.M. Hafz}

\author[label1]{Zsolt L\'ecz}

\affiliation[label1]{organization={ ELI ALPS, ELI-HU Non-Profit Ltd.},
             addressline={Wolfgang Sandner utca 3.},
             city={Szeged},
             postcode={H-6728},
             country={Hungary}}
\affiliation[label2]{organization={Doctoral School of Physics, Faculty of Science and Informatics, University of Szeged},
             addressline={Dóm tér 9.},
            city={Szeged},
            postcode={H-6720},
             country={Hungary}}

%\address[label2]{Doctoral School of Physics, Faculty of Science and Informatics, University of Szeged, Dóm tér 9., Szeged, H-6720, Hungary}

\begin{abstract}
Outstanding advances in solid-state laser technology, employing the optical parametric  chirped-pulse-amplification (OPCPA) technique, have led physicists to focus laser pulses to highly- relativistic intensities which led to novel schemes for charged-particle acceleration and radiation generation in laser-driven plasmas. Microscopic understanding of these highly-nonlinear processes is possible via accurate modeling of the laser-plasma interaction using particle-in-cell (PIC) simulations. Numerous codes are available and they rely on finite difference time domain methods on Yee-grids or on the analytical solution of the Maxwell-equations in spectral space. In this work, we present a solution bridging these two methods, which we call finite difference exponential time domain solution. 
\revise{This scalable method is able to provide very high accuracy even in 3D}, but with improved locality, similar to the pseudospectral analytical methods without relying on transformation to special basis functions.
We verified the accuracy and the convergence of the method in various benchmarks, including laser propagation in vacuum and in underdense plasma. We also simulated electron injection in a non-linear laser-plasma wakefield acceleration and surface high-harmonic generation in the overdense regime. The results are then compared with those obtained from standard PIC codes.

\end{abstract}

\end{frontmatter}

\section{Introduction} \label{sec:introduction}

The invention of the chirped-pulse amplification technique (CPA) \cite{Strickland85} opened the era of relativistic laser-plasma interactions, where the predictions of earlier theoretical works could be tested and confirmed \cite{Tajima79}. Development of the laser-driven particle acceleration schemes \cite{esarey2009laser_plasma_accelerators, lu2007electron_laser_wakefield, albert2016applications_lwfa} rely on accurate modeling with particle-in-cell simulations \cite{BOOK_PLASMA_SIMULATION, fonseca2002OSIRIS, pritchett2003pic_tutorial, arber2015pic_epoch, belayev2015PICsar, derouillat2018pic_smilei}. The ever-increasing quality of laser pulses \cite{Kiriyama18, Kiriyama23} will require high-precision numerical tools for the modeling of the electromagnetic field propagation and the coupled self-consistent plasma fluid motion, especially in the QED regime \cite{Cole18_PRX, Poder18_PRX}. For these, the go-to Maxwell solvers were the finite difference time domain methods on Yee-grids for decades. These latter are proven to be very effective to provide real-life simulation results at high resolutions with efficient computational load and parallelization properties. However, they suffer from various finite difference artifacts, like excessive dispersion during electromagnetic wave propagation and the numerical Cherenkov radiation (NCR) which could be always present independent of the resolution. There have been successful efforts to improve Yee-solvers \cite{lehe2013numerical, pukhov2016pic, bourgeois2020numerical_cherenkov, bourgeois2023improved_pic, sekido2024accuracy_difference} by addressing specific defects the method possesses, and it is possible to get exact wave dispersion along one dimension. However, the simple structure of the Yee-method is not accurate enough to fix all of these deficiencies in two (2D) or three dimensions (3D).

The other approach yielded the most accurate methods available, which are taking advantage of the analytical solution of the Maxwell-equations in spectral space (pseudospectral analytical time domain, PSATD), which usually involves fast Fourier transform on uniform spatial grids \cite{godfrey2014stability_spectral, vay2018warpx, lehe2016fbpic, vincenti2018ultrahigh_pic, li2017maxwell_cherenkov, jalas2017pseudo_spectral} or discrete Hankel-Bessel transforms in the well known FBPIC code \cite{lehe2016fbpic}. The drawback of these methods is that their solutions are non-local, and relies on the boundary conditions that the basis provides, and parallelizing them with domain decompositions requires a difficult and special implementation or a compromise on accuracy.

\revise{In this work we present another type of solution that bridges the two approaches, 
which we call finite difference exponential time domain solution, well-known in quantum mechanics \cite{leforestier1991tdsecomparisom, castro2004kohnshampropagators, bandrauk2013splitting}. 
In plasma physics some of the concepts of this so far has been employed in structure preserving methods by splitting the Hamiltonian of the nonrelativistic Vlasov-Maxwell system \cite{he2015hamiltonian_maxwell_vlasov, kraus2017gempic, campos2024variational_particle}, but we employ these tools to form high order Maxwell propagator for the electromagnetic fields and integrate these into the relativistic PIC method.
}
The spatial differences are represented with high order finite differences (6th-32nd) and the time stepping with exponential operators – in practice a lot of matrix multiplications. We found it worthwhile investigating this approach, partly because of the authors background in quantum mechanics, and to develop real space solutions that are highly accurate and local, and can be applied to any geometry, basis set, or wave propagation problem. Since this work is not an iterative improvement of an existing PIC method, we started from the fundamental literature on plasma and numerical computation \cite{BOOK_PLASMA_SIMULATION, BOOK_NUMERICAL_RECIPIES}, and during development we reviewed the relevant literature at each step, which is also reflected in this work. Here, we present our solution in Cartesian 2D, and 3D geometries. We will also discuss other sources of errors in the PIC code, like the one that arises from the field interpolation, needed for the equation of motion, and from the numerical dispersion which can be corrected by high-order finite differences and can be eliminated by spectral filtering of the current density. We discuss all of these errors and possible solutions to avoid them in the paper.

This paper is organized as follows. In Section \ref{sec:maxwell}, we discuss how the Maxwell-equations can be solved formally with exponential operators, and what these exponential operators mean in practice. In Section \ref{sec:spatial} we discuss how we construct the spatial representation for the field solver, we discuss how the required matrices are calculated for the exponential solution, and we also present the actual numerical algorithm. Then, in Section \ref{sec:PIC} we develop our particle-in-cell routine compatible with our field solution, which mostly follows the conventional Yee-PIC solutions. Finally, in  Section \ref{sec:benchmarks} we benchmark our method in various cases mostly focusing on laser propagation in vacuum, in an underdense plasma, producing accurate wakefields and nonlinear laser wakefield electron acceleration (LWFA). We also tested the generation of surface high harmonics from laser-solid density plasma interaction, and we show that this method \revise{is able to} provide very high accuracy in the spectral range of physical interest. We also included many appendices, in \ref{subsubsec:units}-\ref{subsubsec:deposit_ES} we discuss advanced topics like absorbing layers, coordinate transformations, enhanced finite differences, current deposition approaches which did not warrant another publication.

In the following, we write all quantities in the normalized units commonly used in plasma physics unless stated otherwise. In these units we set the speed of light ($c$), the elementary charge ($e$), the  electron mass ($m_e$) equal to unity. For the precise definitions see \ref{subsubsec:units}.  These will simplify the physical and matrix equations related to the electromagnetic fields and the relativistic particles.

\section{Electromagnetic fields}  \label{sec:maxwell}

\subsection{Maxwell equations in Cartesian coordinates} \label{subsec: maxwell_cartesian}
For a theoretical introduction let us summarize the differential form of the Maxwell equations for the electric  $\bf E$ and magnetic fields $\bf B$, including the total current density $\bf J$ and charge density $\varrho$. What follows is a rather verbose version of their respective differential equations, the forms of which will be useful when we start making numerical approximations.

Let us start by writing out the two set of field evolution equations, namely the Amp\'ere-equation with $\nabla \times {\bf B}$ and source ${\bf J}$:
\begin{align}
\partial_t E_x &= -\partial_z B_y +\partial_y B_z - J_x ,\label{eq:maxwell_Ex}\\
\partial_t E_y &=  \partial_z B_x -\partial_x B_z - J_y ,\label{eq:maxwell_Ey}\\
\partial_t E_z &= -\partial_y B_x +\partial_x B_y - J_z ,\label{eq:maxwell_Ez}
\end{align}
and the Faraday-equation  with $\nabla \times {\bf E}$ :
\begin{align}
\partial_t B_x &=  \partial_z E_y -\partial_y E_z ,\label{eq:maxwell_Bx}\\
\partial_t B_y &= -\partial_z E_x +\partial_x E_z ,\label{eq:maxwell_By}\\
\partial_t B_z &=  \partial_y E_x -\partial_x E_y .\label{eq:maxwell_Bz}
\end{align}
Next, the  expression for $\nabla \cdot {\bf B}$ is:
\begin{equation} \label{eq:maxwell_divB}
\partial_x B_x + \partial_y B_y +\partial_z B_z  =  0,
\end{equation}
and the differential Gauss's law for $\nabla \cdot {\bf E}$ is:
\begin{equation} \label{eq:maxwell_divE}
\partial_x E_x + \partial_y E_y +\partial_z E_z = \varrho.
\end{equation}

We note that solving Eqs. (\ref{eq:maxwell_divB}) or (\ref{eq:maxwell_divE}) has no effect on the  right hand side of evolution equations Eq. (\ref{eq:maxwell_Ex})-(\ref{eq:maxwell_Bz}) and the charge density $\varrho$ does not occur within them - only the current density $\bf  J$. As known these evolution equations only propagate the transverse (divergence free) part of ${\bf E}$, ${\bf B}$ (and ${\bf J}$) which means that the longitudinal (divergence) part of ${\bf E}$ and ${\bf B}$ are left behind if the charges do not move.

If the initial fields ${\bf E}_0$ and ${\bf B}_0$ satisfy the divergence laws Eqs. (\ref{eq:maxwell_divB})-(\ref{eq:maxwell_divE}), then during propagation Eqs. (\ref{eq:maxwell_Ex})-(\ref{eq:maxwell_Bz}) ensure that they are satisfied at all times. Using the time derivative of Eq. (\ref{eq:maxwell_divE}) it can be shown that that the analytical $\bf J$ needs to satisfy the continuity equation:
\begin{equation} \label{eq:maxwell_divJ}
\partial_x J_x + \partial_y J_y +\partial_z J_z =  \nabla \cdot {\bf J} = -\partial_t \varrho.
\end{equation}
If we introduce numerical approximations, this statement will no longer be true - additional care must be taken when constructing $\bf J$. Enforcing the validity of the continuity equation is commonly called charge conserving current deposition schemes \cite{ villasenor1992charge_conservation, esirkepov2001charge_conservation}.  We note that directly solving Eq. ({\ref{eq:maxwell_divE}}) for the longitudinal components are done by solving the Poisson's equation, and it is far from trivial (see \ref{subsubsec:poisson}). In the exponential solution that we outline in the following the particular form of $\bf J$ does not matter.

We introduce the exponential electromagnetic time evolution method in the following way: let us formally combine the Maxwell equations Eqs. (\ref{eq:maxwell_Ex})-(\ref{eq:maxwell_Bz}) into the following first order equation:
\begin{equation} \label{eq:maxwell_formal}
 \partial_t \Psi      = \op{H} \Psi - {\bf J},
\end{equation}
where we introduce $\op{H}$ as linear operator from the right hand side (RHS) of the Maxwell equations, and $\Psi$ is vector describing the whole electromagnetic state as

\begin{equation} \label{eq:maxwell_psi}
\Psi =
\left( \begin{array}{cccccc}
E_x & E_y & E_z & B_x & B_y & B_z
\end{array} \right)  ^T
\end{equation}
\begin{equation} \label{eq:maxwell_psiJ}
{\bf J} =
\left( \begin{array}{cccccc}
J_x & J_y & J_z & 0 & 0 & 0
\end{array} \right)  ^T
\end{equation}
\begin{equation}  \label{eq:maxwell_H}
\op{H} =
\left( \begin{array}{cccccc}
0            & 0         & 0       &      0      & -\partial_z &  \partial_y \\
0            & 0         & 0       &  \partial_z & 0           & -\partial_x \\
0            & 0         & 0       & -\partial_y & \partial_x  & 0           \\
0            &  \partial_z & -\partial_y & 0          & 0         & 0          \\
 -\partial_z & 0           &  \partial_x & 0          & 0         & 0         \\
  \partial_y & -\partial_x  & 0           & 0          & 0         & 0         \\
\end{array} \right) .
\end{equation}

This formalism is a direct analogy to the Schrödinger's equation used in quantum mechanics. The formal solution of Eq. (\ref{eq:maxwell_formal}) for a short $\Delta t$ time step and assuming that $\op{H}$ is constant are given by \cite{dijk2014tdsesource}:
\begin{equation} \label{eq:maxwell_solution}
\Psi(t+\Delta t) = \exp \left( \Delta t \op{H} \right)\Psi(t)- \int_0^{\Delta t} 
                \exp \left((\Delta t-s)  \op{H} \right){\bf J}(t+s) {\rm d}s
\end{equation}
which is exact, since it contains no other approximations. Note that using pure spectral basis with the Maxwell equations \cite{BOOK_PLASMA_SIMULATION} also leads to Eq. (\ref{eq:maxwell_solution}),  which shows that this exponential solution is independent of the geometry, spatial grid or the basis set we might choose for representing Eq. (\ref{eq:maxwell_formal}) in the spatial domain.  In the next section we give an overview on how this expression can be interpreted to provide a numerical solution.  Subsequent time steps are done by repeating step Eq. (\ref{eq:maxwell_solution}).

\subsection{Exponential methods} \label{subsec:exponential}

\revise{Many well developed methods} exist to deal with advanced time evolution and exponential operators\revise{.}
%, the authors of this paper has worked with these in the past \cite{majorosi2016tdsecoulomb}. 
As a prerequisite, however, we need to approximate the integral in Eq. (\ref{eq:maxwell_formal}), which we do using the second order midpoint quadrature:
\begin{equation} \label{eq:exponential_solution2}
\Psi(t+\Delta t) \approx \exp \left( \Delta t \op{H} \right)\Psi(t)- 
                \exp \left(\Delta t \op{H} /2 \right){\bf J}\left(t+\Delta t / 2\right) \Delta t
\end{equation}
We note that if $\op{H}$ depends on time we need to replace it with its average between each $t$ and $t+\Delta t$ time step to achieve second order accuracy, the general method of which is called  Magnus-expansion \cite{blanes2009review_magnus_expansion}. For the Maxwell-problem $\op{H}$ is constant, but $\bf J$ self-consistently depends on the value of the state $\Psi$ through physical laws (PIC particles).  Using Equation (\ref{eq:exponential_solution2}) can circumvent this problem if we temporally stagger the sources (particles) or fields by $\pm \Delta t /2$. We also give other possible approximations and simplifications for the source integrals in \ref{subsubsec:exponential_source}.  In the following, we summarize the exponential operator expressions that can be used in Eq. (\ref{eq:exponential_solution2}) to \revisee{form our numerical method}.

\subsubsection{Explicit Taylor-expansion} \label{subsubsec:exponential_taylor}

The exponential of a linear operator $\op{H}$ can be directly defined via its Taylor-expansion \cite{castro2004kohnshampropagators}:
\begin{equation}  \label{eq:exponential_taylor}
\exp{\left( \Delta t \op{H} \right)}  \approx \sum_{n=0}^{N} \frac{\Delta t^n}{n!}\op{H}^n  = 1 + \Delta t \op{H} + \frac{\Delta t^2}{2}\op{H}^2+\hdots
\end{equation}
where the expansion contains $N+1$ terms to the $N$th order. Since this method contains matrix products with $\op{H}$ it is an explicit method. Therefore it is subject to the stability condition $\Delta t \left|\left| \op{H} \right|\right| < 1$, where $ \left|\left| \op{H} \right|\right|$ is the so-called operator (matrix) norm.\footnote{\revise{In practice we compute this norm by taking the  maximum of absolute column sum of the matrix corresponding to $\op{H}$.}} For the Maxwell equations, this stability limit is related to the Courant–Friedrichs–Lewy time step $t_{\rm CFL}$ \revise{in a given dimension}. The propagation stability also depends on the expansion order $N$: ($i$) the Taylor expansion is only stable for orders divisible by 4, ($ii$) and the value of the stable time step increases if we increase the exponential order due to its higher degree of convergence. %\revise{An alternative way to achieve stability for larger $\Delta t$ steps is to factorize the exponential in Eq. (\ref{eq:exponential_taylor}) into smaller substeps without affecting the rest of the system.}

Notice that the Taylor-expansion does not conserve physical properties, but offers error suppression and convergence by increasing order of the expansion or by decreasing $\Delta t$. In contrast to Yee-methods where one performs 1 matrix multiplication by $\op{H}$ per propagation step, here we could perform 4, 8 or 12... to achieve superior accuracy. We use expansion Eq. (\ref{eq:exponential_taylor}) for solving the Maxwell equations, we continue its analysis using simplified model in Section \ref{subsubsec:exponential_advect} to get its dispersion curve.

\subsubsection{Split-operator methods}

\revisee{For solving partial differential equations, there is an important problem of how to factorize exponentials in a general manner with {\it split operator methods} \cite{bandrauk2013splitting, majorosi2016tdsecoulomb}.} Let us briefly summarize the concepts of these in the following. Let $\op{H} = \op{H}_1 + \op{H}_2$ then
\begin{equation} \label{eq:exponential_split1}
\exp \left(\Delta t \op{H} \right) =  
\exp \left(\Delta t \op{H}_1 \right) 
\exp \left(\Delta t \op{H}_2 \right), 
{\rm  \ if \ \ }
\left[\op{H}_1,\op{H}_2 \right] = 0,
\end{equation}
where the expression $\left[\op{H}_1,\op{H}_2 \right]$ stands for the $\op{H}_1\op{H}_2-\op{H}_2\op{H}_1$  commutator. If $\op{H}_1$ and $\op{H}_2$ commute then Eq. (\ref{eq:exponential_split1}) is exact, otherwise it is first order accurate in $\Delta t$. \revisee{This also shows we could just factorize any exponential to arbitrary number of little timesteps when $\op{H}_1=\op{H}_2$. A useful second order splitting formula is given by:}  
\begin{equation} \label{eq:exponential_split2}
\exp \left(\Delta t \op{H} \right) \approx
\exp \left(\Delta t \op{H}_1/2 \right) 
\exp \left(\Delta t \op{H}_2 \right)
\exp \left(\Delta t \op{H}_1/2 \right).
\end{equation}
\revisee{Using this method we can split any $\op{H}$ operator into pieces to improve stability, and make individual exponentials easy to evaluate in multidimensional systems. High order generalizations of Eq. (\ref{eq:exponential_split2}) do exist even for temporal splittings for time-dependent $\op{H}$ operators \cite{bandrauk2013splitting, majorosi2016tdsecoulomb}. In plasma physics, this approach are already used for the Hamiltonian splitting of the nonrelativistic Vlasov-Maxwell system \cite{he2015hamiltonian_maxwell_vlasov}. We use exponential splitting in the implementation of absorbing layers at the boundaries (see \ref{subsubsec:absorbing_layer}).}

\subsubsection{Other exponential methods} 
\revisee{In this subsection we would like to mention a couple of exponential methods that might be of interest for electromagnetic field solver codes. First, if we combine one forward and one backward $\Delta t$ step of Eq. (\ref{eq:exponential_taylor}), then we get such an expression  \cite{leforestier1991tdsecomparisom}  for $\Psi(t+\Delta t)$ where all even orders of $\op{H}$ vanish - in the second order limit this yields the conventional Yee-method if ${\bf E}$, ${\bf B}$ are temporally staggered.}

%\revisee{In this subsection we would like to mention a couple of exponential methods that might be of interest for electromagnetic %field solver codes. First, if we combine one forward and one backward $\Delta t$ step of Eq. (\ref{eq:exponential_taylor}), then in %the resulting expansion  for $\Psi(t+\Delta t)$ \cite{leforestier1991tdsecomparisom} all the even orders of $\op{H}$ vanish - in %the second order limit this yields the conventional Yee-method if ${\bf E}$, ${\bf B}$ are temporally staggered.}
\revisee{It is also possible to expand the exponential operator implicitly in such a way that it will be unconditionally stable using the {\it diagonal Pad\'e approximant} \cite{dijk2007tdseaccurate} of the exponential, which in the second order is the Crank-Nicolson (or Cayley) formula:
\begin{equation} \label{eq:exponential_CN2}
\exp{\left( \Delta t \op{H} \right)}  \approx \left(1-\Delta t \op{H}/2   \right)^{-1} \left(1+\Delta t \op{H}/2   \right).
\end{equation} 
As evaluating of this formula requires operator inversion (or the solution of system of linear equations), it is weakly parallelizable and non-local. It can be used in small or one dimensional problems effectively. In plasma physics, this type of exponential can be found in the Boris, Higuera-Cary and similar particle pushers (see Section \ref{subsec:particles_push}).  
}

\subsection{Dispersion relation for the Taylor-expansion} \label{subsubsec:exponential_advect}
For our general explicit solver, which we base on Taylor expansion Eq. (\ref{eq:exponential_taylor}) we proceed with a quantitative analysis on how this series expansion behaves on the temporal domain. The quantity of interest is the $\omega(\kk)$ dispersion curve of the one dimensional (1D) wave equation. Since the Taylor-expansion has complicated matrix form for the wave equation, we model the problem using $\partial _t \psi = -\partial_x \psi$ advection equation. This models the wave propagation in one direction for a single (complex) field $\psi$. It can be shown that in spectral space the 1D wave equation can be rewritten as two complex 1D advection equations.

To perform the Neumann wave analysis we transform the advection equation to spectral space
($\partial_x \rightarrow \ii \kk'$)  first:
\begin{equation}  \label{eq:exponential_advect}
\partial_t \psi = -\ii \kk'\psi,
\end{equation}
then we use its exponential solution and temporal Fourier-transform to get the dispersion relation
\begin{equation}  \label{eq:exponential_advectom}
\omega = \ii \ln \left( \exp (-\ii \Delta t \kk') \right)/\Delta t,
\end{equation}
where $\kk'=g(\kk)$ designates the spectral function of the differential, $\Delta t$ is the time step. For the exact derivative $\kk' = \kk$. In Section \ref{subsec:differences}, we will show the spectral functions of various finite difference approximations.

For the analysis, we substitute the exp function in Eq. (\ref{eq:exponential_advectom}) with a given order of its Taylor-expansion  while keeping the logarithm analytic.

\begin{figure*}[tp!]
\centering
\includegraphics[width=5.5cm]{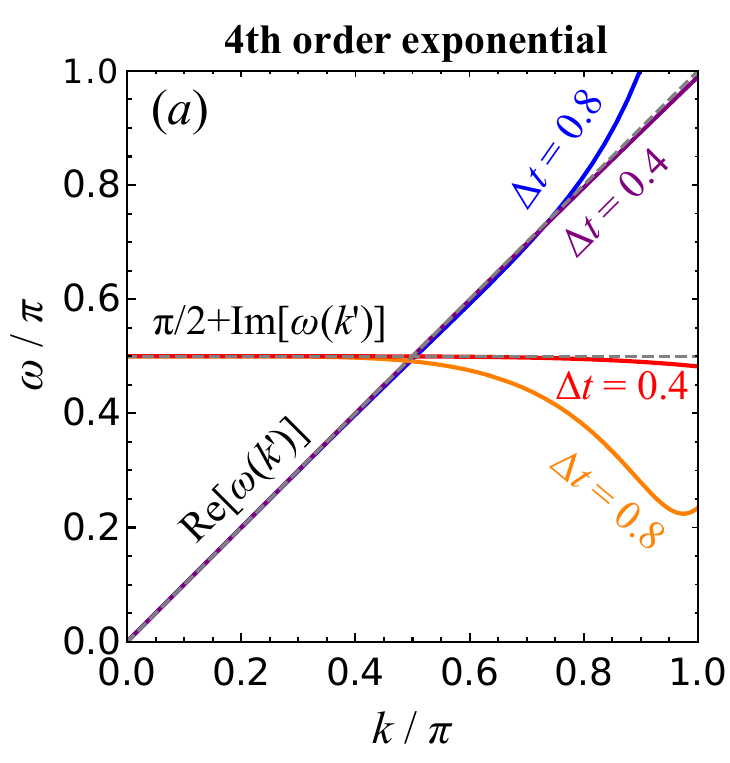}
\includegraphics[width=5.5cm]{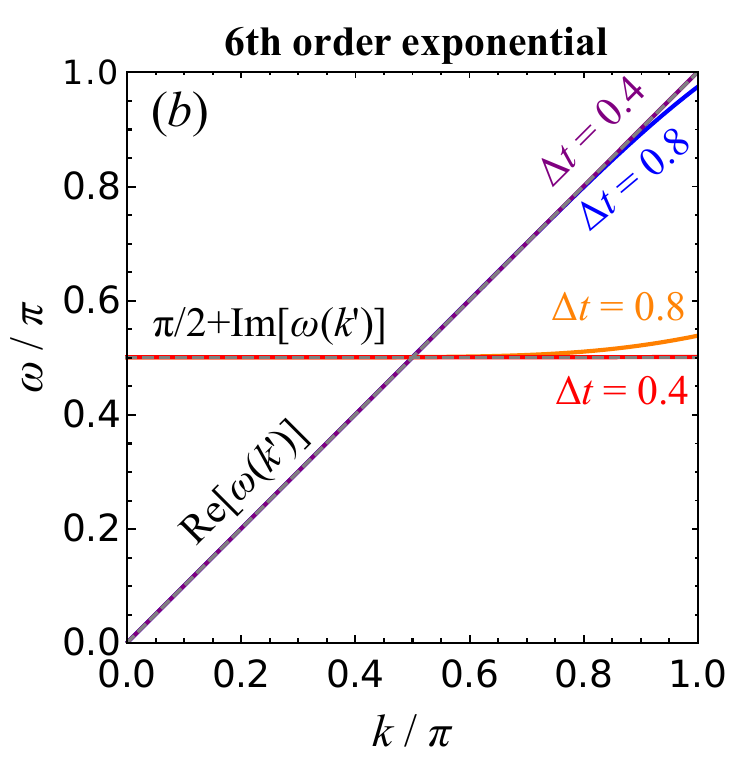}
\includegraphics[width=5.5cm]{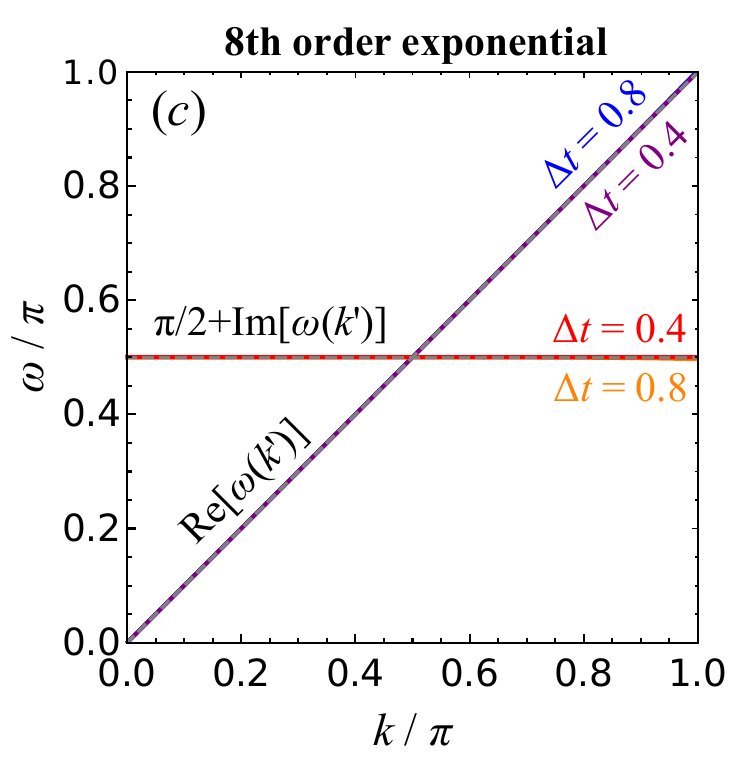}
\caption{ The $\omega(\kk')$ dispersion curves of the spectral advection equation Eq. (\ref{eq:exponential_advect}) using 4th $(a)$, 6th $(b)$ and 8th $(c)$ order Taylor expansion of the exponential with time steps $\Delta t = 0.8$ (blue and orange curves) and $\Delta t = 0.4$ (purple and red curves). For the derivative, the ideal spectral representation $\kk' = \kk$ was used. We plot two sets of curves on each figure where one set is corresponding to the real part (in blue and purple) and to imaginary part (in orange and red) of $\omega(\kk)$, the latter is up shifted by the value of 0.5. %The $k$ axis is shown in normalized $\pi/\Delta x$ units and the $\omega$ values are shown in normalized $\pi/\Delta t$ units.
\label{fig:exponential_dispersion}}
\end{figure*}

In Fig. \ref{fig:exponential_dispersion} we plot the real and imaginary parts of dispersion curve $\omega(\kk)$ for 4th, 6th, 8th order Taylor-expansion of the exponential using $\kk'=\kk$ and with two different time steps near the stability limit. We can see that the real part of the dispersion is close to the ideal $\omega=\kk$ line but the 4th order has slight overshoot at the highest frequencies. We can immediately see that the primary problem will be the imaginary part of the dispersion which means we have wave attenuation or amplification at higher frequencies. We can see that the 6th order expansion has wave amplification for the highest frequencies which indicates an unstable solution (converges downward from positive values). In contrast the 4th and 8th order exponentials has high frequency attenuation (converges upwards from negative values), which results in an inaccurate 4th order solver. These curves do show the expected degree of temporal converge by decreasing $\Delta t$, and increasing expansion order by 4 almost eliminates attenuation issues. This shows that the Taylor expansion is only useful in the high order limit. 
\revise{We note that the high order expansion of the exponential propagator is also responsible for correcting the anisothropic field propagation issues in multiple dimensions \cite{sekido2024accuracy_difference}.}

Overall, the high-order exponential propagation will inherit its (real) dispersion curve from the spectral function $\kk'=g(\kk)$ of the derivative. High frequency attenuation can even viewed as beneficial, it could reduce noise or artifacts at those frequencies.

\section{Spatial representation}  \label{sec:spatial} 

\subsection{Banded diagonal matrices} \label{subsec:bandedmatrices}
 
The most important mathematical tool that we use during actual computing of the field propagation are banded diagonal matrices \cite{BOOK_NUMERICAL_RECIPIES, majorosi2016tdsecoulomb}. First, we discretize the coordinates on uniform grids in each direction in 3D as:
\begin{equation} \label{eq:spatial_grid}
x_k = k \Delta x+x_{0}, \quad y_j = j \Delta y+y_{0}, \quad z_i = i\Delta z+z_{0}, 
\end{equation}
where $x_{0}, y_{0}, z_{0}$ are the grid offsets in each dimension. The indices $k\in[0, M_x-1]$, $j\in[0, M_y-1]$ and $i\in[0, M_z-1]$, where $M_x, M_y, M_z$ are the grid resolution per dimension.

A banded diagonal operator $\rm A$ corresponding to each direction $x,y,z$ has a matrix product of the form for a discretized field $\phi_{i,j,k}$:
\begin{align}
\left( {\rm A}_x \phi \right)_{i,j,k} &=  \sum_{k' = -N_{A}}^{N_{A}} {\rm A}_{x,k,k+k'}\phi_{i,j,k+k'} \quad \rightarrow \quad  {\rm A}_x \phi, \label{eq:bandedmatrices_Ax} \\
\left( {\rm A}_y \phi \right)_{i,j,k} &=  \sum_{j' = -N_{A}}^{N_{A}} {\rm A}_{y,j,j+j'}\phi_{i,j+j',k} \quad \rightarrow \quad  {\rm A}_y \phi, \label{eq:bandedmatrices_Ay} \\
\left( {\rm A}_z \phi \right)_{i,j,k} &=  \sum_{i' = -N_{A}}^{N_{A}} {\rm A}_{z,i,i+i'}\phi_{i+i',j,k} \quad \ \ \rightarrow \quad  {\rm A}_z \phi, \label{eq:bandedmatrices_Az} 
\end{align}
where $N_A$ denotes the half width of the banded diagonal matrix such that $N_A=0$ is diagonal, $N_A=1$ tridiagonal, $N_A=2$ is pentadiagonal etc. (Of course $N_A$ can be chosen to be different for each dimension.) The total number of matrix elements that are allowed to be nonzero is $2N_A+1$ in each row. Note that we can also form (banded) diagonal operators from the grids themselves, for example ${\rm x}_{k,k+k'} = \delta_{k',0} x_k$.  Note that in the sum (\ref{eq:bandedmatrices_Ax}) the index $k+k'$ could go out of bounds from $[0, M_x-1]$ (same for Eqs. (\ref{eq:bandedmatrices_Ay}), (\ref{eq:bandedmatrices_Az}) ). In actual calculations any of such terms should be ignored. 

We list two important mathematical properties of the banded diagonal operators: the matrix product of two banded diagonals      ${\rm A}_x {\rm B}_x = ({\rm AB})_x$ is a banded diagonal of half width $N_A+N_B$. The inverse of a banded diagonal matrix, however, is not a banded diagonal matrix. For the calculation of the matrix product with the inverse as ${A_x}^{-1}\phi$  we can store, however, the lower-upper triangular (LU) decomposition \cite{BOOK_NUMERICAL_RECIPIES} of $A_x$ in the form of a banded diagonal matrix of the same half width ($N_A$). Using the precomputed LU decomposition for calculating ${A_x}^{-1}\phi$ with forward and backward substitution algorithms has the same computational complexity as the matrix product ($\sim M_x N_A$).
    
Using banded diagonal matrices for any explicit (see Section \ref{subsubsec:exponential_taylor}) of iterative solver method (see \ref{subsubsec:poisson}) is very convenient because of the fast matrix multiplications. In  3D, these matrix products can be evaluated in sequence according to Eqs. (\ref{eq:bandedmatrices_Ax})-(\ref{eq:bandedmatrices_Az}). In the subsequent parts of the paper, we cast all of the one dimensional spatial operators into banded diagonal matrices.

\subsection{High-order finite differences} \label{subsec:differences}

%For the spatial representation of the derivatives we use high-order finite difference formalism \cite{BOOK_NUMERICAL_RECIPIES} which is usually derived from spatially offset  Taylor-expansion:
%\begin{equation} \label{eq:spatial_taylor}
%f_{k+k'} = f_{k} + (k+k') \Delta x  f'_{k}  + \frac{(k+k')^2}{2} \Delta x^2 f^{(2)}_{k} + \ldots,
%\end{equation}
%where we usually take a couple ($2 N_D$) of $k'$ instances  of Eq. (\ref{eq:spatial_taylor})  around point $k$ (from $-N_D$ to $N_D$)
%and express $f'_k$ with $f_{k+k'}$ by eliminating its high order derivatives $f^{(n)}_k$ to a given order (limited by $N_D$). The coefficients of $f_{k+k'}$ from the finite difference formula  are placed into the rows of the banded diagonal (in $k+k'$ column for each $k$ row). The same formula are true in dimensions $y,z$ with grid spacings $\Delta y, \Delta z$ and indices $i,j$.
%In \ref{subsubsec:lagrange} we summarize their generalized derivation using Lagrange-interpolation basis polynomials which yield the same finite difference formulas. 

\revise{For the spatial representation of the derivatives we use high-order finite differences and Lagrange-interpolation polynomial basis \cite{BOOK_NUMERICAL_RECIPIES}. We place the resulting finite difference coefficients into the columns of the banded diagonals. A $2 N_D$ order difference requires band diagonal half width of $N_D$.
} 

For the Maxwell-equations there are two distinct representations for first order derivatives that are possible, the first one is what we call \emph{centered} formulas, in the banded diagonal matrix form of:
\begin{equation} \label{eq:spatial_derivativeC}
\left( \partial_x   \phi  \right)_k \approx \sum_{k' = -N_{D}}^{N_D} {\rm D}^{(C)}_{k,k+k'} \phi_{k+k'},
\end{equation}
where its left-hand side (LHS) and right-hand side (RHS) are evaluated on the same grid. Its also true that this is anti symmetric $ {\rm D}^{(C)} = - \left( {\rm D}^{(C)}\right)^T$ when transposed\revise{.} %At first, this one seems to be a good discretization method.

The second representation differentiates the right hand side and staggers the grid on the left hand side (others also call this a centered formula, but we refer to this as \emph{staggered} ):
\begin{equation} \label{eq:spatial_derivativeSt}
\left( \partial_x   \phi  \right)_{k\pm1/2} \approx \sum_{k' = -N_{D}}^{N_D} {\rm D}^{(\pm)}_{k,k+k'} \phi_{k+k'},
\end{equation}
This version is used in conventional Yee Maxwell-solvers  \cite{arber2015pic_epoch, derouillat2018pic_smilei} while the other one is used by FBPIC with domain decomposition \cite{jalas2017pseudo_spectral, kirchen2020spectral_galilean}. We also note that Eq. (\ref{eq:spatial_derivativeSt}) is not applicable using Galilean-frame in the Maxwell-solver.  Is is also true for its transposition that $ {\rm D}^{(\pm)} = - \left( {\rm D}^{(\mp)}\right)^T$. We list the finite difference matrix coefficients corresponding to 10th order approximation in Table \ref{tab:spatial_deriv}.

We also provide form of the second order derivative for analysis purposes. It can be defined, however, by two different ways:
\begin{equation} \label{eq:spatial_derivative2}
\left( \partial_x^2   \phi  \right)_k 
\approx \sum_{k' = -N_{D}}^{N_D} \left( {\rm D}^{2} \right) _{k,k+k'} \phi_{k+k'}
\approx \sum_{k' = -2N_{D}}^{2N_D} \left(  {-\rm D}^{T} {\rm D}^{} \right) _{k,k+k'} \phi_{k+k'} 
\end{equation}
where the RHS and LHS are always defined on the same grid. Here ${\rm D^2}$ is calculated from the finite-differences of the second order derivative, whereas the other one ${-\rm D}^{T} {\rm D}^{}$ is the "square" of the first order derivative operators (this formula is suitable for the staggered derivatives). This latter occurs during the solution of the Maxwell-equations Eqs. (\ref{eq:maxwell_Ex})-(\ref{eq:maxwell_Bz}).

To decide between the first order difference types, we need to analyze the spectral behavior of Eq. (\ref{eq:spatial_derivative2}) using  $\widetilde{\rm D}^2={-\rm D}^{T} {\rm D}$ with ${\rm D} = {\rm D}^{(C)}$ and ${\rm D} = {\rm D}^{(\pm)}$.  Here, we get the spectral function from the square root of $\widetilde{\rm D}_2$. Each row of $\widetilde{\rm D}^2$ is symmetric, and in $\exp(\ii \kk ( k \Delta x) )$ Fourier basis it has spectral function of 
\begin{equation} \label{eq:spatial_spectrum2}
g^2(\kk) = \widetilde{{\rm D}}^{2}_{k,k}+2\sum_{k'=1}^{2N_D}  \widetilde{\rm D}^2_{k,k+k'}\cos(k' \kk\Delta x),
\end{equation}
where the symbol $\kk$ refers to the spectral wave number, and the $k, k'$ are indices in the $x$ dimension. (The imaginary unit is denoted by the symbol $\ii$, and grid indices in the $z$ dimension are by $i,i'$. In most cases the exact meaning can be inferred from the context.) As we have shown in Section \ref{subsubsec:exponential_advect} the real part of the 1D dispersion curve inherits the curve of the $g^2(\kk)$ spectral function of the finite differences.

\begin{figure*}[tp!]
\centering
\includegraphics[height=5.5cm]{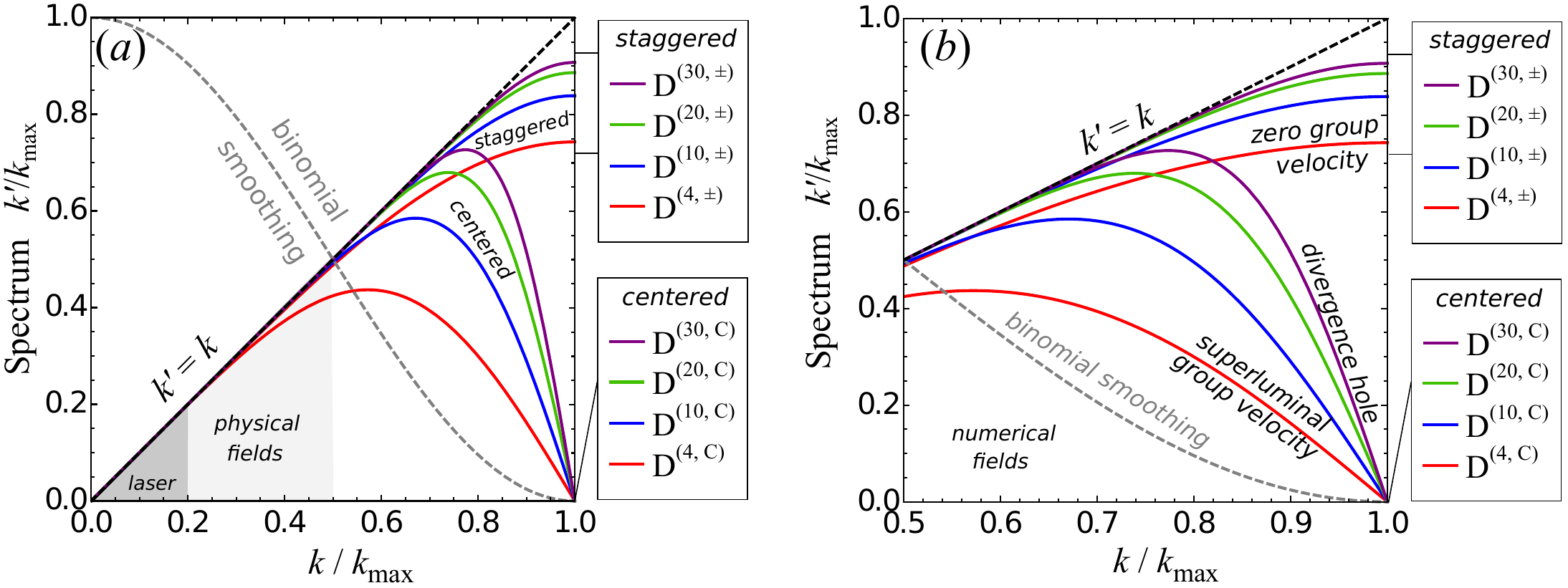}
\caption{ The $\kk' = \sqrt{g^2(\kk)}$ spectral curves Eq. (\ref{eq:spatial_spectrum2}) for centered Eq. (\ref{eq:spatial_derivativeC}) and staggered Eq. (\ref{eq:spatial_derivativeSt}) finite difference operators using 4th (red), 10th (blue), 20th (green) and 30th (purple) order accurate formulas. We also show the attenuation function of the binomial filter with dashed gray lines for reference. Here $\kk_{\max}=\pi/\Delta x$ (i.e. $\lambda_{\max}= 2 \Delta x$) corresponds to the largest frequency representable on the grid. The laser pulse is typically discretized below $0.2\kk_{\max}$ (which equals $\Delta x = \lambda_0/10$).  We show the spectral range $ > 0.5\kk_{\max}$ on the left ($b$) which could contain various numerical defects depending on the finite-difference formula. Temporally, the exponential  propagation inherits the real part of $\omega(\kk)$ curve from the  $\kk'(\kk)$ spectral functions of the differences shown here.
\label{fig:spatial_deriv}}
\end{figure*}

On Fig. \ref{fig:spatial_deriv} we show the spectral functions using Eq. (\ref{eq:spatial_spectrum2}) for centered Eq. (\ref{eq:spatial_derivativeC}) and staggered Eq. (\ref{eq:spatial_derivativeSt}) finite difference operators with various orders of  accuracy. We can see that the formulas show the expected degree of convergence with the accuracy order.   However, the centered differences go to zero at $\kk_{max}$ which is a bad behavior. For the solution of Maxwell-equations Eqs. (\ref{eq:maxwell_Ex})-(\ref{eq:maxwell_Bz}), this formula renders the highest spatial frequencies invisible in the solver, resulting two kind of defective behavior: first there will be spurious numerical longitudinal  (divergence) components for the ${\bf E}, {\bf B}$ fields that are invisible to Poisson's equation; and any numerical noise at the highest frequencies propagate much faster than the speed of light (i.e. near instantaneous boundary-boundary reflections). This erroneous behavior makes it evident that we need to use the staggered representation. However, due to the non-exactness even of the staggered finite differences, the highest frequencies should be filtered out from the sources (${\bf J}$) to suppress artifacts (numerical Cherenkov radiation).

To use the staggered spatial representation, \revise{we can also make a banded diagonal matrix that staggers or destaggers the fields (grids) using the high-order Lagrange-interpolation basis polynomials \cite{BOOK_NUMERICAL_RECIPIES}, } 
%(\ref{subsubsec:lagrange}), 
in the form of:
\begin{equation} \label{eq:spatial_stagger}
\phi_{k\pm1/2} \approx \sum_{k' = -N_{D}}^{N_D} {\rm S}^{(\pm)}_{k,k+k'} \phi_{k+k'}.
\end{equation}
We list their matrix coefficients using 10th order approximation in Table \ref{tab:spatial_deriv}. Due to the high-order nature of the this interpolation, we can do this  with minimal accuracy loss (using the same order as the finite difference). This can be convenient when providing input or output for the field solvers. An important property is that $ {\rm S}^{(\pm)} = \left( {\rm S}^{(\mp)}\right)^T$, using which we can approximately reverse its effect as ${\rm S}^{(+)} {\rm S}^{(-)} \approx 1$. The spectral functions of these staggering operators have the same kind of hole at $\kk_{\max}$ as the centered differences (we show the spectral function of the 20th order stagger operator in Fig. \ref{fig:spatial_filter}). However, for consecutively switching back and forth between staggered and non-staggered representation, one must use the inverse of ${\rm S}^{(-)}$ to prevent the accumulation of numerical errors. We set the convention here  that the operators ${\rm D}^{(-)},{\rm S}^{(-)}$ (which stagger down by $-\Delta x /2$) act on non-staggered coordinates, while the other ones ${\rm D}^{(+)},{\rm S}^{(+)}$ act on staggered coordinates.

% To complete the representation, we discuss the boundary conditions that can be locally incorporated to the edge values of banded diagonal matrices ${\rm D}^{(\pm)}$,  ${\rm S}^{(\pm)}$ that stagger (see Table \ref{tab:spatial_boundary}, which summarizes the boundary conditions for the lower edge. 
\revise{We represent the boundary conditions by extending the fields outside the domain (mapping them to the interior points) then modifying the edge values of banded diagonal matrices ${\rm D}^{(\pm)}$,  ${\rm S}^{(\pm)}$ that stagger, using the former local conditions.}
If we do not modify any edge values, we get the Neumann "zero" boundary condition (which is reflective). We can also extend the functions analytically using symmetric and anti symmetric boundary conditions.
% \revise{, which makes them continuosly differentiable}.  To incorporate \revise{the latter} into the banded diagonals, 
\revise{To do this, }however, we need to store two separate instances for each banded diagonal matrix discussed here, for example ${\rm D}^{(\pm,S)}$ denotes the symmetric and ${\rm D}^{(\pm,A)}$ the anti symmetric boundary variant of ${\rm D}^{(\pm)}$. It is also a difficulty that these latter boundary conditions are different for down (-) and for up (+) staggering operators (${\rm D}^{(\pm)}$, ${\rm S}^{(\pm)}$) due to the fact \revise{that} one acts on a non-staggered fields and the other one act on the staggered fields. The latter determines the exact form of their symmetric boundary conditions. \revise{In Table \ref{tab:spatial_boundary}, we summarize the boundary conditions for the lower edge.}

\revise{We note that it is possible to construct an entirely distinct spatial representation using the general theory of mimetic differences \cite{palha2014physics_discretization} which are based on the integral relations of the physical systems. These are the methods to optimally handle complex geometries and general curvilinear coordinates.}

\begin{table*}[tp!]
\scriptsize
%\small
\begin{tabular*}{\linewidth}{@{\extracolsep{\fill}} cccccccccccc }

Matrix & $-5$ & $-4$ & $-3$ & $-2$ & $-1$ & $0$ & $1$ & $2$ & $3$ & $4$ & $5$  \tabularnewline
\hline 
$\Big. {\rm D}^{(10,C)}  \Delta x$   
& $-0.000794$  & $ 0.009921$ & $-0.059524$ & $ 0.238095$ & $-0.833333$ 
& $ 0$
& $ 0.833333$  & $-0.238095$ & $ 0.059524$ & $-0.009921$ & $ 0.000794$
\tabularnewline
$\Big. {\rm D}^{(10,-)}  \Delta x$   
& $-0.000119$  & $ 0.001766$ & $-0.013843$ & $ 0.089722$ & $-1.211243$ 
& $ 1.211243$
& $-0.089722$  & $ 0.013843$ & $-0.001766$ & $ 0.000119$ & $ 0$
\tabularnewline
$\Big. {\rm D}^{(10,+)}  \Delta x$   
& $ 0$ & $-0.000119$  & $ 0.001766$ & $-0.013843$ & $ 0.089722$ & $-1.211243$ 
& $ 1.211243$
& $-0.089722$  & $ 0.013843$ & $-0.001766$ & $ 0.000119$ 
\tabularnewline
$\Big. {\rm D}^{2,(10)}  \Delta x^2$   
& $ 0.000318$  & $-0.004960$ & $ 0.039683$ & $-0.238095$ & $ 1.666667$ 
& $-2.927222$
& $ 1.666667$  & $-0.238095$ & $ 0.039683$ & $-0.004960$ & $ 0.000318$
\tabularnewline
$\Big. {\rm S}^{(10,-)} $   
& $ 0.000534$  & $-0.006180$ & $ 0.034607$ & $-0.134583$ & $0.605621$ 
& $ 0.605621$
& $-0.134583$  & $ 0.034607$ & $-0.006180$ & $ 0.000534$ & $ 0$
\tabularnewline
$\Big. {\rm S}^{(10,+)} $   
& $ 0 $  & $ 0.000534$ & $-0.006180$ & $ 0.034607$ & $-0.134583$ 
& $0.605621$
& $0.605621$  & $-0.134583$ & $ 0.034607$ & $-0.006180$ & $ 0.000534$
\tabularnewline
\hline
\end{tabular*}

\caption{Table of the matrix coefficients of the central finite difference formula Eq. (\ref{eq:spatial_derivativeC}) in the 1st row, the staggered finite difference formula Eq. (\ref{eq:spatial_derivativeSt}) in the 2nd and 3rd rows, the second order finite difference formula Eq.  (\ref{eq:spatial_derivative2}) in the 4th row, and staggered interpolator Eq. (\ref{eq:spatial_stagger}) in the 5th and 6th rows using 10th order polynomial approximation ($N_D=5$) up to seven digits. The first row shows the $k'$ index offset in relation to the main diagonals of the respective banded matrices.   \label{tab:spatial_deriv}}
\end{table*}

\begin{table}[ht]
\begin{tabular*}{\linewidth}{@{\extracolsep{\fill}} rcr }

 Boundary name  &  Definition &  Derivative \quad \quad \quad \tabularnewline
\hline 
anti symmetric   & $ \phi_{-k'} = - \phi_{k'} $  &  $  {\rm D}_{k,k'}^{(-)} -  {\rm D}_{k,-k'}^{(-)}  = {\rm D}_{k,k'}^{(-, A)}$      \\
symmetric   & $ \phi_{-k'} =  \phi_{k'} $  &  $  {\rm D}_{k,k'}^{(-)}+ {\rm D}_{k,-k'}^{(-)} = {\rm D}_{k,k'}^{(-,S)}$       \\
anti symmetric (st)   & $ \phi_{-k'} = - \phi_{k'-1} $  &  $  {\rm D}_{k,k'-1}^{(+)} -  {\rm D}_{k,-k'}^{(+)} = {\rm D}_{k,k'-1}^{(+,A)}$       \\
symmetric (st)  & $ \phi_{-k'} =  \phi_{k'-1} $  &  $ {\rm D}_{k,k'-1}^{(+)}+ {\rm D}_{k,-k'}^{(+)} = {\rm D}_{k,k'-1}^{(+,S)} $       \\
zero   & $ \phi_{-k'} = 0 $  &  not modified 
\\
constant   & $ \phi_{-k'} =  \phi_{0} $  &  $   {\rm D}_{k,0}+ \sum_{k'>0} {\rm D}_{k,-k'} =  \tilde{{\rm D}}_{k,0}$       \\
\hline
\end{tabular*}
\caption{List of local boundary conditions that can be incorporated into banded diagonal matrices for the lower edge (near $k = 0$) for $k'>0$ the definitions of which are shown in the second column for discretized $\phi_{k'}$ function.  Note that the symmetric boundaries have different formulas for staggered (st) and non-staggered grids. The third column lists the modification of the edge elements of  ${\rm D}^{(\pm)}$ banded matrices. {\it Note to the 3rd column}: We store the banded diagonal matrices as $M_x \times (2N_D+1)$ arrays which contain the full finite differences at lower (upper) edges as extra elements. These extra elements correspond to the matrix indices $k,-k'$ (at the lower edge) but are not taken into account during the calculation of a matrix product or inversion. For local boundary conditions, we "fold" those extra values back to the actual banded diagonal matrix elements.   The upper edge boundaries near $k = M_x$ with $k+k'$ is in direct analogy with the formulas showed here.
\label{tab:spatial_boundary}}
\end{table} 

\subsection{Current filtering} \label{subsec:spatial_current}

As we have shown in Fig. \ref{fig:spatial_deriv} using staggered finite differences always result in erroneous behavior near the highest frequency $\kk_{\max}$ at finite polynomial orders. Any such a high frequency  wave has nearly zero group velocity, which could lead to numerical Cherenkov like radiation during field propagation. At reasonably high orders this is not an issue below $0.5 \kk_{\max}$: for example the laser pulse will not exhibit dispersion errors (using high order expansion of the exponentials). However, this issue is intriguing because it cannot be fixed within spatial representation using arbitrary high-order (local) finite difference approximations.

To address this issue the usual (and here the seemingly only) method is to apply digital filters (smoothing) \cite{vay2011method_instability, godfrey2014stability_spectral, li2017maxwell_cherenkov} to the sources, in particular, to the current density ${\bf J}$. One might argue that just use higher resolution, but that only shifts physical waves towards lower frequencies relative to $\kk_{\max}$ - so the flaw remains in the spectral functions. Our strategy is use high enough order derivatives (10th-32nd) to get the physical waves in the spectral range $\lesssim0.5 \kk_{\max}$ right. Then we apply a lowpass spectral filter that dampens the frequencies $\gtrsim0.5 \kk_{\max}$ (if the physical problem demands it).  These filters are, again, represented as banded diagonal matrices: 
\begin{equation} \label{eq:spatial_filter}
\left( {\rm F}   \phi  \right)_k \approx \sum_{k' = -N_{F}}^{N_F} {\mathcal{\rm F}}^{}_{k,k+k'} \phi_{k+k'}.
\end{equation}
They can be specified independently in each direction, then we can apply them using the matrix product formulas Eqs. (\ref{eq:bandedmatrices_Ax})-(\ref{eq:bandedmatrices_Az}). They are usually symmetric, and should satisfy $\sum_{k'} {\rm F}_{k,k+k'}=1$ for  each $k$ row. They also need to be stored in two instances to include the proper symmetric and anti symmetric boundary conditions as ${\rm F}^{(S)}$ and ${\rm F}^{(A)}$.

In Table \ref{tab:spatial_filter} we list the matrix elements of couple different current filters. We listed three of our custom developed lowpass filters, named lowpass A, B and C - the suppression bandwidth widen in this order. We can see that they have much increased computational cost compared to binomial smoothing filter - but this latter cost is consistent using high-order finite differences.

\begin{figure}[ht]
\centering
\includegraphics[height=6.0cm]{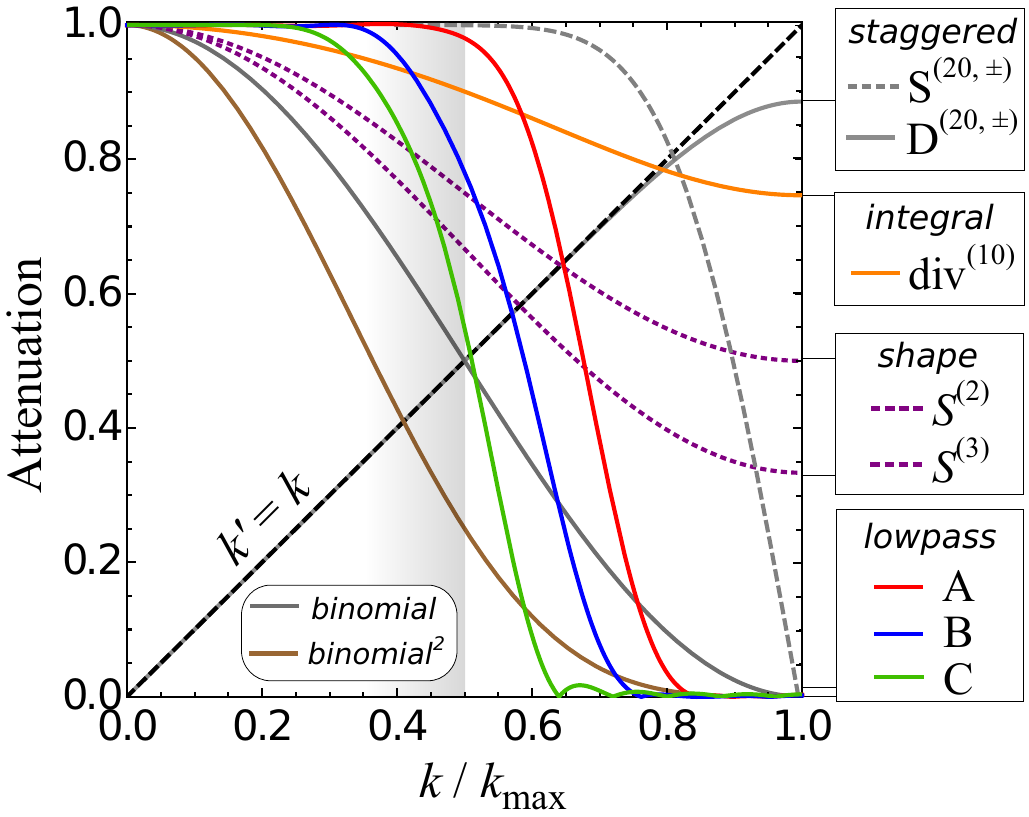}
\caption{ Spectral attenuation functions of the current filters from Table \ref{tab:spatial_filter}: binomial and squared binomial filter (gray and brown), our lowpass filters (red, blue, green), divergence correction filter (orange).
The worst case spectral functions of the 2nd and 3rd order PIC particle shapes are also shown (purple dotted lines). For reference, we also show the spectral functions of the 20th order staggering (dashed gray) and difference operator (gray).
\label{fig:spatial_filter}}
\end{figure}

We show the spectral functions of the filters on Fig. \ref{fig:spatial_filter}. We can see that the  binomial smoothing filter significantly affects our target spectral range $\lesssim0.5 \kk_{\max}$. Our lowpass filters affect this range less and they do not affect the very low frequency components. We tailored these filters especially to dampen the spectral components where the high-order difference is erroneous. Because of this, these are effective tools to suppress high frequency numerical Cherenkov radiation within PIC implementation. Of course any filter can be combined (matrix multiplied) with any other to increase the effectiveness of the damping. For reference, we also plotted the 2nd or 3rd order PIC particle shapes (see \ref{subsubsec:shapes}) which have some spectral dampening built in - the discussed current filters will be applied on the top of these.\footnote{More precisely, the spectrum of the deposited current of the particles depends on their position relative to the grid points, which is the consequence of the aliasing error of the discretized particle shape function.}

\begin{table*}[tp!]
\scriptsize
%\small
\begin{tabular*}{\linewidth}{@{\extracolsep{\fill}} rcccccccccc }

Matrix & $0$ & $\pm1$ & $\pm2$ & $\pm3$ & $\pm4$ & $\pm5$ & $\pm6$ & $\pm7$ & $\pm8$ & $\pm9$  \tabularnewline
%\myarrow & $\pm10$ & $\pm11$ & $\pm12$ & $\pm13$ & $\pm14$ & $\pm15$ & $\pm16$ & $\pm17$ & $\pm18$ & $\pm19$  
%\tabularnewline
\hline 
$\Big.$ binomial   
& $ 0.5$  & $ 0.25$ & $ 0$ & $ 0$ & $0$ & $ 0$ & $ 0$  & $0$ & $0$ & $0$
\tabularnewline
$\Big.$ lowpass A
& $ 0.6705591$   & $ 0.2659328$ 
& $-0.1250962$   & $ 0.0047488$ 
& $ 0.0404772$   & $-0.0284749$
& $ 0.0046729$   & $ 0.0064292$ 
& $-0.0061203$   & $ 0.0021509$ 
\tabularnewline
$\Big.$ lowpass B 
& $ 0.5765593$   & $ 0.2960557$ %
& $-0.0638091$   & $-0.0506896$ %
& $ 0.0347840$   & $-0.0000579$ %
& $-0.0086651$   & $ 0.0070747$ %
& $-0.0024724$   & $-0.0029338$ %
\tabularnewline
$\Big.$ lowpass C   
& $ 0.4963604$   & $ 0.3074442$ %
& $ 0.0005223$   & $-0.0776277$ & $ 0.0057427$ %
& $ 0.0265352$   & $-0.0094575$ 
& $-0.0073647$   & $ 0.0081750$ 
& $ 0.0001465$   
\tabularnewline 
$\Big.$ div-10   
& $ 0.8871747$   & $ 0.0625199$ 
& $-0.0069063$   & $ 0.0008591$ 
& $-0.000060$   & $0$
& $0$   & $0$ 
& $0$   & $0$ 
\tabularnewline
$\Big.$ shape-3*  
& $ 0.6666667$   & $ 0.1666667$ 
& $0$   & $0$ 
& $0$   & $0$
& $0$   & $0$ 
& $0$   & $0$ 
\tabularnewline
\hline
$\Big.$ Matrix [\myarrow] & $\pm10$ & $\pm11$ & $\pm12$ & $\pm13$ & $\pm14$ & $\pm15$ & $\pm16$ 
& $\pm17$ & $\pm18$ & $\pm19$ 
\tabularnewline
$\Big.$ lowpass B
& $ 0.0024339$   & $0$ 
& $0$   & $0$ 
& $0$   & $0$
& $0$   & $0$ 
& $0$   & $0$ 
\tabularnewline
$\Big.$ lowpass C
& $-0.0038714$ & $0.0017115$   & $-0.0001363$ 
& $0$   & $0$ 
& $0$   & $0$
& $0$   & $0$ 
& $0$    
\tabularnewline
\hline
\end{tabular*}

\caption{Table of the matrix coefficients (kernels) of current filter types for Eq. (\ref{eq:spatial_filter}) up to seven digits. The 1st and 8th rows show the $k'$ index offset in relation to the main diagonals of the respective banded matrices.  The div-10 filter refers to 10th order integral  filter used during divergence correction of the current (see Section \ref{subsec:particles_deposit}). Though it is not a current filter, we listed an evaluated third order particle shape for reference (*). The coefficients may differ slightly in our final code implementation.   \label{tab:spatial_filter}}
\end{table*}

\subsection{Discrete Maxwell equations} \label{subsec:spatial_maxwell}

If the fields ${\bf E}^{(C)}$ and ${\bf B}^{(C)}$ are not staggered initially, then we stagger them to be on the Yee-grid (downward)  with the following operation:
\begin{align} \label{eq:spatial_stagger_E}
E_x &= {\rm S}_x^{(-)} E_x^{(C)},\quad  E_y = {\rm S}_y^{(-)} E_y^{(C)}, \quad E_z = {\rm S}_z^{(-)} E_z^{(C)}  \\
B_x &= {\rm S}_y^{(-)}{\rm S}_z^{(-)} B_x^{(C)},\  B_y = {\rm S}_x^{(-)} {\rm S}_z^{(-)} B_y^{(C)},\  B_z = {\rm S}_x^{(-)} {\rm S}_y^{(-)} B_z^{(C)},  \label{eq:spatial_stagger_B}
\end{align}
where ${\rm S}_{x}^{(-)}, {\rm S}_{y}^{(-)}, {\rm S}_{z}^{(-)}$ are the stagger matrices with anti symmetric boundary conditions (same for ${\rm S}_{x}^{(+)}, {\rm S}_{y}^{(+)}, {\rm S}_{z}^{(+)}$). To (approximately) reverse staggering we change all ${\rm S}^{(-)}$ to ${\rm S}^{(+)}$ and matrix multiply them with the Yee staggered components. We keep the ${\bf E}, {\bf B}$ fields staggered for the Maxwell-solver and only destagger them in a separate instance when needed. 

Before placing the current density $\bf J$ into the numerical solution we  filter it with ${\rm F}_x$, ${\rm F}_y$ or ${\rm F}_z$ filter matrices, in the general form of:
\begin{align} \label{eq:spatial_filter_Jxy}
\overline{J}_x =   {\rm F}_z^{(S)}  {\rm F}_y^{(S)} {\rm F}_x^{(A)} & {\rm F}_x^{(\text{div},A)}  J_x , \quad 
\overline{J}_y =   {\rm F}_z^{(S)} {\rm F}_x^{(S)} {\rm F}_y^{(S)} {\rm F}_y^{(\text{div},A)} J_y, \\
\overline{J}_z &=   {\rm F}_x^{(S)} {\rm F}_y^{(S)} {\rm F}_z^{(A)} {\rm F}_z^{(\text{div},A)} J_z.  \label{eq:spatial_filter_Jz}
\end{align}
These also only involve matrix multiplication with banded diagonal matrices.
As we have seen in Section \ref{subsec:spatial_current} filtering the current may be necessary to dampen numerical artifacts. The optional integral filter ${\rm F}_x^{(\text{div},A)}$ may augment PIC charge conserving deposition that we use (see Section \ref{subsec:particles_deposit}). Filtering the current do not affect other parts of the solver.

To discretize the Maxwell-equations Eqs. (\ref{eq:maxwell_Ex})-(\ref{eq:maxwell_Bz}) we replace the analytical derivatives in them as $\partial_x \rightarrow {\rm D}_x$,  $\partial_y \rightarrow {\rm D}_y$, $\partial_z \rightarrow {\rm D}_z$, and carefully taking into account the Yee-staggering. Regarding the boundary conditions, we mark the derivatives in staggered dimensions as anti symmetric, and we mark them as symmetric in the other dimensions. Then, the Amp\'ere-equations become:
\begin{align}
\partial_t E_x &= -{\rm D}_z^{(+,A)} B_y +{\rm D}_y^{(+,A)} B_z - \overline{J}_x ,\label{eq:spatial_maxwell_Ex}\\
\partial_t E_y &=  {\rm D}_z^{(+,A)} B_x -{\rm D}_x^{(+,A)} B_z - \overline{J}_y ,\label{eq:spatial_maxwell_Ey}\\
\partial_t E_z &= -{\rm D}_y^{(+,A)} B_x +{\rm D}_x^{(+,A)} B_y - \overline{J}_z ,\label{eq:spatial_maxwell_Ez},
\end{align}
where $\overline{\bf J}$ must have the same the staggering and symmetry as $\bf E$.
The discrete Faraday-equations become:
\begin{align}
\partial_t B_x &=  {\rm D}_z^{(-,S)} E_y -{\rm D}_y^{(-,S)} E_z ,\label{eq:spatial_maxwell_Bx}\\
\partial_t B_y &= -{\rm D}_z^{(-,S)} E_x +{\rm D}_x^{(-,S)} E_z ,\label{eq:spatial_maxwell_By}\\
\partial_t B_z &=  {\rm D}_y^{(-,S)} E_x -{\rm D}_x^{(-,S)} E_y .\label{eq:spatial_maxwell_Bz}
\end{align}
Interestingly, even though ${\bf E}$ and ${\bf B}$ are defined on the spatial Yee-grid, they are temporally synchronous. This avoids temporal interpolation of ${\bf B}$ fields, which is known to cause accuracy loss during the PIC field interpolation step in the traditional PIC codes \cite{lehe2016fbpic}.
In summary, Eqs. (\ref{eq:spatial_maxwell_Ex})-(\ref{eq:spatial_maxwell_Bz}) yield the matrix form of the $\op{H}$  operator of Eq. (\ref{eq:maxwell_H}) as:
\begin{equation}  \label{eq:spatial_maxwell_H}
%{\rm H} =
\left( \begin{array}{cccccc}
0            & 0         & 0       &      0      & -{\rm D}_z^{(+,A)} &  {\rm D}_y^{(+,A)} \\
0            & 0         & 0       &  {\rm D}_z^{(+,A)} & 0           & -{\rm D}_x^{(+,A)} \\
0            & 0         & 0       & -{\rm D}_y^{(+,A)} & {\rm D}_x^{(+,A)}  & 0           \\
0            &  {\rm D}_z^{(-,S)} & -{\rm D}_y^{(-,S)} & 0          & 0         & 0        \\
 -{\rm D}_z^{(-,S)} & 0           &  {\rm D}_x^{(-,S)} & 0          & 0         & 0         \\
 {\rm D}_y^{(-,S)} & -{\rm D}_x^{(-,S)}  & 0           & 0          & 0         & 0         \\
\end{array} \right) .
\end{equation}

For completeness, we provide the discrete expressions for $\nabla \cdot {\bf E}$ and for the gradient of a  scalar potential $\phi$:
\begin{align} \label{eq:spatial_maxwell_divE}
{\rm D}_x^{(+,A)} E_x + {\rm D}_y^{(+,A)} E_y +{\rm D}_z^{(+,A)} E_z &= \varrho, \\
\left( {\rm D}_x^{(-,S)}  \phi \right) {\bf \hat{e}_x} + \left({\rm D}_y^{(-,S)}  \phi \right) {\bf \hat{e}_y} + \left( {\rm D}_z^{(-,S)}  \phi \right) {\bf \hat{e}_z}  &= \nabla \phi  = - {\bf E}^{(L)}.  \label{eq:spatial_maxwell_gradphi}
\end{align}
Both the scalar potential $\phi$  and the charge density  $\varrho$ are completely non-staggered. Thus, the same is true for the Poisson's equation. \revisee{In this formalism imposing reflectional symmetry on the Maxwell-system is handled only by selecting the indicated operator matrix variants with the $S,A$ symmetric boundary pre-applied.} %(This halves the number of grid points needed to propagate linearly polarized laser fields). 

% In Eqs. (\ref{eq:spatial_filter_Jxy})-(\ref{eq:spatial_maxwell_Bz}) the right-hand sides are fully discretized, and contain only matrix products with banded diagonal matrices, the elements of which we calculate 

\revisee{{\it Our exponential solution can be summarized as follows}: At the initialization step, we calculate all banded diagonal matrices in Eqs. (\ref{eq:spatial_stagger_E})-(\ref{eq:spatial_maxwell_gradphi}) in accordance with descriptions in Sections \ref{subsec:differences}-\ref{subsec:spatial_current} at a selected spatial order.
To step forward in time, we compute the midpoint solution Eq. (\ref{eq:exponential_solution2}) with the filtered current from Eqs. (\ref{eq:spatial_filter_Jxy}), (\ref{eq:spatial_filter_Jz}) using 4th, 8th, 12th or higher order Taylor-expansion of the exponentials as Eq. (\ref{eq:exponential_taylor}). The definition of $\op{H}\Psi$ using Eqs. (\ref{eq:spatial_maxwell_Ex})-(\ref{eq:spatial_maxwell_H}) is known. Any matrix-vector product we encounter we compute in accordance with Eqs. (\ref{eq:bandedmatrices_Ax})-(\ref{eq:bandedmatrices_Az}).
We call this method finite difference exponential time domain (FDETD), or finite difference exponential solution.}

\revisee{For completeness, in \ref{subsubsec:spectral_maxwell} we derive the equivalent spectral exponential solution - which gives the same results if it is combined with the spectral forms of high-order finite differences and the exponential Taylor-expansion.}

\revise{We designed this algorithm with two levels of parallelization in mind when deploying it on high performance computers. At the lowest level (only this is implemented at the time of writing this paper) everything is stored in shared memory and the computation of the above exponential solution are done in 3D by a large number of parallel threads. In practice, we implemented this using OpenMP parallelization on CPU and Nvidia CUDA language on GPU. At this level the computations are
nearly perfectly parallellelized. We checked the compute scaling on the Hungarian KIFÜ Komondor HPC and various desktop PC-s and we found the performance scaling on the number of CPU cores \revise{seems to be bound by memory bandwidth, and it is sensitive to non-uniform memory access (NUMA) effects. On a desktop computer, running it on an Nvidia RTX 3080 Ti GPU the computation was about 10x faster than CPU calculations with an Intel 11700K 8 core processor. On GPU-s, however, the performance is bound by double precision floating point performance.}}

\revise{At the top parallelization level we plan to use domain decomposition using an MPI library to enable our code to be used on multiple NUMA nodes effectively and on larger clusters. To do this, we could use the ideas of localized spectral solutions   \cite{jalas2017pseudo_spectral}: during each $\Delta t$ exponential step only finite number of neighboring grid points are taken into account depending on the spatial and temporal expansion orders. For example, using the 30th order finite differences we need approximately 30 ghost cells at the domain interfaces to be synchronized to reach $10^{-14}$ accuracy. This perfectly fits the usual Yee domain decomposition methods, except we get more accurate result with more work. Unlike true spectral methods \cite{godfrey2014stability_spectral}, we avoid performing basis transform integrals across many processors this way.}

%\noindent\hrulefill

\revisee{We note that the above base algorithm can be optimized in several ways to improve simulation runtimes in practice. For example, using the combined formula of Eqs. (\ref{eq:exponential_sourceT1}), (\ref{eq:exponential_sourceT}) for the source terms removes the cost of evaluating the source exponential. It is also a possibility to include any linear combination of finite differences in the above matrices: interestingly, as suggested by \cite{li2017maxwell_cherenkov, li2021solver} the accuracy of group and phase velocities can be improved further this way (\ref{subsubsec:enhance}) without additional computational cost. To variably reduce the resolution of a simulation domain  we can straightforwardly incorporate coordinate stretching (i.e. coordinate transforms, see \ref{subsubsec:transform}) into the matrix elements of ${\rm D}^{(\pm)}_x$, ${\rm D}^{(\pm)}_y$ and ${\rm D}^{(\pm)}_z$ in such a way that Eqs. (\ref{eq:spatial_stagger_E})-(\ref{eq:spatial_maxwell_gradphi}) remain valid without any change. Alternatively, absorbing layers can be easily included in the exponential formalism as well (\ref{subsubsec:absorbing_layer}). For example, using these two  techniques, we could contain a diffracting laser pulse or absorb reflecting waves at boundaries at a reduced cost. We consider all of these as extensions to our main method described in this paper, but our code implements all these features.}

%%%%%%%%%%%%%%%%%%%%%%%%%%%%%%%%%%%%%%%%%%%%%%%%%%%%%%%%%%%%%%%%%%%%%%%%%%%%%%%%%%%%%%%%%%%%%%%%%%%%%%

\newpage

%%*********************************************************************************************

\section{Particle in cell routine}    \label{sec:PIC} 

\subsection{Quasi particle model of plasmas} \label{subsec:particles}

The kinetic description of a collisionless charged continuous medium, which we refer simply as \emph{plasma} in the following, are given by Vlasov-Maxwell system of equations \cite{shadwick2005fluid_vlasov, derouillat2018pic_smilei}. The state of species $\sss$ of particles with charge $q_\sss$ and mass $m_\sss$, that constitute plasma,
is described by its phase space distribution function $\ff_{\sss}({\bf r}, {\bf p},t)$. (We still assume normalized units, see \ref{subsubsec:units}). 
The distribution $\ff_\sss$ satisfies the relativistic Vlasov's equation:
\begin{equation} \label{eq:particles_vlasov}
\partial_{t}\ff_\sss = - \frac{\bf p} {m_s \gamma} \nabla \ff_\sss - \left[ q_\sss {\bf E} + \frac{q_s}{m_s \gamma} {\bf p} \times {\bf B} \right]  \cdot \nabla_{\bf p}  \ff_\sss,
\end{equation}
where $\gamma = \sqrt{1+ {\bf p}^2/m_\sss^2}$ and the vector field in the brackets is the Lorentz-force  ${\bf F}_L$. The electric ${\bf E}({\bf r})$, and magnetic ${\bf B}({\bf r})$ fields are considered in the mean field approximation and they evolve evolve according to the Maxwell Eqs. (\ref{eq:maxwell_Ex})-(\ref{eq:maxwell_divE}). \revise{Interestingly, He et. al. \cite{he2015hamiltonian_maxwell_vlasov} realized that the nonrelativistic Vlasov-Maxwell system can be formally described with a single Vlasov-Maxwell state, which is governed by an exponential evolution. In practice, though, this involves splitting the Hamiltonian to provide a $\sim\Delta t ^{2}$ accurate solution. However, treating the relativistic case is much more difficult, it is better just to couple the Vlasov and Maxwell subsystems through nonlinear self-consistent interactions, and provide a high quality solution for each. However, even propagating the $\ff_{\sss}({\bf r}, {\bf p},t)$ distribution directly is extremely challenging because of the high dimensionality of the problem and the possible shock propagation \cite{zalesak1979fluid_transport, filbet2001vlasov_conservative} due to the relativistic dynamics we want to enable. 
 }

%Though the solution of Eq. (\ref{eq:particles_vlasov}) is  possible with exponential operators but it would pose additional challenge because of the high dimensionality of the problem and the possible shock propagation \cite{zalesak1979fluid_transport, filbet2001vlasov_conservative} due to the relativistic dynamics we want to enable. 

Instead, the Vlasov-equation (\ref{eq:particles_vlasov}) is usually solved in the quasi particle approximation \cite{ derouillat2018pic_smilei, kraus2017gempic} with the ansatz  (also referred to as "macro-particles"): 
\begin{equation} \label{eq:particles_vlasovf}
 \ff_\sss({\bf r}, {\bf p}, t) {\rm d} V = \sum_{\pp = 1}^{M_\sss} w_{\pp} \hat{\mathcal{S}}({\bf r} - {\bf r}_\pp(t))  \delta ({\bf p} - {\bf p}_\pp(t)) {\rm d} V,
\end{equation}
here $w_\pp$ denotes the quasi-particle "weight", which in the above form is the number of actual particles in the quasi-particle. ${\rm d}V$ denotes the elementary volume the quasi-particle occupies.
After discretization, this latter is referred to the {\it cell size}. Here, $\hat{\mathcal{S}}(\bf r)$ denotes the 3D shape function of the quasi-particle, which must be local and normalized as $\int \hat{\mathcal{S}}({\bf r} -{\bf r}_\pp) {\rm d}V= 1$.
For the simplification of calculations, we assume the particle shape to be \emph{separable}: 
\begin{equation} \label{eq:particles_shape3}
\hat{\mathcal{S}}({\bf r}-{\bf r}_\pp) = \hat{S}_x(x-x_\pp) \hat{S}_y(y-y_\pp) \hat{S}_z(z-z_\pp),
\end{equation}
where  $\hat{S}$ denotes one dimensional shape function (they can be different for each direction $x,y,z$), which is also assumed to be normalized $\int \hat{S}_x(x-x_\pp) {\rm d}x= 1$. We overview the relevant formulas of the conventional PIC particle shapes in \ref{subsubsec:shapes}.

\paragraph {\bf Equations of motion} By combining Eqs. (\ref{eq:particles_vlasov}) and (\ref{eq:particles_vlasovf}) it can be derived that the quasi-particles must satisfy the relativistic equations of motion  (for species $\sss$):
\begin{align} \label{eq:particles_motionr}
\frac{{\rm d} {\bf r}_\pp}{{\rm d} t} &=  \frac{{\bf u}_\pp}{\gamma_\pp} \quad \text{with} \quad \gamma_\pp = \sqrt{1+{\bf u}_\pp^2},\\
\frac{{\rm d} {\bf u}_\pp}{{\rm d} t} &= r_\sss \left(  {\bf E}_\pp + \frac{{\bf u}_\pp}{\gamma_\pp}\times {\bf B}_\pp \right), \label{eq:particles_motionu}
\end{align}
where   $r_\sss = q_\sss/m_\sss$ the charge-over-mass ratio, ${\bf u}_\pp = {\bf p}_\pp/m_s$ is the quasi particle reduced momentum. For simplicity, we assume that the {\bf particle index $\pp$} always refers to species $\sss$.

\paragraph {\bf Field interpolation} The fields ${\bf E}_\pp$ and ${\bf B}_\pp$ are interpolated at the particle position  using the normalized shape function:
\begin{equation} \label{eq:particles_interpolateEB}
{\bf E}_\pp = \int \hat{\mathcal{S}}( {\bf r }-{\bf r}_\pp) {\bf E} ( {\bf r} ) {\rm d}V, \quad
{\bf B}_\pp = \int \hat{\mathcal{S}}( {\bf r }-{\bf r}_\pp) {\bf B} ( {\bf r} ) {\rm d}V.
\end{equation}
Using these, it is also possible to define the particle shapes from an interpolation formula.

\paragraph {\bf Charge deposition} 
The charge density $\varrho_\sss$ and current density ${\bf J}_\sss$ are reconstructed from Eq. (\ref{eq:particles_vlasovf}) for species $\sss$ as:
\begin{align} \label{eq:particles_charge}
 \varrho_\sss({\bf r}, {\bf p}, t)  = \sum_{\pp = 1}^{N_\sss} q_\sss w_{\pp} \hat{\mathcal{S}}({\bf r} - {\bf r}_\pp(t)), \\
{\rm J}_\sss({\bf r}, {\bf p}, t)  = \sum_{\pp = 1}^{N_\sss} q_\sss {\bf u}_{\pp}/{\gamma_{\pp}} w_{\pp} \hat{\mathcal{S}}({\bf r} - {\bf r}_\pp(t)), \label{eq:particles_chargeJ}
\end{align}
To get the total charge density $\varrho_\sss$ and current density ${\bf J}$ we need sum to all the $\varrho_\sss$, ${\bf J}_\sss$ for all species $\sss$. The analytical forms Eqs. (\ref{eq:particles_charge}) and (\ref{eq:particles_chargeJ}) satisfy the continuity equation of Eq.  (\ref{eq:maxwell_divJ}).

These are the fundamental quasi-particle equations. The form of quasi-particle formalism we outline in this paper separates the exponential Maxwell-solver and the plasma kinetics such that the field solver only interacts with the quasi-particles through the source $\bf J$ in the midpoint rule Eq. (\ref{eq:exponential_solution2}). This separation allows us to fine tune the numerical features of field solver and the discretized quasi-particles method independently.  In the following we meticulously go over the latter step-by-step to develop our particle-in-cell method. The particle-in-cell approximation does not take into account microscopic effects like inter-particle collisions or ionization, we plan to implement these later in adhoc modules like in EPOCH \cite{arber2015pic_epoch} or SMILEI \cite{derouillat2018pic_smilei} codes.

\subsection{Particle pusher} \label{subsec:particles_push}

\subsubsection*{Particle storage}
The particles have 7 core properties: their $w_\pp$ weight, their ${\bf r}_\pp$ position and their ${\bf u}_\pp$ reduced momenta, which we always assume to be in 3D Cartesian coordinates. We store these particle data for each species $\sss$ in a \revisee{container (class) that allows random access of any particle with index $\pp$}.  The methods described in this paper do not depend on exact implementation details, those become only relevant for particle insertion and deletion.\footnote{Our current particle storage implementation stores particles in 2 to 8  continuous arrays in the memory, which may change in the future as our code developed further.}  Any particle with $w_\pp = 0$ is completely ignored during computations - during particle insertion they could be overwritten.
%We also specify that any particle with $w_\pp = 0$ is deleted, and completely ignored during calculations - during particle insertion or sorting they are overwritten. 

\subsubsection* {Assumptions}
In every time step we need to solve the relativistic equations of motion Eqs. (\ref{eq:particles_motionr}) and (\ref{eq:particles_motionu}), the step of which is called the \emph{particle push}. We do this for each $\pp$ particle of species $\sss$ independently from each other using 3D Cartesian coordinates. For simplicity, we drop the subscript $\pp$ in the following.

%We do this in 3D Cartesian coordinates during the solution of the equations of motion. 
%We do not stray away from the conventional wisdom here. 
%We always store the particle positions ${\bf r}_\pp$, and reduced momenta ${\bf u}_\pp$ in 3D Cartesian form and

We assume that the fields $\overline{{\bf E}}$, $\overline{{\bf B}}$ represent temporal averages along the selected particle's trajectory when we advance its kinetics from time step $t_{n}$ to $t_{n+1}$ with time step $\Delta t$. For example, the fields are interpolated at temporal midpoints using the notation:
\begin{equation} \label{eq:pusher_interpolate2}
\overline{{\bf E}}_{n} = r_{\sss} \frac{\Delta t}{2} {\bf E} \left( {\bf r} \left( t_{n} +\frac{\Delta t}{2} \right) \right), \ \
\overline{{\bf B}}_{n} = r_{\sss} \frac{\Delta t}{2} {\bf B}\left({\bf r} \left(t_{n}+\frac{\Delta t}{2}\right)\right),
\end{equation}
where  we included the extra factors $r_\sss \Delta t /2$ which will simplify the formulas that follow. %In the following we work with the conventional assumption that these interpolated average fields are constant within each time step interval $n$.

\subsubsection* {Semi-implicit pushers}

Due to the possible sharp motion that may occur during the relativistic physics implicit particle pushers are used conventionally because they possess superior accuracy properties, compared to, for example, a Runge-Kutta integrator. To get the most accuracy as possible one must take an approximation which is \emph{time reversible}. Usually, these are called semi-implicit schemes because ultimately they can be written in an explicit form.  

The most important step is how to advance the reduced momentum Eq.  (\ref{eq:particles_motionu}).
Let us denote the magnetic rotation matrix $- \overline{\bf B}_n \times $ that occurs in the right hand of the momentum Eq. (\ref{eq:particles_motionu}) side as:
\begin{equation}  \label{eq:pusher_matrixR}
\overline{\rm R}_n = -\overline{\bf B}_n \times =
\left( \begin{array}{cccccc}
0                     & \overline{B}_{n,z}      & -\overline{B}_{n,y}    \\
-\overline{B}_{n,z}    & 0                       & \overline{B}_{n,x}    \\
\overline{B}_{n,y}    & -\overline{B}_{n,x}      & 0                     \\
\end{array} \right).
\end{equation}

We prefer the use of the particle pusher developed by Higuera and Cary \cite{higuera2017pusher}, which takes the form of:
\begin{equation} \label{eq:pusher_solutionHC}
 \left[\mathbbm{1} - \overline{\gamma}_{n}^{-1} \overline{\rm R}_{n} \right] {\bf u}_{n+1} =
  \left[\mathbbm{1}+\overline{\gamma}_{n}^{-1} \overline{{\rm R}}_{n} \right] {\bf u}_{n} +\\
 2 \overline{\bf E}_{n},
\end{equation}
where $\mathbbm{1}$ denotes the identity matrix,  and $\overline{\gamma}_n = \sqrt{1+ ({\bf u}_n + {\bf u}_{n+1})^2/4}$ is the average gamma in the time step,  the closed form of which are determined consistently with the above scheme. This is an implicit  scheme because the the inverse of the matrix needs to be computed to get ${\bf u}_{n+1}$. However, it can be expressed in an explicit form as:
\begin{equation} \label{eq:pusher_inverseR}
 \left[\II - \overline{\gamma}_{n}^{-1} \overline{\rm R}_{n} \right]^{-1} = \II +
 \left( 1+ \overline{\gamma}^{-2}_{n} \overline{\bf B}_n^2  \right)^{-1}
 \left[\overline{\gamma}^{-1}_{n}  \overline{\rm R}_n + \overline{\gamma}^{-2}_{n}  \overline{\rm R}_n\overline{\rm R}_n  \right]
\end{equation}
For completeness, its average gamma can also be given in closed form:
\begin{equation} \label{eq:pusher_solutionHCgamma}
 \overline{\gamma}_{n}^2 = \frac{1}{2} \left(
 \gamma_{-}^{2} - \overline{\bf B}^2_n + \sqrt{ 
 \left( \gamma_{-}^2-\overline{\bf B}^2_n \right)^2+
 4 \left( \overline{\bf B}^2_n + \left|\overline{\bf B}_n \cdot {\bf u}_{-}\right|^2 \right)
 } \right),
\end{equation}
where  ${\bf u}_{-} = {\bf u}_{n}+\overline {\bf E}_n$, $\gamma_{-} = \sqrt{1+{\bf u}_{-}^2}$. 
This reduced momentum pusher is norm (volume) conserving because the same $\overline{\gamma}$ appears on the left and right hand side of Eq. (\ref{eq:pusher_solutionHC}). We could see this when we express ${\bf u}_{n+1}$, because  the diagonal Pad\'e approximant (see Eq. (\ref{eq:exponential_CN2})) will appear on the right hand side,  the formula of which is norm (and gamma) conserving.

A second type of pusher that we wanted to consider is one that is derived from an exponential solution of the reduced momentum equation (\ref{eq:particles_motionu}). To our surprise, the exponential pusher we were looking for turned out to be the relativistic Boris pusher (see \ref{subsubsec:pusherexp}), which does explain its robust properties \cite{qin2013boris}. We restored this pusher to its complete exponential form, and based on that we succeeded to derive an improved formula. A reminder, the usual Boris pusher actually has the form: 
\begin{equation} \label{eq:pusher_solutionB}
 \left[\mathbbm{1} - \overline{\gamma}_{n}^{-1} \overline{\rm R}_{n} \right] ( {\bf u}_{n+1}-\overline{\bf E}_n ) =
  \left[\mathbbm{1}+\overline{\gamma}_{n}^{-1} \overline{{\rm R}}_{n} \right] ({\bf u}_{n} +\\
  \overline{\bf E}_{n}),
\end{equation}
the choice for the average gamma is $\overline{\gamma}_{n} = \sqrt{1+{\bf u}_{-}^2}$, with  ${\bf u}_{-} = {\bf u}_{n}+\overline {\bf E}_n$. From the restored exponential form Eq. (\ref{eq:pusherexp_boris}) we propose the following improved Boris pusher ("Boris 2"):
\begin{equation} \label{eq:pusher_solutionB2}
 {\bf u}_{n+1} = \left[\mathbbm{1} - \frac{1}{2} \overline{\gamma}_{n}^{-1} \overline{\rm R}_{n} \right]^{-2}
  \left[\mathbbm{1}+\frac{1}{2}\overline{\gamma}_{n}^{-1} \overline{{\rm R}}_{n} \right]^{2} ({\bf u}_{n} +\\
  \overline{\bf E}_{n}) +\overline{\bf E},
\end{equation}
Here, what we can see is that the rotation term evaluated with two substeps, each using a norm conserving Pad\'e approximant (its explicit form can be found at Eq. (\ref{eq:pusherexp_pade2})). Unlike the Boris-pusher, however, here the gamma of the midpoint actually equals Boris' choice $\gamma_{n+1/2} = \overline{\gamma}_{n}$. Vay and others \cite{vay2008pusher} have shown that disadvantage of the Boris-pusher is that, it is subject to a spurious force when  $\overline{\bf E}_n = -\gamma_{n}^{-1} \overline{\rm R}_n$, unlike, for example the Vay or the Higuera-Cary pusher. We also found that this is indeed the case, however, the Boris 2 pusher is at least a factor of 2 more accurate than the usual Boris pusher in this scenario depending on the parameters. Aside from this edge case, Ripperda \cite{ripperda2018comparison_pusher} found that the various semi-implicit pushers perform really similar, each having its  own strength and weakness.

Finally, the formal solution of Eq. (\ref{eq:particles_motionr}) is
\begin{equation} \label{eq:pusher_solutionr}
{\bf r}_{n+1} = {\bf r}_n + \Delta t \overline{\bf v}_n, \quad \text{with } \quad
\overline{\bf v}_n = \frac{1}{\Delta t} \int_{s = 0}^{\Delta t} {\bf v} (t_n +s) {\rm d}s,
\end{equation}
where ${\bf v}(t) = {\bf u}(t)/\gamma(t)$ denotes the velocity of the particle and $\overline{\bf v}_n$ the particle's average velocity in the time interval. There are couple of possible choices in the second order, we used the midpoint formula:
\begin{equation} \label{eq:pusher_solutionrv}
\overline{\bf v}_{n} \approx \frac{1}{2} \overline{\gamma}^{-1}_{n} \left( {\bf u}_{n+1}+{\bf u}_{n} \right),
\end{equation}
with $\overline{\gamma}_{n} = \sqrt{1+\left({\bf u}_{n} + {\bf u}_{n+1}  \right)^2/4}$ which is already used in  the Higuera-Cary pusher.

In conclusion, our method of choice for the particle pusher is the Higuera-Cary method and the Boris 2 pusher.

\subsection{Field interpolation} \label{subsec:particles_interpolate}
The interpolation of field values for each particle $\pp$ of species $\sss$ can be done independently.

\subsubsection*{Predictor step}
We  stagger  the electromagnetic state $\Psi$ in time compared to the particle dynamics such that it is available at midpoints as $\Psi_{n+1/2} = \left( {\bf E}_{n+1/2},\  {\bf B}_{n+1/2} \right)^T$, which is also consistent with the field propagation formula Eq. (\ref{eq:exponential_solution2}).

We do not use leapfrog style integrator for the particle positions therefore the particle positions ${\bf r}_{\pp,n}$ and reduced momenta ${\bf u}_{\pp, n}$ coincide temporally. This means that before field interpolation we have to push the particle positions to the temporal midpoint as:
\begin{equation} \label{eq:particles_interpolater}
{\bf r}_{\pp,n+1/2} = {\bf r}_{\pp,n} + \frac{1}{2} \Delta t \gamma_{\pp,n}^{-1}{\bf u}_{\pp,n},
\end{equation}
this is a first order predictor step. 
We note that only choosing trapezoid rule for  $\overline{\bf v}$ average velocity in Eq. (\ref{eq:pusher_solutionr}) is equivalent using to the usual leapfrog scheme. In the following, we assume that the particle position ${\bf r}_{\pp}$ at field interpolation step always are at the temporal midpoint $n+1/2$ without explicitly indicating it.

\subsubsection*{Interpolation formula}
Let us denote the discretized normalized particle shapes evaluated at particle positions $x_{\pp},y_{\pp}, z_{\pp} $ as 
\begin{align} \label{eq:particles_interpolateSx}
\hat{S}_{x, k} (x_{\pp}) = \hat{S}_x(x_k-x_{\pp})&, \quad
\sum_k \hat{S}_{x, k} (x_{\pp}) \Delta x_k = 1, \\ \label{eq:particles_interpolateSy}
\hat{S}_{y, j} (y_{\pp}) = \hat{S}_y(y_j-y_{\pp})&, \quad
\sum_j \hat{S}_{x, j} (y_{\pp}) \Delta y_j = 1, \\ \label{eq:particles_interpolateSz}
\hat{S}_{z, i} (z_{\pp}) = \hat{S}_z(z_i-y_{\pp})&, \quad
\sum_i \hat{S}_{z, i} (z_{\pp}) \Delta z = 1.
\end{align}
where $\Delta x_k$ and $\Delta y_j$ denote the local cell volume if coordinate transformations applied for the physical coordinates (see \ref{subsubsec:transform}). However, in the interpolation step \emph{we ignore the variations in local cell volume}\footnote{To illustrate why we do so, suppose that the local volume is larger left to the particle's center than to the right. This means that the nearest grid point to the left is farther from particle's center than the one to the right, and during interpolation it should be weighted no more than the right one. This approach is also used in cylindrical PIC codes.}, such that $\Delta x_k = \Delta x$, $\Delta y_j = \Delta y$ always. The normalization conditions are satisfied with the standard PIC particle shapes (see \ref{subsubsec:shapes}).

The interpolation formula for a discrete 3D scalar field $\phi_{i,j,k}$ is written as:
\begin{equation} \label{eq:particles_interpolate3S}
{\bf \phi}_{\pp} = \sum_{i,j,k}  {\phi}_{i,j,k} \hat{S}_{x,k}(x_{\pp}) \hat{S}_{y,j}(y_{\pp}) \hat{S}_{z,i}(z_{\pp})  \Delta x \Delta y \Delta z.
\end{equation}
We use the 3rd order (cubic) particle shape Eq. (\ref{eq:shapes_S3}) as default \cite{godfrey2014stability_spectral}, which extends to approximately two grid points in each direction from the grid point closest to ${\bf r}_{\pp,n}$. The field interpolation can be done for each particle in parallel and as such, it is a GPU compatible, domain local computation.

\subsubsection*{Field destaggering}
 Because in our spatial representation the fields ${\bf E}$ and ${\bf B}$ are Yee-staggered, we choose to destagger them before the PIC field interpolation step. We can do this using the ${\rm S}^{(\pm)}$  matrices from Eq. (\ref{eq:spatial_stagger}) and approximately reversing Eqs. (\ref{eq:spatial_stagger_E}), (\ref{eq:spatial_stagger_B}) as:
\begin{align} \label{eq:interpolate_destagger_E}
&{E}_x^{(C)} \approx {\rm S}_x^{(+)} E_x,\quad  {E}_y^{(C)} \approx {\rm S}_y^{(+)} E_y, \quad {E}_z^{(C)} \approx {\rm S}_z^{(+)} E_z  \\
{B}_x^{(C)} \approx &{\rm S}_y^{(+)}{\rm S}_z^{(+)} B_x,\  {B}_y^{(C)} \approx {\rm S}_x^{(+)} {\rm S}_z^{(+)} {B}_y,\  {B}_z^{(C)} \approx {\rm S}_x^{(+)} {\rm S}_y^{(+)} B_z,  \label{eq:interpolate_destagger_B}
\end{align}
For interpolating the field components ${\bf E}$ and ${\bf B}$ the formula Eq. (\ref{eq:particles_interpolate3S}) has to be repeated 6 times, but the shape functions for particle $\pp$ has to be discretized only once. Also extra care needed to be taken for the symmetric and anti symmetric boundary conditions, which we do by extending the field values at the outer edges with appropriately assigned ghost cells.

Using the above non-staggered interpolation scheme helps us to avoid inaccuracies during the interpolation of the Lorentz-force  \cite{lehe2016fbpic}.

\begin{figure}[ht]
\centering
\includegraphics[width=7.8cm]{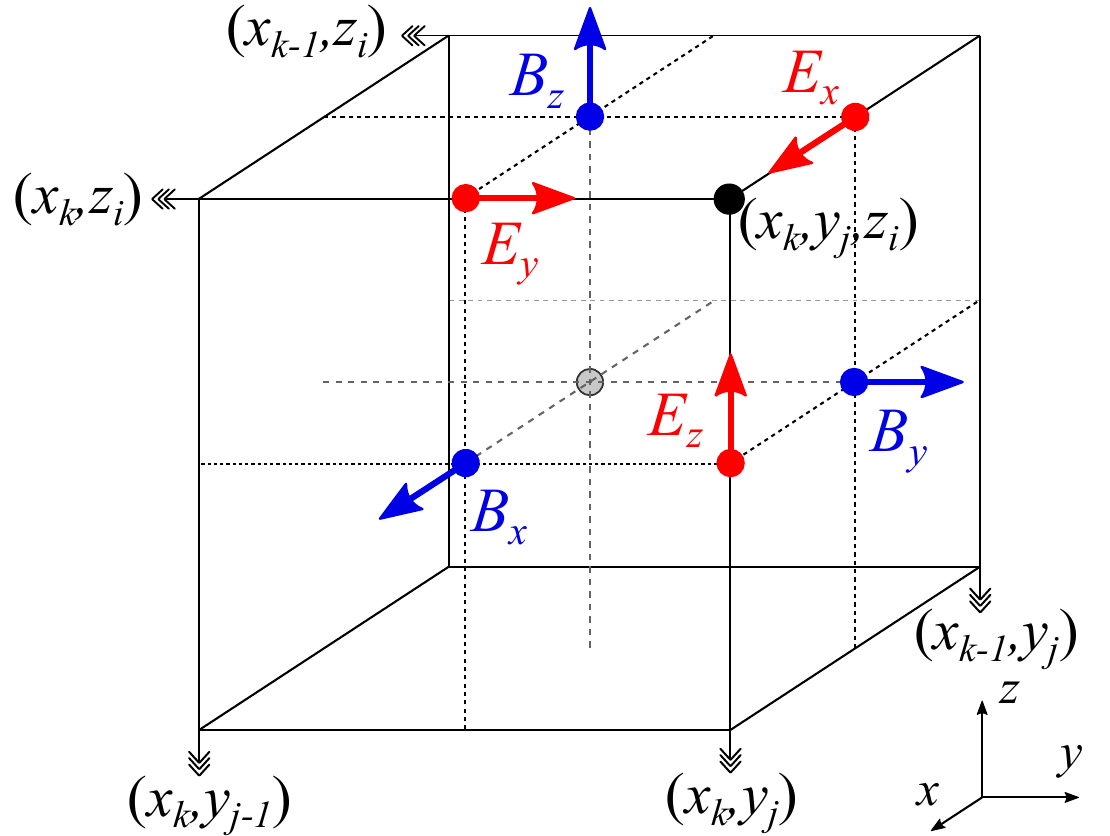}
\caption{ Illustration of Yee-staggering near the grid point $(x_k,y_j,z_i)$. We show subcell corresponding to staggered coordinates with dashed lines. Our $2\times$ supersampling reconstructs all field values at all 8 corners of the subcell (marked with dots of various color).
\label{fig:interpolation_Yee}}
\end{figure}

\subsubsection*{Field supersampling}

\begin{table*}[tp!]
\newcommand{\Sxp}{\ensuremath{\rm S}^{(+)}_x}
\newcommand{\Sx }{\ensuremath{\rm S}^{(-)}_x}
\newcommand{\Syp}{\ensuremath{\rm S}^{(+)}_y}
\newcommand{\Sy }{\ensuremath{\rm S}^{(-)}_y}
\newcommand{\Szp}{\ensuremath{\rm S}^{(+)}_z}
\newcommand{\Sz }{\ensuremath{\rm S}^{(-)}_z}

\begin{tabular*}{\linewidth}{@{\extracolsep{\fill}} r|ccccccccccc }

\emph{ } & $i,j,k$ & $i,j,k-\frac{1}{2}$, & $i,j-\frac{1}{2},k$ & $i,j-\frac{1}{2}, k-\frac{1}{2}$ & $i-\frac{1}{2},j,k$ & $i-\frac{1}{2}, j, k-\frac{1}{2}$, & $i-\frac{1}{2}, j-\frac{1}{2}, k$ & $i-\frac{1}{2}, j-\frac{1}{2}, k-\frac{1}{2}$ \tabularnewline 
\hline
$\phi^{(2\times)}$ & $1$ & $\Sx$  & $\Sy$ & $\Sy\Sx$  & $\Sz$ & $\Sz\Sx$ & $\Sz\Sy$ & $\Sz\Sy\Sx$ 
\tabularnewline
$E_x^{(2\times)}$ & $\Sxp$ & $1$  & $\Sy\Sxp$ & $\Sy$  & $\Sz\Sxp$ & $\Sz$ & $\Sz\Sy\Sxp$ & $\Sz\Sy$ 
\tabularnewline
$E_y^{(2\times)}$ & $\Syp$ & $\Syp\Sx$  & $1$ & $\Sx$  & $\Sz\Syp $ & $\Sz\Syp\Sx$ & $\Sz$ & $\Sz\Sx$
\tabularnewline
$E_z^{(2\times)}$ & $\Szp$ & $\Szp\Sx$  & $\Szp\Sy$ & $\Szp\Sy\Sx$  & $1$ & $\Sx$ & $\Sy$ & $\Sy\Sx$
\tabularnewline
$B_x^{(2\times)}$ &$\Szp\Syp$ & $\Szp\Syp\Sx$ & $\Szp$ & $\Szp\Sx$ & $\Syp$ & $\Syp\Sx$ & $1$ &$\Sx$ 
\tabularnewline
$B_y^{(2\times)}$ &$\Szp\Syp$ & $\Szp$ & $\Szp\Sy\Syp$ & $\Szp\Sy$ & $\Sxp$ & $1$ &$\Sy\Sxp$&$\Sy$ 
\tabularnewline
$B_z^{(2\times)}$ &$\Syp\Sxp$ & $\Syp$ & $\Sxp$ & $1$ & $\Sz\Syp\Sxp$ & $\Sz\Syp$ &$\Sz\Sxp$&$\Sz$ 
\tabularnewline\hline
\end{tabular*}

\caption{Table of the sequence of stagger matrices ${\rm S}^{(\pm)}$ that we need to apply in order to double the resolution  of the field components (supersample) in each direction, as the first part of PIC field interpolation step. The first row shows the supersampling for a non-staggered $\phi$ scalar field in 3D, and the rest for the Yee-staggered ${\bf E}$, ${\bf B}$. As can be seen the supersampling of these fields in a single pass results in a rather complicated algorithm. The first column shows how to remove the Yee-staggering.   \label{tab:interpolation_supersample}}
\end{table*}

Even with the above destaggered interpolation scheme, we found the accuracy of the interpolation bottlenecking the accuracy of our laser wakefield acceleration (LWFA) simulations. We use 3rd order particle shape Eq. (\ref{eq:shapes_S3}) to suppress numerical artifacts, but the latter did not really result in  as  accurate interpolation as we hoped for. This is problematic considering the efforts we made for the high accuracy exponential field solver. We modeled the interpolation accuracy in \ref{subsubsec:shapes_alias} which indeed confirms that this is a problem, and suggests that one solution is to increase the spatial resolution of the fields.\footnote{An alternative solution is to use high order Lagrange interpolation as discrete shape functions, but that could increase the per particle computational cost to unfeasible levels, and it makes the number density not positive definite during  deposition. However, this method is used in quasi 3D PIC codes for the angular variable with fully spectral discrete interpolator.}

We propose to use the high order staggering operators ${\rm S}^{(\pm)}$ not only to destagger the fields, but double their resolution ($2\times$ supersample) in each direction (this octuples the field data points in 3D). We implemented an algorithm that performs this from the Yee-staggered ${\bf E}$, ${\bf B}$ directly, and we summarize the required sequence of one dimensional ${\rm S}^{(\pm)}$ matrix products to achieve this in Table \ref{tab:interpolation_supersample} with cell index $i,j,k$. We implemented this supersampling process as the following procedure per field component in 3D:
\begin{enumerate}
\item  \revisee{Stagger the field in $z$ direction to get the missing data point at subcell $i'$ along the line where the field value is known. To $E_x, E_y, B_z$ apply ${\rm S}^{(-)}_z$, to the rest ${\rm S}^{(+)}_z$.}

\item \revisee{Stagger the field in $y$ direction to get the two missing data points at subcells $j'$ in the corresponding 2D plane. \newline To $E_x, E_z, B_y$ apply ${\rm S}^{(-)}_y$, to the rest ${\rm S}^{(+)}_y$.} %for $i-\tfrac{1}{2} ,i$

\item \revisee{Stagger the field in $x$ direction to get the four missing data points at subcells $k'$ in the 3D cell.
\newline To $E_y, E_z, B_x$ apply ${\rm S}^{(-)}_x$, to the rest ${\rm S}^{(+)}_x$.}
\end{enumerate}
This involves calculating 7 matrix products with ${\rm S}^{(\pm)}$ per field component overall - which has similar cost as the 4th order Taylor expansion of the main exponential. Although this is significant, this process is independent of the number of particles. This supersampling method is GPU compatible, and can be done independently in each subdomain if MPI domain decomposition is involved. In our code we store the supersampled fields as a separate, more tightly packed data structure using 32 bit floating point format to reduce the memory costs. The interpolation is done with full double precision which provides 7 digits of interpolation accuracy.

 In \ref{subsubsec:shapes_alias} we have shown that  spatial waves of $\kk_{\max}/10$ frequency could be subject to around 5\% interpolation error near their peaks which is much larger than that we specified for our field solver. Each $2\times$ supersampling (or doubling the resolution) only reduces the error   approximately by $1/4$.
We can easily see why this directly improves interpolation accuracy: if we have more than one particle per cell, then this kind of supersampling results in nearly identical forces felt by the particles as real double resolution simulations - for all spatial field wavelengths that present in both simulations. This includes both the oscillating laser fields and plasma waves.

\subsubsection*{Temporal substepping}
It is also possible to implement substepping (temporal supersampling) for the combined PIC field interpolation and particle pusher steps alone. This directly improves the accuracy of the particle dynamics at the cost of repeating the particle interpolation step (with spatial supersampling or destaggering).

An important point which is worth mentioning is that using low exponential Taylor-expansion it is possible to get the fields in the $\delta$ vicinity of the temporal midpoint $t+\Delta t /2$:
\begin{equation} \label{eq:particles_interpolatePsi}
\Psi_{n+1/2}(\delta) = \Psi_{n+1/2}+\delta \left[ \op{H} \Psi_{n+1/2} - {\bf J}_{n}\right] +\frac{\delta^2}{2}\op{H}^2\Psi_{n+1/2}.
\end{equation}
and we can do this with minimal overhead because $\op{H} \Psi_{n+1/2}$ terms are shared between this and the main Taylor expansion.  This temporal substepping inherits the parallelization properties of the PIC interpolation we discussed in this section: it is GPU compatible and domain local.

We implemented the option of using two substeps at the current stage, however, we found it more important  to use our spatial supersampling feature first for the interpolated fields, which makes the computational cost of this not really worth it. In our experience this substepping have notable impact when the PIC particles interact with high intensity oscillating fields.

\subsection{Deposition} \label{subsec:particles_deposit}
In the following we describe our density and current deposition algorithm and other auxiliary methods linked to it. These are to be done after particle push.

\subsubsection*{Particle map}
We designed the parallelization of the deposition method around the concept of a particle map. This  object provides map of particle indices to be accessed from cell indices $(i,j,k)$ in which the particles belong to. This particle map could be used for physics modules in the future, for example particle-particle collisions.

 Let us denote the number of particles for species $\sss$ as $M_\sss$. We implemented this map as two arrays that store the particle indices $\pp \in [0, M_\sss-1]$: 
\begin{itemize}
\item[$(A)$] Cell map which is a 3D array of size $M_z\times M_y \times M_x$.
\item[$(B)$] Particle offset table  which is an array of the size  $M_\sss$. 
\end{itemize}
\revisee{Initially, both arrays are filled with an invalid particle index value of -1. For each $(i,j,k)$ cell index we make a linked list starting from array $(A)$ using only the $\pp$  particle indices.  The purpose of each stored $\pp$ particle index is to point to the next $\pp'$ particle in array $(B)$ that belongs to the same list (same cell). Any invalid index value mark the end of these lists.}

\revisee{We make this as the following. For each particle $\pp$ do: first compute its $i,j,k$ cell index from its position; then swap $\pp$ index with the $\pp'$ stored in array $(A)$ at the same cell; finally, store the acquired $\pp'$ at $\pp$ in array $(B)$.}

\revisee{ The advantage of particle map that we do not actually need to sort the particle data to build this structure, only determine which particle belongs to which cell. This can be done in parallel using \texttt{atomic} operations, such as OpenMP \texttt{atomic capture} directive or CUDA \texttt{atomicExch} functions. We found that building this particle map has barely any performance impact on CPU. During the make of this structure we also apply 'remove' and 'reflective' boundary conditions for each particle.}

\begin{figure*}[tp!]
\centering
\includegraphics[width=17.0cm]{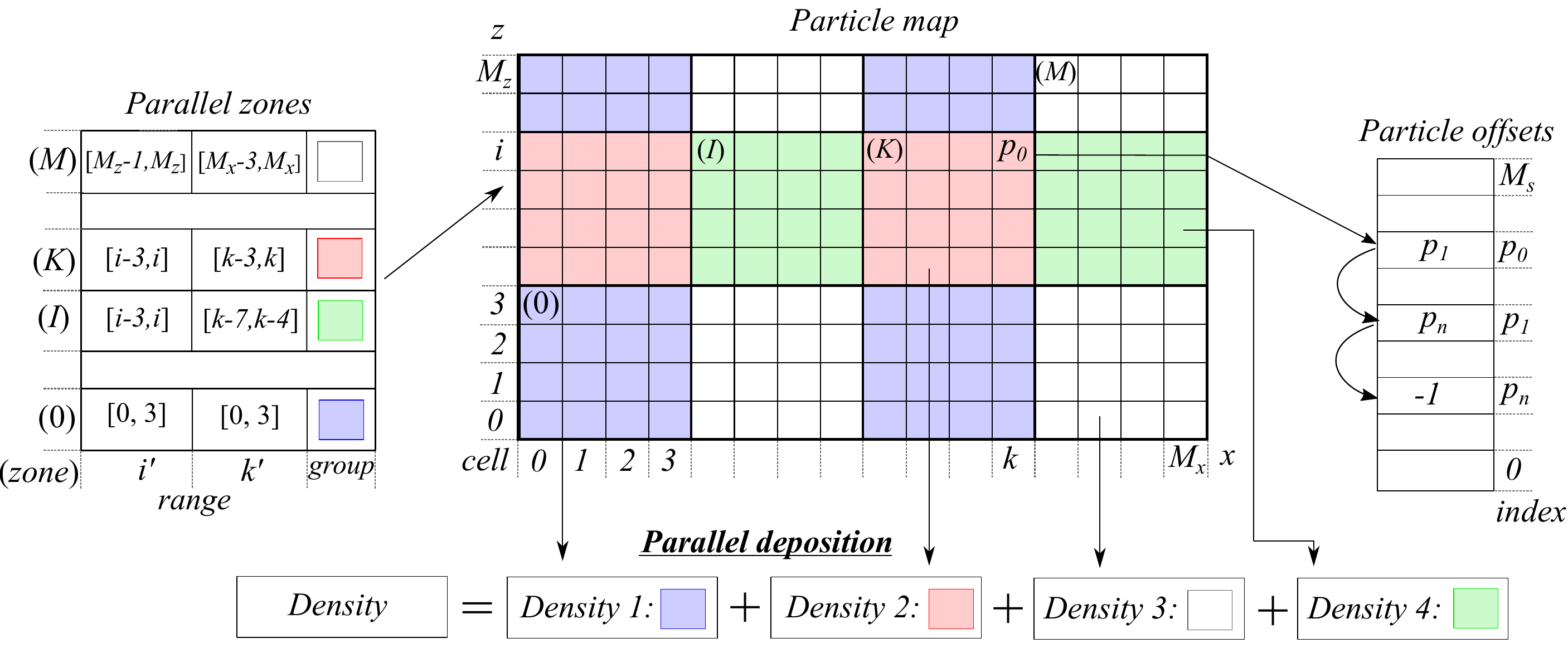}
\caption{Sketch of our 2D checkered parallel deposition algorithm. At the beginning of the simulation we calculate table of \emph{parallel zones} which are $4\times4$ cell blocks, and we assign each zone to $2\times2$ density groups (shown here by blue, red, white and green). The particles corresponding to each cell are accessed through the \emph{particle map} which provides the starting point of a linked list. In the example we show here that the cell $(i,k)$ stores the particle index $\pp_0$,  and the particle at $\pp_0$ points to $\pp_1$, and the latter points to $\pp_n$, which terminates the list with an invalid particle index. Parallel density deposition can be done independently in each zone by depositing into \revisee{one of the 4 components in the density buffer with matching group color} and summing the density afterwards.
\label{fig:particles_depositmap}}
\end{figure*}

\subsubsection*{Density deposition formula}

The discrete normalized particle shapes during deposition follow the same definition as Eqs. (\ref{eq:particles_interpolateSx})-(\ref{eq:particles_interpolateSz}) at particle positions $x_{\pp},y_{\pp}, z_{\pp}$. The notable differences are that the deposition is done at the actual time point  the particles are  after particle push, and when coordinate transformations (\ref{subsubsec:transform}) are used the local cell volume must be taken into account during normalization.\footnote{The local cell sizes become $\Delta x_k = g_{x,k}\Delta x$ and $\Delta y_j = g_{y,j} \Delta y$, where $g_{x,k}$ and $g_{y,j}$ are the discretized transformation functions.} 

The deposition formula for a discrete charge density $\varrho_{i,j,k}$ for species $\sss$ are written as:
\begin{equation} \label{eq:particles_deposit}
{\varrho}_{i,j,k} = \sum_{\pp} q_{\sss} w_\pp \hat{S}_{x,k}(x_{\pp}) \hat{S}_{y,j}(y_{\pp}) \hat{S}_{z,i}(z_{\pp}),
\end{equation}
where $q_\pp$ is the charge of the particle. We use the 3rd order particle shape Eq. (\ref{eq:shapes_S3}) as default, same as during interpolation. Even though a the current density ${\bf J}$ for species $\sss$ can be deposited the same way, we choose to do this with a modified Esirkepov current deposition scheme which we discuss in the next section. Another problem is the efficient thread based parallelization, for which we outline our solution using particle maps.

\subsubsection*{Current deposition}

We contemplated how to deposit current in our code, because in our testing we found no obvious self-consistent artifacts related to the lack of charge conserving current deposition (\ref{subsubsec:deposit_conserve}), however, we also filtered the current (Section \ref{subsec:spatial_current}). We mandate that any current deposition method to be used with our exponential solver must be local and separable according to Cartesian directions: this would ensure reasonable computational complexity and avoiding of the introduction of unexpected numerical artifacts. We choose to adopt modified Esirkepov current deposition scheme \cite{esirkepov2001charge_conservation} which we derived for the integral continuity equations (\ref{subsubsec:deposit_ES}).
 
To do this kind of deposition, first we need to introduce the Cartesian particle offset positions:
\begin{equation} \label{eq:particles_depositxyz}
z_\pp^{(\pm)}  = z_\pp \pm \frac{\Delta t}{2} {v_{z,\pp}}, \ 
y_\pp^{(\pm)}  = y_\pp \pm \frac{\Delta t}{2} {v_{y,\pp}}, \ 
x_\pp^{(\pm)}  = x_\pp \pm \frac{\Delta t}{2} {v_{x,\pp}},
\end{equation}
where ${\bf v}_\pp = \gamma^{-1}_\pp {\bf u}_\pp$ denotes the particle velocity.

The staggered current components $J_x,J_y,J_z$ are deposited by:
\begin{align} \label{eq:particles_depositJx}
{J}_{x,i,j,k-\frac{1}{2}} =   \sum_{\pp}   q_\sss  w_\pp \delta \hat{S}_{x} (x_k,x_\pp) \overline{S}_{y,z}(y_j, z_i, y_\pp, z_\pp), \\ \label{eq:particles_depositJy}
{J}_{y,i,j-\frac{1}{2},k} =   \sum_{\pp}   q_\sss  w_\pp \delta \hat{S}_{y} (y_j,y_\pp) \overline{S}_{x,z}(x_k, z_i, x_\pp, z_\pp), \\
{J}_{z,i-\frac{1}{2},j,k} =   \sum_{\pp}   q_\sss  w_\pp \delta \hat{S}_{z} (z_i,z_\pp) \overline{S}_{x,y}(x_k, y_j, x_\pp, y_\pp), \label{eq:particles_depositJz}
\end{align}
where we introduced the following spatially integrated one dimensional differential particle shapes as:
\begin{align} \label{eq:particles_depositSJx} 
\delta \hat{S}_{x} (x_k,x_\pp) &= \sum_{k' = 0}^{k-1} \left[ \hat{S}\left(x_{k'}-x_\pp^{(+)} \right)-\hat{S}  \left(x_{k'}-x_\pp^{(-)} \right)  \right] \frac{\Delta x_k}{\Delta t}, \\  \label{eq:particles_depositSJy} 
\delta \hat{S}_{y} (y_k,y_\pp) &= \sum_{j' = 0}^{j-1} \left[ \hat{S}\left(y_{j'}-y_\pp^{(+)} \right)-\hat{S}  \left(y_{j'}-y_\pp^{(-)} \right)  \right] \frac{\Delta y_j}{\Delta t}, \\
\delta \hat{S}_{z} (z_i,z_\pp) &= \sum_{i' = 0}^{i-1} \left[ \hat{S}\left(z_{i'}-z_\pp^{(+)} \right)-\hat{S}  \left(z_{i'}-z_\pp^{(-)} \right)  \right] \frac{\Delta z}{\Delta t}.  \label{eq:particles_depositSJz} 
\end{align}
The symbols $\overline{S}_{x,y}$, $\overline{S}_{y,z}$, and $\overline{S}_{x,z}$ denote temporally averaged particle shape in the perpendicular plane according to Esirkepov's paper \cite{esirkepov2001charge_conservation}. For example:
\begin{multline} \label{eq:particles_depositSAxy}
\overline{S}_{y,z}(y_j, z_i, y_\pp, z_\pp) =  \\
\frac{1}{6}\hat{S}\left(y_{j}-y_\pp^{(-)} \right) \left[
2\hat{S}\left(z_{i}-z_\pp^{(-)} \right) + \hat{S}\left(z_{i}-z_\pp^{(+)} \right) \right] + \\
\frac{1}{6}\hat{S}\left(y_{j}-y_\pp^{(+)} \right) \left[
2\hat{S}\left(z_{i}-z_\pp^{(+)} \right) + \hat{S}\left(z_{i}-z_\pp^{(-)} \right) \right].
\end{multline}

In this scheme, the current values $J_x$ are actually computed from second order integral continuity equation of form (ignoring the transversal density):
\begin{equation} \label{eq:particles_depositcont}
J_{x,k+1/2}-J_{x,k-1/2} = q_\pp w_\pp  \frac{\Delta x_k}{\Delta t} \left[ \hat{S}\left(x_{k'}-x_\pp^{(+)} \right)-\hat{S}  \left(x_{k'}-x_\pp^{(-)} \right)  \right].
\end{equation}

In the current deposition algorithm we use Eqs. (\ref{eq:particles_depositJx})-(\ref{eq:particles_depositJz}). These are actually identical to the Esirkepov formulas in standard second order Yee PIC codes. One other feature of this method that it results in a properly staggered current. 
 
Our modification comes from the fact we may substitute higher order integral quadratures of $\int_{-\Delta x /2}^{\Delta x/2} {\rm d}x$ in the right hand side of Eq. (\ref{eq:particles_depositcont}). Let us fill the rows of the banded diagonal matrix ${\rm F}^{({\rm div})}$ with the coefficients of this integral quadrature (we list the 10th order quadrature in Table \ref{tab:spatial_filter} by as div-10). Then the high order integral continuity equation can be enforced by the matrix multiplications
\begin{equation}  \label{eq:particles_depositJN}
\overline{J}_{x} ={\rm F}^{({\rm div})}_x J_x, \quad
\overline{J}_{y} ={\rm F}^{({\rm div})}_y J_y, \quad
\overline{J}_{z} ={\rm F}^{({\rm div})}_z J_z.
\end{equation}
We implemented these latter after the deposition is done,  in the current filtering step (Section \ref{subsec:spatial_current}), which we also indicated earlier. The continuity equation for the deposited current can be enforced to arbitrary order in Cartesian coordinates.

\subsubsection*{Parallel deposition}

We developed a combined charge density and current deposition method that is efficient for shared memory parallelization. In this method we divide the spatial domain into mutually exclusive deposition zones such that the all deposition can be done in each zone in parallel. Using 3rd order $\hat{S}$ particle shape we found that the minimal width of each zone has to be 4 cells in each dimension.\footnote{The exact width depends on how many grid points the particle shape could extend from its center during current deposition. The 3rd order particle shape could extend by 2 grid point's from its center.  Esirkepov's current deposition also widens the width of the current shape by at least 1 grid point.}

Our parallel combined density and current deposition algorithm consists of the following steps in $n$ dimension: 
\begin{enumerate}
\item \revisee{Build a list of \emph{deposition zones}. Each zone listed is a block of $4^n$ number of adjacent cells (4 per dimension), such that the zones do not overlap and they include all the cells.} 
\item  \revisee{We categorize each zone into one of the $2^n$ number \emph{deposition groups} (2 per dimension), such that no adjacent zone belongs to the same deposition group.} 
\item \revisee{We need to allocate and zero-initialize a \emph{deposition buffer} which contains $2^n$ density and current density fields.} %to which we will deposit to.
\item \revisee{We build the \emph{particle map} object.}
\item \revisee{We do the density and current deposition for all deposition zones we listed, for all cells each zone contains, and for all particles each cell contains, such that each particle deposits into the density and current density component in the deposition buffer that matches its deposition group.}
\item \revisee{We sum the $2^n$ density and current density values in the deposition buffer to get the total density and current density values (for species $\sss$).}
\end{enumerate}

\revisee{We note that we can perform step 5 independently in parallel in all deposition zones, but not within a deposition zone. For this kind of parallelization dynamic thread management is needed (like OpenMP's \texttt{schedule(dynamic)}) to compensate for load balancing issues, since the number of particles per deposition zone could significantly vary.}

\revisee{This parallel deposition method yields in one and two dimensions an "interlaced" and "checkered pattern".} Even more interestingly, it is also possible to use 1D  or 2D version of this parallel deposition in 3D by selecting the dimensions in which we form the deposition zones.
We illustrate this method in 2D in Fig. \ref{fig:particles_depositmap}.  This algorithm is GPU compatible because it could provide independent work for large number of threads depending on the number of deposition zones. In the 3D version there is only memory overhead related to allocating the 8 field density buffers, the only computational overhead happens because we have to sum all of these. We reduce this overhead during CPU computations using the above parallelization technique only along the dimension of the highest resolution $M_z$ if the required thread count less than $M_z/4$.

\subsection{Summary of the core PIC algorithm}

\begin{figure}[ht]
\centering
\includegraphics[width=8.0cm]{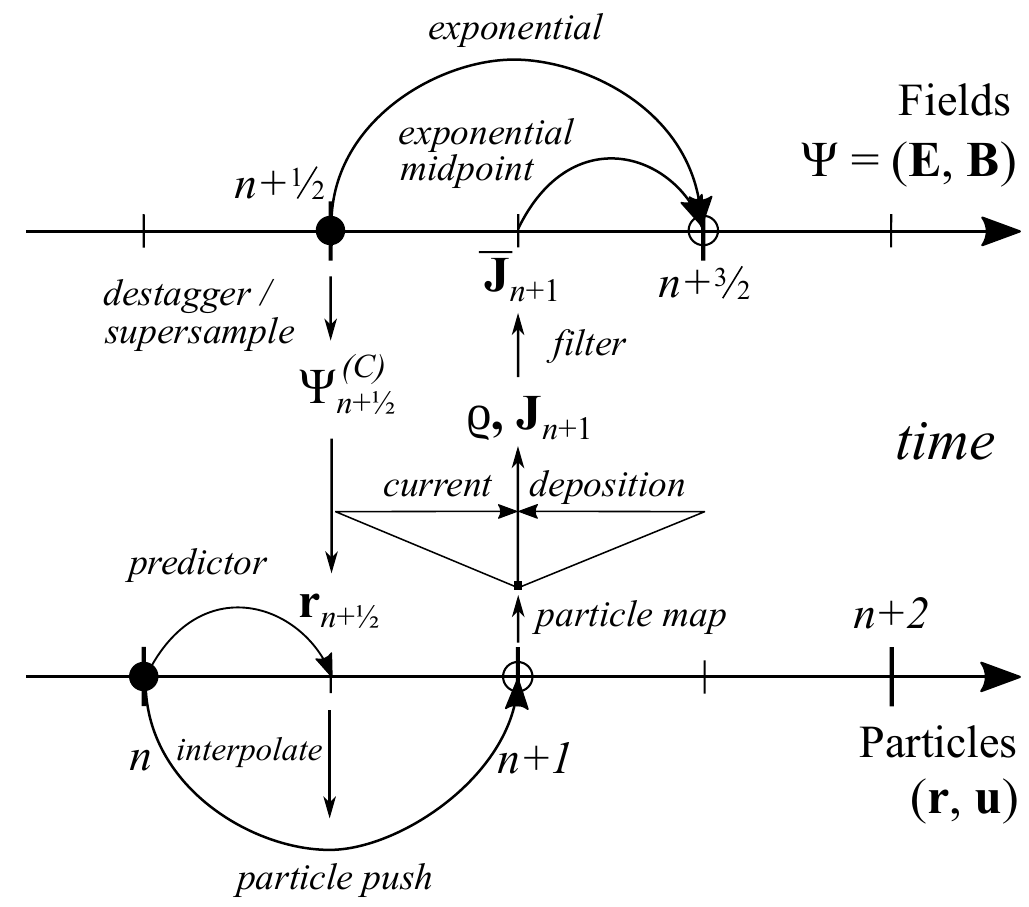}
\caption{Sketch of a time step in our particle-in-cell algorithm. The fields and particle data are temporally staggered and their starting position are indicated with solid dots. The steps are as follows: particle predictor step, field destaggering, field interpolation, particle push; particle map calculation, charge density and current deposition, current filtering and finally the  exponential propagation of the fields.
\label{fig:particles_PIC}}
\end{figure}

Here, let us summarize our core PIC algorithm. We summarize its main steps in Table \ref{tab:particles_PIC}, where we link also each step with the relevant Sections and Appendices present in this paper. The steps are categorized in two categories: the ones that had to be done at the initialization of the simulation and the ones that have to be repeated in the PIC loop. We also show a sketch of the latter in Fig. \ref{fig:particles_PIC}.

We designed this algorithm foremost with massive thread-based parallelization in mind. It is currently implemented with OpenMP with standard compliant C++ code with minimal external dependencies. According to our testing on nodes of Komondor HPC with $2\times64$ CPU cores the parallelization performance  in compute scales mostly as expected for the algorithms described here on a single CPU - we unfortunately experienced performance degradation due to 
%the saturation of the memory access paths
\revise{NUMA effects in} multi CPU configuration. This only can be improved by implementing domain decomposition. Though this is currently work in progress we believe this algorithm  already has all the pieces to incorporate domain decompositions based on MPI libraries, GPU acceleration using Nvidia CUDA language and incorporating other physics modules in the future, like field ionization and QED phenomena.

%We regard these hardware specific limitations as not really important in the scientific context of this paper, those may be investigated elsewhere.

\begin{table*}
\newcommand{\textitm}[1]{\textbf{\upshape #1}}
\newcommand{\texthdr}[1]{{\textbf{{\upshape  #1:}}}}

\begin{minipage}[b]{1.\textwidth}
\small

\noindent\makebox[\linewidth]{\rule{\linewidth}{.4pt}}

\subsubsection*{\texthdr{Initialization}}
%\noindent\makebox[0.88\linewidth]{\rule{0.88\linewidth}{.4pt}}

\paragraph*{\textitm{Domain setup}}  Sets up the the simulation grids and data structures ($\rightarrow$ Section \ref{subsec:bandedmatrices}). Partition the simulation box $x_k$, $y_j$, $z_i$ into overlapping subdomains along dimension $z$ if able (planned in the future).

\paragraph*{\textitm{Compute banded matrices}} Initialize and precompute all banded diagonal matrices that might occur during simulation\revisee{, at a specified spatial approximation order,} these include the finite difference and the stagger operators ($\rightarrow$ Section \ref{subsec:differences}) and current filters ($\rightarrow$ Section \ref{subsec:spatial_current}). We need to include local boundary conditions in these matrices ($\rightarrow$ Table \ref{tab:spatial_boundary}) and create symmetric ($S$) and anti symmetric ($A$) versions if needed. Formulas for coordinate transformations ($\rightarrow$ \ref{subsubsec:transform}) must be incorporated into the matrices if used.

\paragraph*{\textitm{Particle map setup}} Allocate and build the particle map object for all species  ($\rightarrow$ Section \ref{subsec:particles_deposit}). Initialize \revisee{deposition} parallelization zones.

\paragraph*{\textitm{Particle loading}} Load all $\pp$ particles from density or input file at $n=0$ : $w_\pp$ weights, ${\bf r}_\pp$ positions, ${\bf u}_\pp$ reduced-momenta. Conversion to internal units from SI is needed to be done ($\rightarrow$  \ref{subsubsec:units}). Do initial charge density deposition.

\paragraph*{\textitm{Field loading}} Load initial laser fields and other user defined fields at time point $n=1/2$.
Stagger the fields with ${\rm S}^{(-)}$ operators ($\rightarrow$ Section \ref{subsec:spatial_maxwell}) to make them Yee-staggered.

\paragraph*{\textitm{Compute static fields}} If required, solve Poisson's equation with BiCGSTAB iterative method ($\rightarrow$ \ref{subsubsec:poisson}) to get static electric fields.  

\noindent\makebox[\linewidth]{\rule{\linewidth}{.4pt}}
\subsection*{\texthdr{Summary of the PIC  Loop}}
%\noindent\makebox[0.9\linewidth]{\rule{0.9\linewidth}{.4pt}}

\paragraph*{\textitm{Move window or insert particles}} We move the simulation window by shifting the field data by specified number of grid points if needed. Additional particles might be inserted at specified boundaries. 

\paragraph*{\textitm{Destagger fields}} Remove the Yee-staggering with ${\rm S}^{(\pm)}$ operators, which creates a separate field instance $\Psi^{(C)}_{n+1/2}$ at the temporal midpoint that is no longer spatially staggered. Optionally, we might supersample the fields instead ($\rightarrow$ Section \ref{subsec:particles_interpolate}).

\paragraph*{\textitm{Interpolate field values}} First, we push the particle positions to the temporal midpoint in a predictor step and then interpolate the field values ($\rightarrow$ Section \ref{subsec:particles_interpolate}) for all particles $\pp$.

\paragraph*{\textitm{\bf Push particles}} We push all $\pp$ particles from time point $n$ to $n+1$ by computing new reduced momenta ${\bf u}_{\pp}$ and position ${\bf r}_{\pp}$  ($\rightarrow$ Section \ref{subsec:particles_push}). It is also an option to do this with an exponential pusher, like Boris 2 ($\rightarrow$ \ref{subsubsec:pusherexp}).

\paragraph*{\textitm{Construct particle map}} Determine which particle belongs to which cell and construct particle map object ($\rightarrow$ Section \ref{subsec:particles_deposit}). Handle 'reflective' and 'remove' particle boundary conditions. 

%\paragraph*{Optional physics modules (planned)}

\paragraph*{\textitm{Parallel density deposition}}  Simultaneously deposit the charge density and current density ($\rightarrow$ Section \ref{subsec:particles_deposit}) using our parallelized deposition algorithm. We do the current deposition with Esirkepov's method of  second order Yee-PIC solvers ($\rightarrow$ \ref{subsubsec:deposit_ES}), which also yields the properly staggered current from the one dimensional continuity equations.

\paragraph*{\textitm{Filter the currents}} Apply current filters to suppress numerical Cherenkov radiation at high grid frequencies $\gtrsim 0.5 \kk_{\max}$ ($\rightarrow$ Section \ref{subsec:spatial_current}). Filter the current to satisfy high order continuity equations with ${\rm F}^{({\rm div})}$ integral quadrature matrices ($\rightarrow$ \ref{subsubsec:deposit_ES}).  

\paragraph*{\textitm{Absorbing layer effects}} Compute PML source terms and apply the  splitted exponential absorbing layer functions ($\rightarrow$ \ref{subsubsec:absorbing_layer}).

\paragraph*{\textitm{Exponential propagation}} Propagate the Maxwell fields $\Psi_{n+1/2}$ from time point $n+1/2$ to $n+3/2$ using \revisee{4th, 8th or larger order Taylor expansions of the exponential operators ($\rightarrow$ Section \ref{subsubsec:exponential_taylor}) within the midpoint formula Eq. (\ref{eq:exponential_solution2}). We can optimize away the evaluation of the source exponential with a combined formula ($\rightarrow$  \ref{subsubsec:exponential_source}). }

\paragraph*{\textitm{Prepare output}} At time points $n$ that are divisible by a user specified number prepare particle and field data to be saved. Destagger the fields to get a separate field instance $\Psi^{(C)}_{n+1/2}$. Conversion from internal units to SI is also needed to be taken care of ($\rightarrow$  \ref{subsubsec:units}).

\paragraph*{\textitm{ Repeat the loop}}

\quad

\noindent\makebox[\linewidth]{\rule{\linewidth}{.4pt}}

\end{minipage}
\caption{
Overview of the main steps of our particle-in-cell code. The first part of calculations is done at initialization where we are at time step $n = 0$ and time $t = 0$, and we initialize the relevant mathematical objects and the initial state of the system. During the \emph{PIC loop} we propagate from time index $n$ to $n+1$ with time step $\Delta t$. Other steps may be implemented in the future.
\label{tab:particles_PIC}
}
\end{table*}

%%****************************************************************************************************************************

\newpage %!!!
\section{Benchmarks}  \label{sec:benchmarks} 

To determine the accuracy characteristics of our PIC method we performed a series of tests, most of them are fundamental tests that reveal the inner workings of our method, which are heavily inspired by FBPIC paper \cite{lehe2016fbpic}, such as vacuum propagation and linear wake fields. We primarily focus on laser interaction with underdense plasma, where the the laser propagation direction is $z$, and is linearly polarized in the $x$ direction. We perform some of the tests only in 2D Cartesian $(y,z)$ geometry, because our high order exponential Maxwell solver does not discriminate between the geometry chosen or the propagation direction - all error quantities depend on the resolution $\Delta z$, $\Delta t\sim\Delta z/c$ relative to the target field (laser) wavelength $\lambda$ to be resolved during field propagation. For the laser we use \emph{Gaussian beams} and \emph{Gaussian pulses}\footnote{Gaussian pulse length is not FWHM.} for simplicity. Usually, we also use a moving window that moves in direction $z$ with speed of light $c$. We also solve the Poisson equation for the initial laser fields to get them as correct as numerically possible. We indicate if we use stretched coordinates transversally in $(x,y)$ \footnote{Stretched coordinates in $x,y$ (see \ref{subsubsec:transform}) for some of the tests: stretching factor 2, inner zone $10\mum$, transition zone $15\mum$. \label{foot:trf1}}.

\subsection{Laser pulse propagation in vacuum  (2D)} \label{subsubsec:test_2d_vacuum}

\begin{figure*}[t]
\centering
\includegraphics[height=6.2cm]{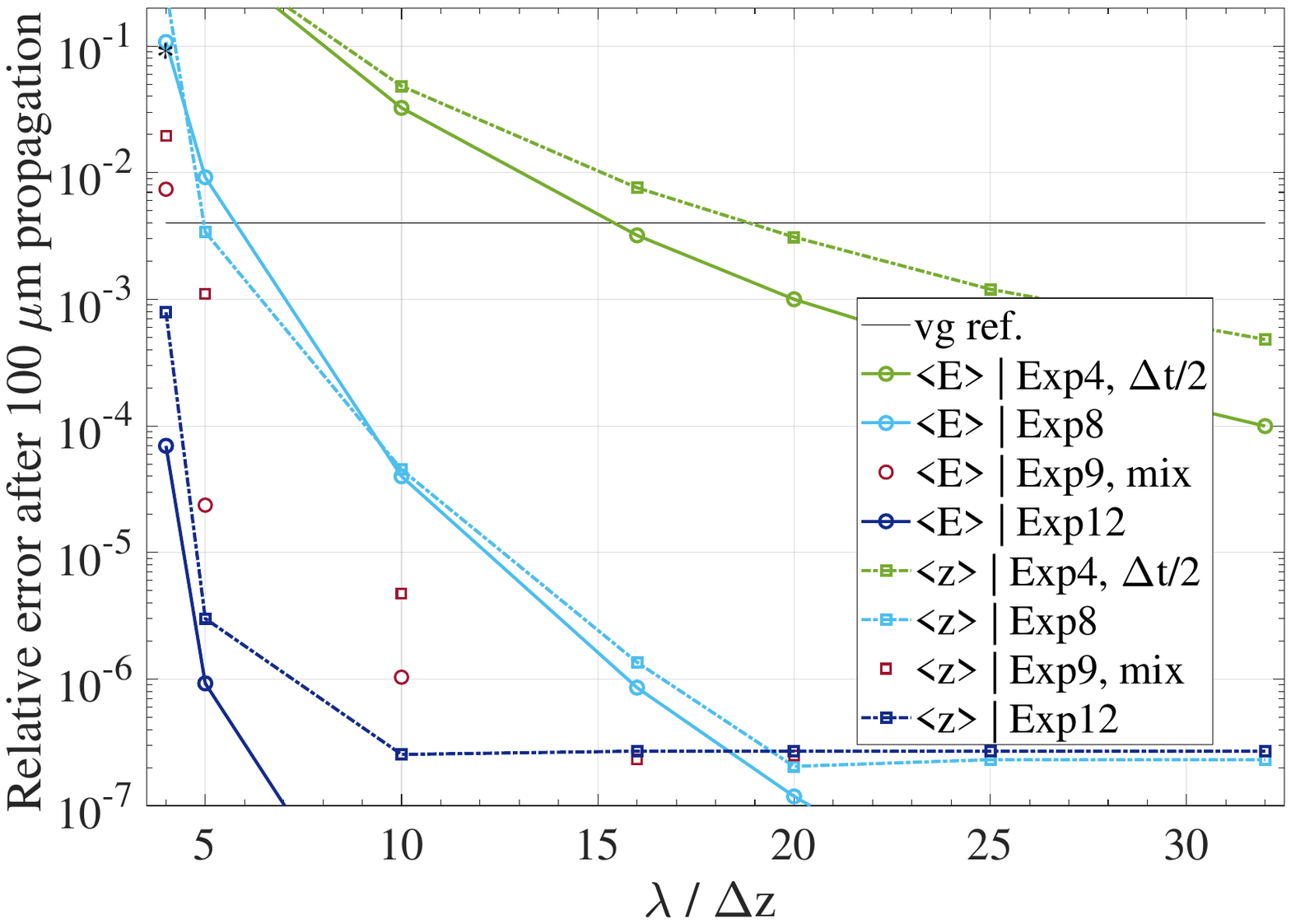}
\includegraphics[height=6.2cm]{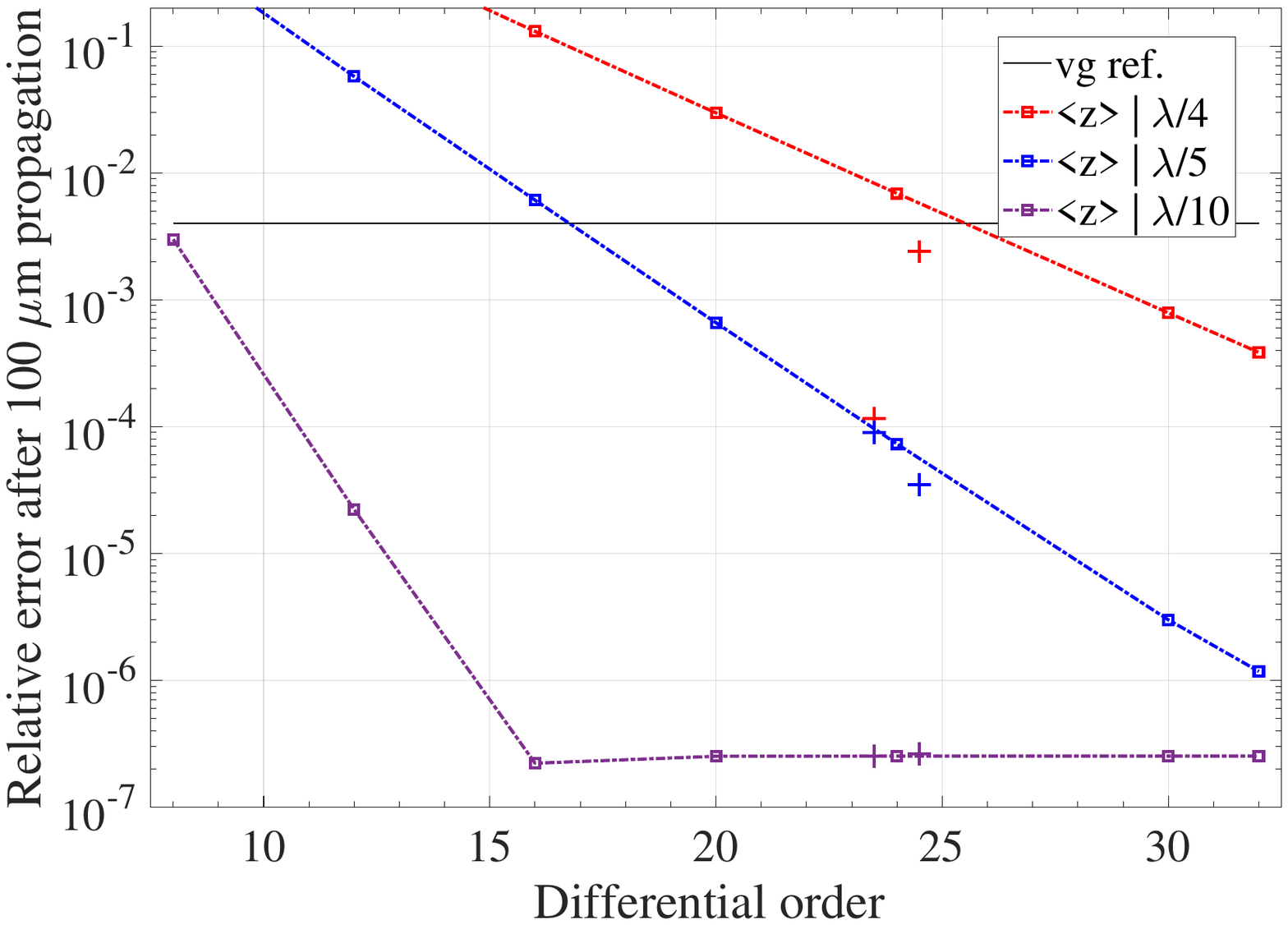}
\caption{ Error characteristics of a laser pulse of \revise{$\lambda=0.8\mu m$} wavelength after $100\mu m$ propagation in vacuum. \emph{On the left} we show the relative energy loss $\delta\left<E \right>$ (solid lines, circle markers) and the mean distance error $\delta\left<z\right>$ of the pulse centroid (in $\lambda$ units, dotted lines, square markers) versus the \revise{$N = \lambda/\Delta z$} resolution with various exponential orders at  $\Delta t = \Delta z /c$ using 30th order finite differences. (The 4th order exponential (green) was run at $\Delta t = 0.5 \Delta z/c$. We ran the 9th order exponential method (dark red) by mixing it with 8th order exponential steps to become stable.) \emph{On the right} we show the mean distance error $\delta\left<z\right>$ of the pulse centroid using 12th order exponential versus the finite difference order at $\lambda/4$ (red), $\lambda/5$ (blue) and $\lambda/10$ (purple) resolution. The + symbols denote the results of our two enhanced 24th order finite difference formulas (see \ref{subsubsec:enhance}). To assess the dispersion errors with specific exponential and differential orders one needs to take the largest error value on both figures. We show the value of the analytical centroid shift $t_{\max}(c-v_g)/\lambda$ of the laser pulse with solid lines ($0.004\lambda$).  *The 8th order exponential (light blue) was unstable at $\lambda/4$ resolution, the result were extrapolated using $\Delta t = 0.5 \Delta z/c$. \revise{This is due to $\Delta x < \Delta z$ in the 2D CFL value.}
\label{fig:test_vacuum_2d}} 
\end{figure*}

In our first test we propagate a laser pulse of $\lambda=0.8\mum$ carrier wavelength in vacuum over a temporal duration of $100\mum/c$, we place this pulse such that it is initially at $z_0=-50\mum$ from its focus position ($z=0$) and after the propagation it should be at position $-z_0$.  We perform this test in 2D Cartesian geometry with varying $\Delta z=\lambda/N$ resolution. We performed numerical simulations in $140\mu m\times60\mum$ moving box with $w_0=16\mum$ laser spot size at the focus, and laser pulse length of $\tau=10\mum/c$. We used fixed transverse resolution $\Delta x = 0.2\mum$ with stretched coordinates\footref{foot:trf1}, the presence of the latter did not affect the accuracy. We choose $\Delta t = \Delta z/c$ unless otherwise stated.

In this test we analyze our Maxwell solver's accuracy properties by analysing the  $\left< E \right>$ pulse energy and the error of its conservation property $\delta\left< E \right>$:
\begin{equation} \label{eq:test_2d_vacuum_E}
\left< E \right> = \int \left(  \left| {\bf E} \right|^2 + c^2\left| {\bf B} \right|^2  \right) {\rm d}V,  \quad
\delta \left< E \right> = \frac{\left< E \right>-\left< E \right>_0}{\left< E \right>_0}
\end{equation}
and for the dispersion error the position of the $\left<z \right>$ pulse centroid and its error $\delta \left<z \right>$:
\begin{equation} \label{eq:test_2d_vacuum_Z}
\left< z \right> = \left< E \right> ^{-1} \int z \left(  \left| {\bf E} \right|^2 + c^2 \left| {\bf B} \right|^2 \right)  {\rm d}V, \ \
\delta \left< z \right> = \frac{\left< z \right>+\left< z \right>_0}{\lambda}.
\end{equation}
It is worth noting that there exists an analytical solution to the $v_g$ group velocity  for the pulse centroid \cite{esarey1995tightly_focused_pulses}:
\begin{equation} \label{eq:test_2d_vacuum_vg}
\frac{c-v_g}{c} = \frac{1}{2} \left(\frac{\lambda}{2\pi w_0} \right)^2 \quad  \text{in 2D},
\end{equation}
which is approximately $-3.17\times10^{-5}$ in our case, which means $0.004\lambda$ centroid delay for this $100\mum$ propagation test. We note that initially the pulse centroid $\left< z \right>_0$ is not exactly at $-50\mum$, it is such that $v_g \approx -2c \left<z \right>_0/100\mu{\rm m}$ with negligible error.

On Fig. \ref{fig:test_vacuum_2d} we summarize the error characteristics of our method in this test scheme, and these errors scale linearly with the propagation distance (for example, $10^2\times$ for 1cm). On Fig. \ref{fig:test_vacuum_2d} (left panel) we can see the errors of the exponential expansions with 30th order finite differences versus spatial resolution (and the temporal resolution $\Delta t = \Delta z /c$). We can see that the error curves of the $\delta\left< E\right>$ energy loss and the relative pulse centroid delay $\delta\left<z\right>$ are of same order for a given exponential. Unfortunately, in this frequency range our primary concern is $\delta\left< E\right>$ energy loss (solid lines). We can see that the exponential expansions follow the expected convergence with $\Delta t$. We can see that the 4th order exponential is of very low accuracy (this was run with $\Delta t = 0.5 \Delta z /c$), and the 8th order exponential is the one that is actually useful. However, even the latter turns out to be inaccurate at higher frequencies, which suggests that 12th order expansion is needed for near analytical precision. These exponentials converge using smaller $\Delta t$, so by doubling only the temporal resolution, the errors reduce by $1/16$, $1/256$ and $1/4096$ respectively.
 
We also indicated a special case of a hand-tuned 9th order exponential expansion (using the properties discussed in Section \ref{subsubsec:exponential_advect} and Fig. \ref{fig:exponential_dispersion}). In this we followed 14 time steps of 9th order exponential propagation (wave amplification) with an 8th order one (wave attenuation), we show these with dark red squares and circle markers on Fig. \ref{fig:test_vacuum_2d}.  This sequence caused this method to become stable, it also suggests that it is possible to customize the exponential Taylor expansion coefficients to improve accuracy. In this case we improved the energy conservation by 2 orders of magnitude at $\Delta z = \lambda/5$ resolution. We also note that the exponential expansion itself causes additional wave dispersion error: we calculated decreased group velocity for 4th, 8th, 12th order exponential, but increase for 9th order exponential.

In addition to the exponential expansion the choice of the finite differences also cause dispersion errors which affect the centroid $\delta\left<z\right>$ (dashed lines, square markers). We show this latter on Fig. \ref{fig:test_vacuum_2d} (right panel) versus the finite difference orders using 12th order exponential expansion. For reference we show the value analytical centroid delay $0.004\lambda$ with a horizontal line, if any of the $\delta\left< z \right>$ error curves go below the latter this effect can be numerically resolved (in 2D). We can also see that the rate of convergence of high order differences slows down dramatically for short $\lambda =5\Delta z$, $4\Delta z$ wavelengths. We also show our dispersion enhanced 24th order finite differences with + markers (see \ref{subsubsec:enhance} , ${\rm D}_2^{(\pm,24)}$, and ${\rm D}_3^{(\pm,24)}$ are shown at 23.5 and 24.5 orders). We can see that by customizing the finite difference coefficients we can drastically improve the high frequency $\approx1/(4\Delta z)$ dispersion behavior. Unfortunately, to propagate the highest frequencies ($>1/(4\Delta z)$) accurately pure spectral solution is required.

%In practice, to use this exponential method we suggest the following. First, choose a target wavelength  $\lambda = 10 \Delta z$, $\lambda = 5 \Delta z$, $\lambda = 4 \Delta z$ or to be resolved. Then choose a finite difference order (for these, approximately 12th, 24th, 30th orders or more). Overall, to assess the dispersion and norms errors with specific exponential and differential orders you need to take the largest error value on both panels of Fig. \ref{fig:test_vacuum_2d} and multiply it with a chosen propagation distance. By properly selecting these, negligible dispersion and norm loss can be achieved for long distances. The high accuracy has a cost, it is 10-100 times the computation than the Yee-method, but it is reliable and much more accurate. However, it is still less accurate than the pure spectral methods.

\revise{
In practice, to use this exponential method we suggest the following. All numerical errors that we show in Fig. \ref{fig:test_vacuum_2d} scale linearly with the propagation distance, and they correspond to $125\lambda$ distance wave propagation. When setting up a simulation one knows the smallest $\lambda$ wave of physical interest, and how far that wave needs to propagate (or the total temporal duration). Then one chooses the spatial resolution, for example such that $\lambda = 10 \Delta z$, $\lambda = 5 \Delta z$, $\lambda = 4 \Delta z$. Then the errors on Fig. \ref{fig:test_vacuum_2d} are needed to be scaled appropriately to the problem at hand. Afterwards one chooses a finite difference order based on how much dispersion error is tolerated (in the three examples approximately 12th, 24th, 32th orders differences are needed for $10^{-4}\lambda$ error after $125\lambda$ distance). Finally, one chooses the order of the exponential and the temporal step $\Delta t$ the same way such that the dissipation error are within tolerance. Unlike Yee-methods choosing a smaller $\Delta t$ always improves the result. In this way it is possible to scale the numerical complexity to a particular problem, or just use very high order expansions and compute more. By properly selecting these, negligible dispersion and norm loss can be achieved for long distances. 
}

\subsection{Fields of a relativistically travelling particle (2D)} \label{subsubsec:test_2d_ncr}

\begin{figure*}[tp!]
\centering
\includegraphics[trim={115px 0 100px 12},clip,width=5.5cm]{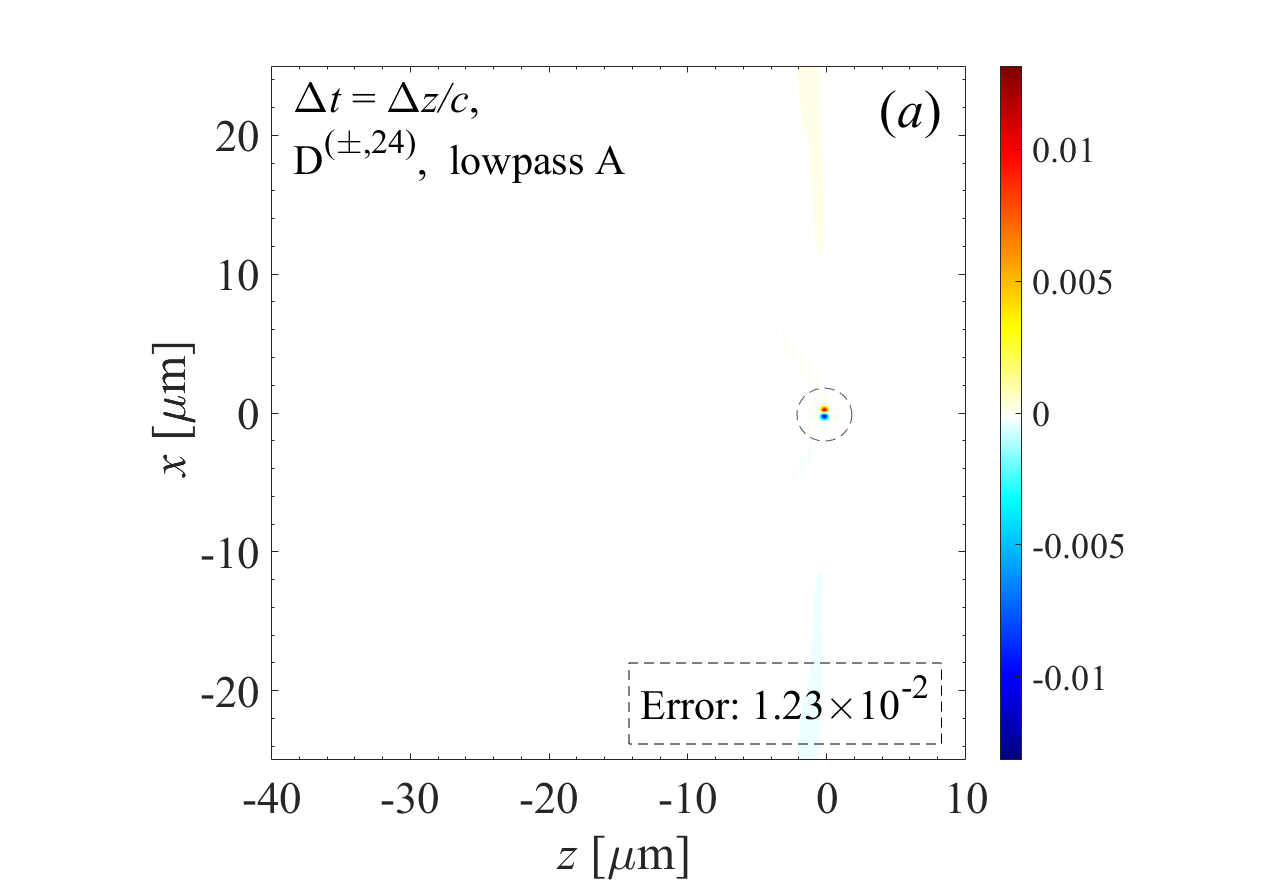}
\includegraphics[trim={115px 0 100px 12},clip,width=5.5cm]{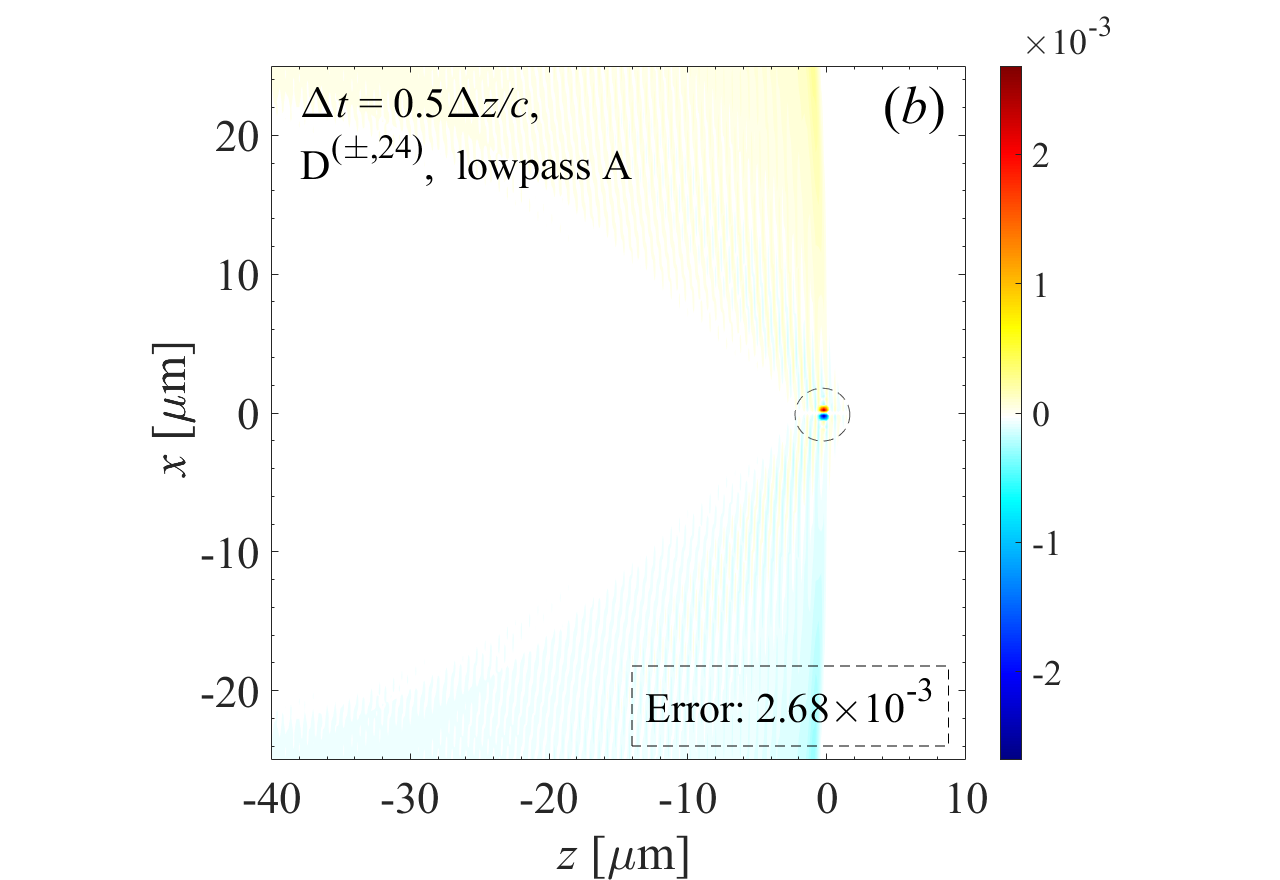}
\includegraphics[trim={115px 0 100px 12},clip,width=5.5cm]{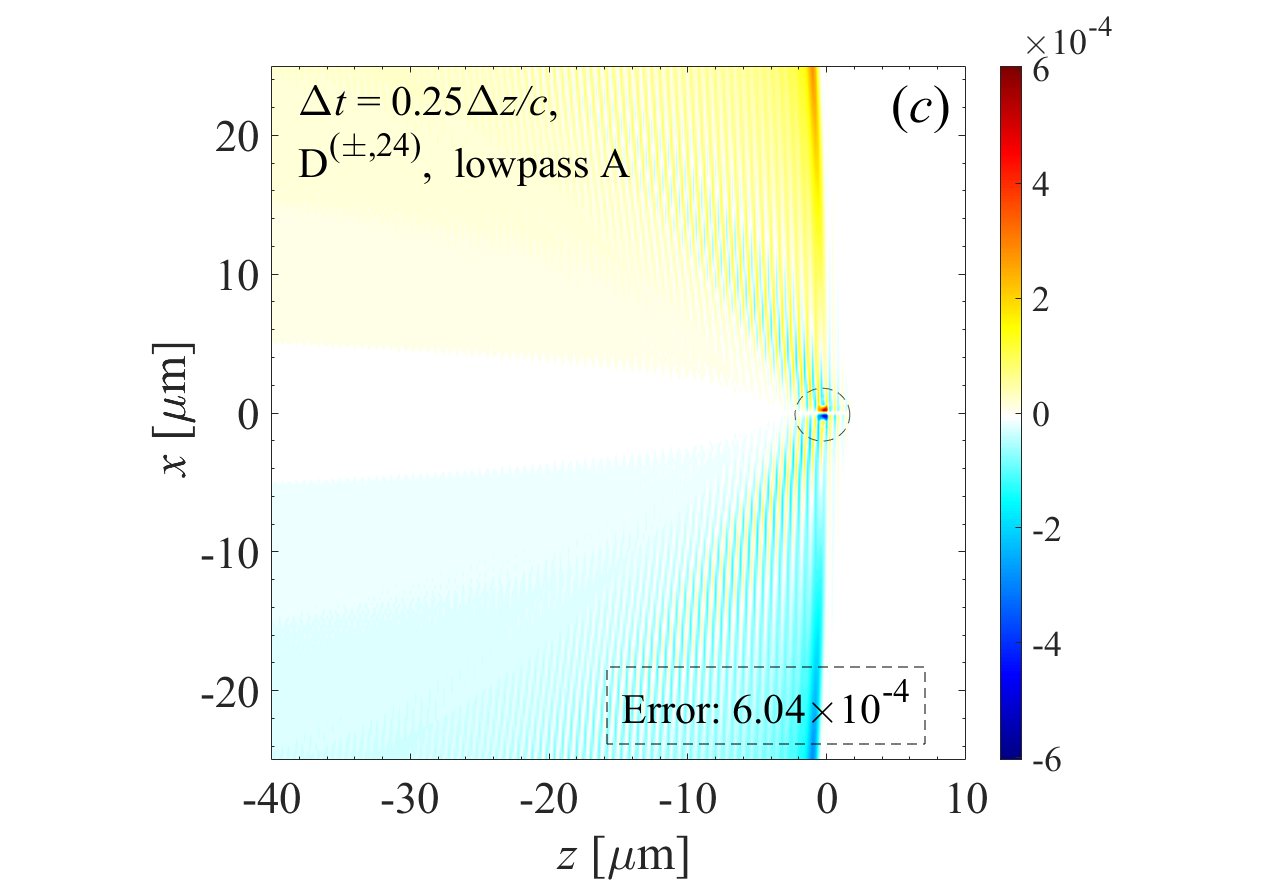}
\caption{The normalized field $E_x-cB_y$ of a relativistically travelling particle bunch after $500\mu m$ propagation with   $\Delta t = \Delta z/c$ $(a)$, $\Delta t = 0.5\Delta z/c$ $(b)$, and $\Delta t = 0.25\Delta z/c$ $(c)$ using 24th order differentials with lowpass A current filter. We indicated the approximate amount of erroneous fields in the particle bunch in the lower right corners. The NCR fields can also be observed. 
\label{fig:test_ncr_fields}}
\end{figure*}

\begin{figure*}[tp!]
\centering
\includegraphics[trim={100px 0 224px 35},clip,height=4.8cm]{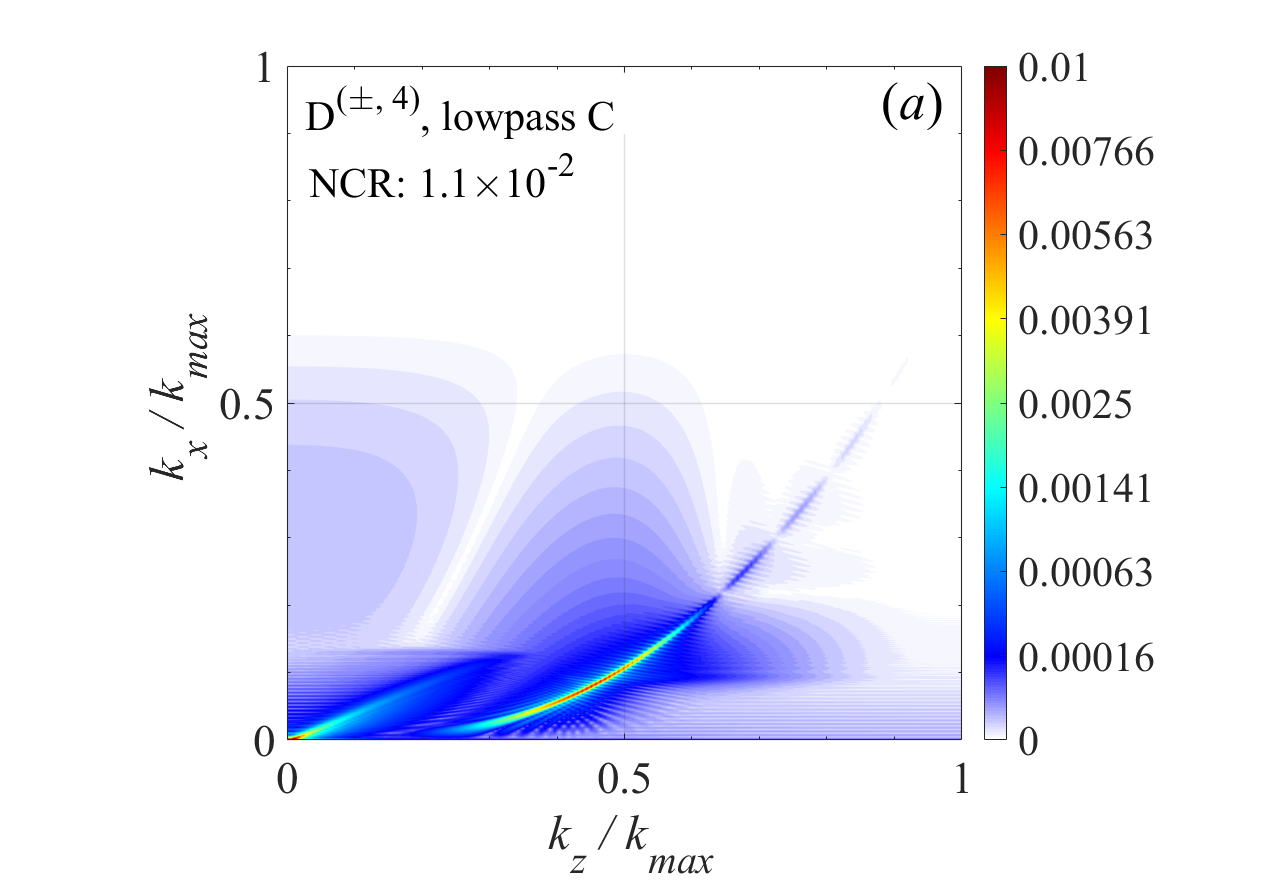}
\includegraphics[trim={198px 0 224px 35},clip,height=4.8cm]{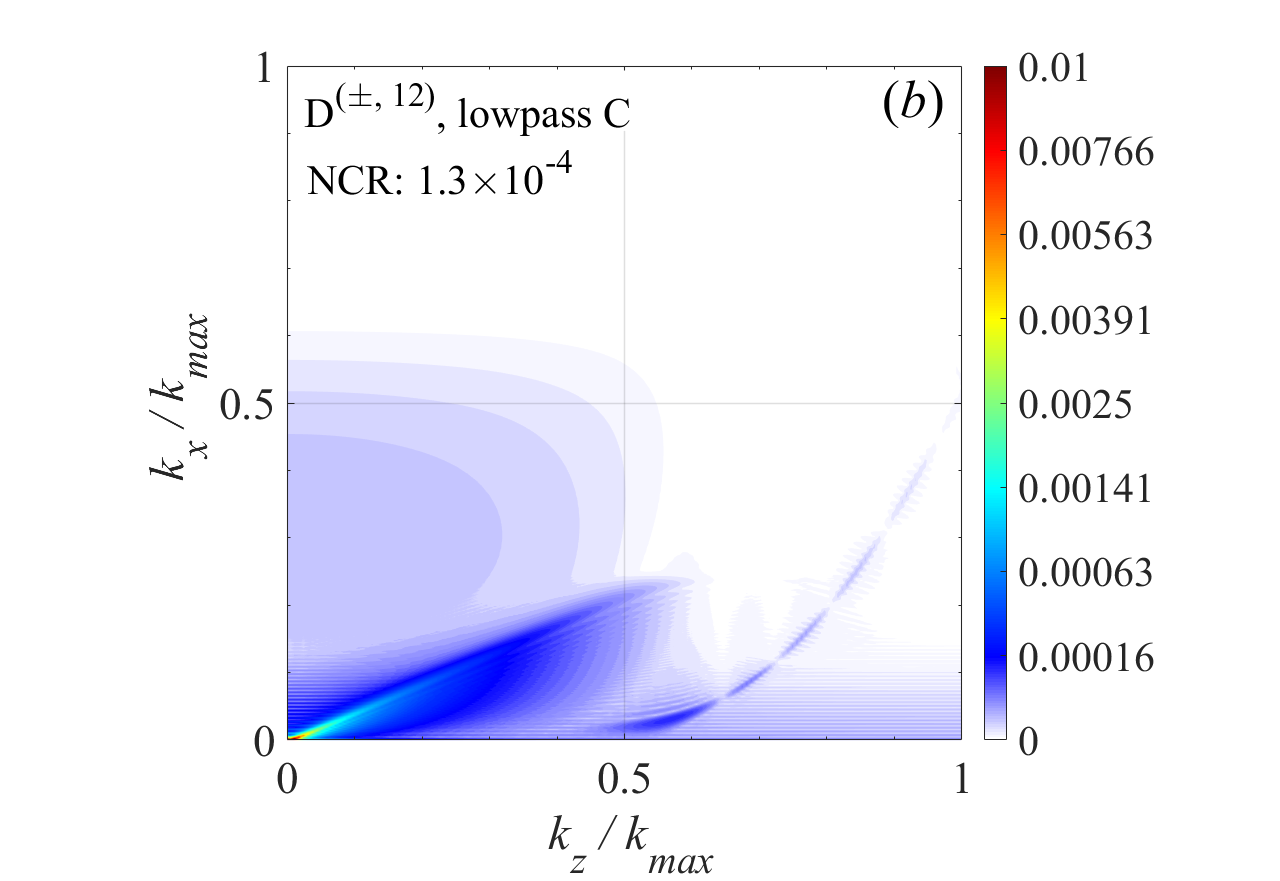}
\includegraphics[trim={198px 0 224px 35},clip,height=4.8cm]{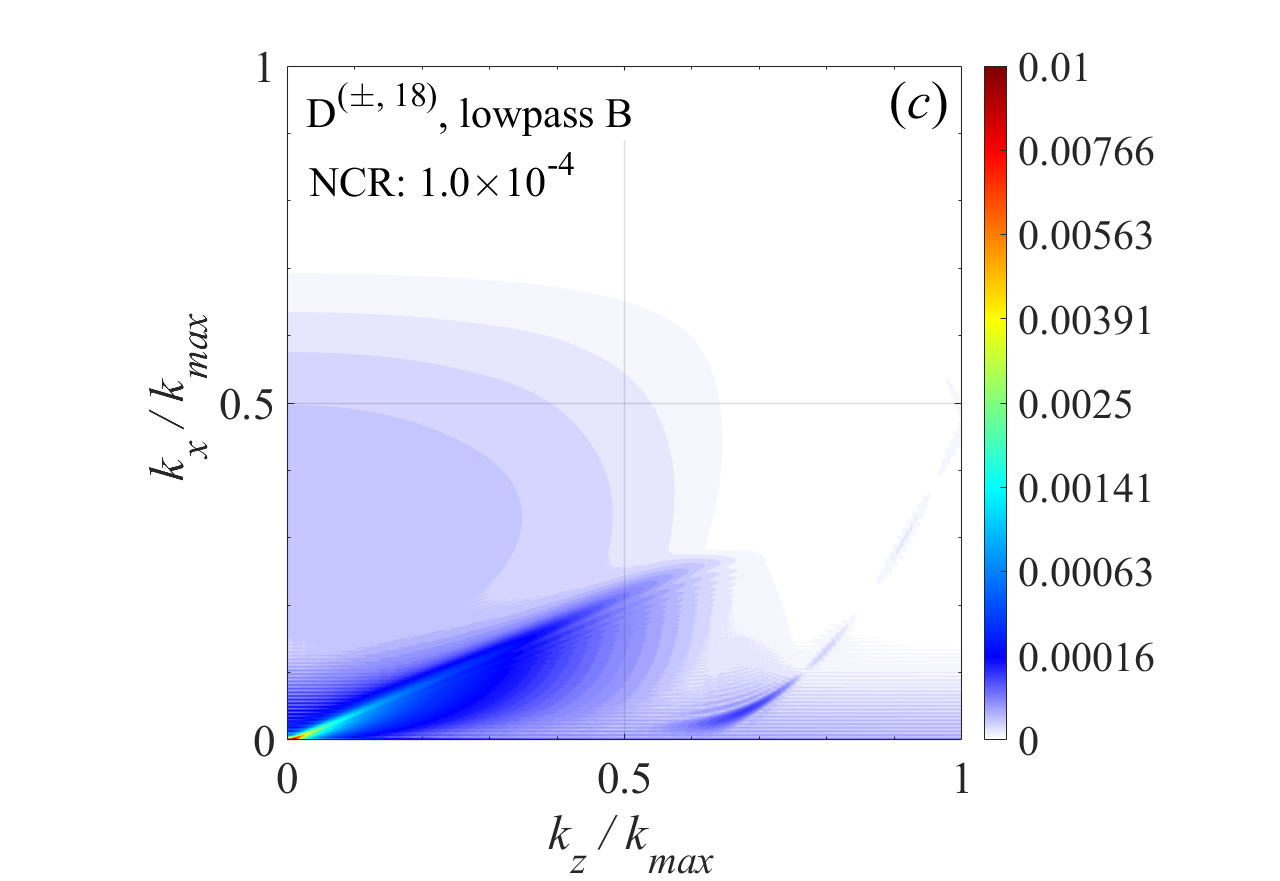}
\includegraphics[trim={196px 0 80px 35},clip,height=4.8cm]{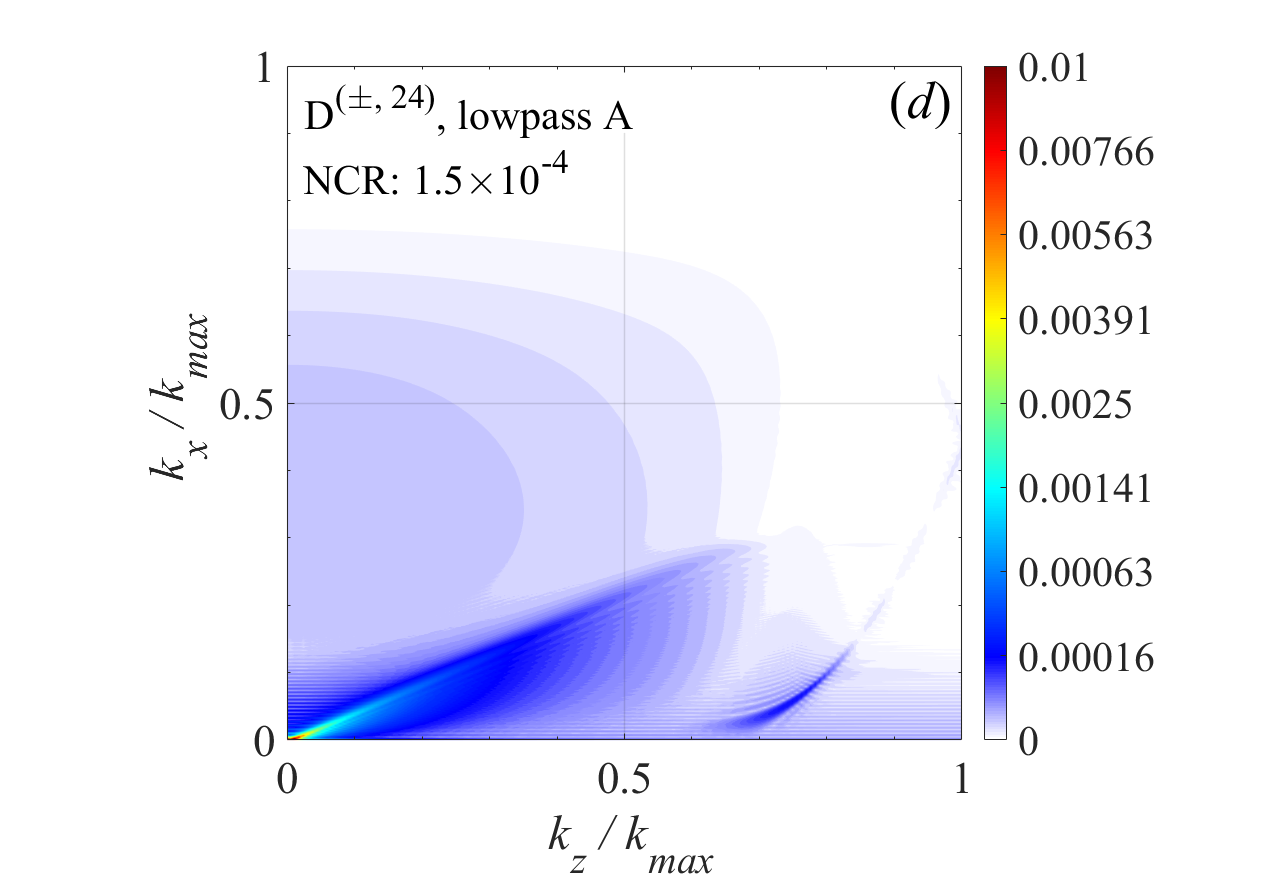}
\includegraphics[trim={100px 0 224px 30 },clip,height=4.8cm]{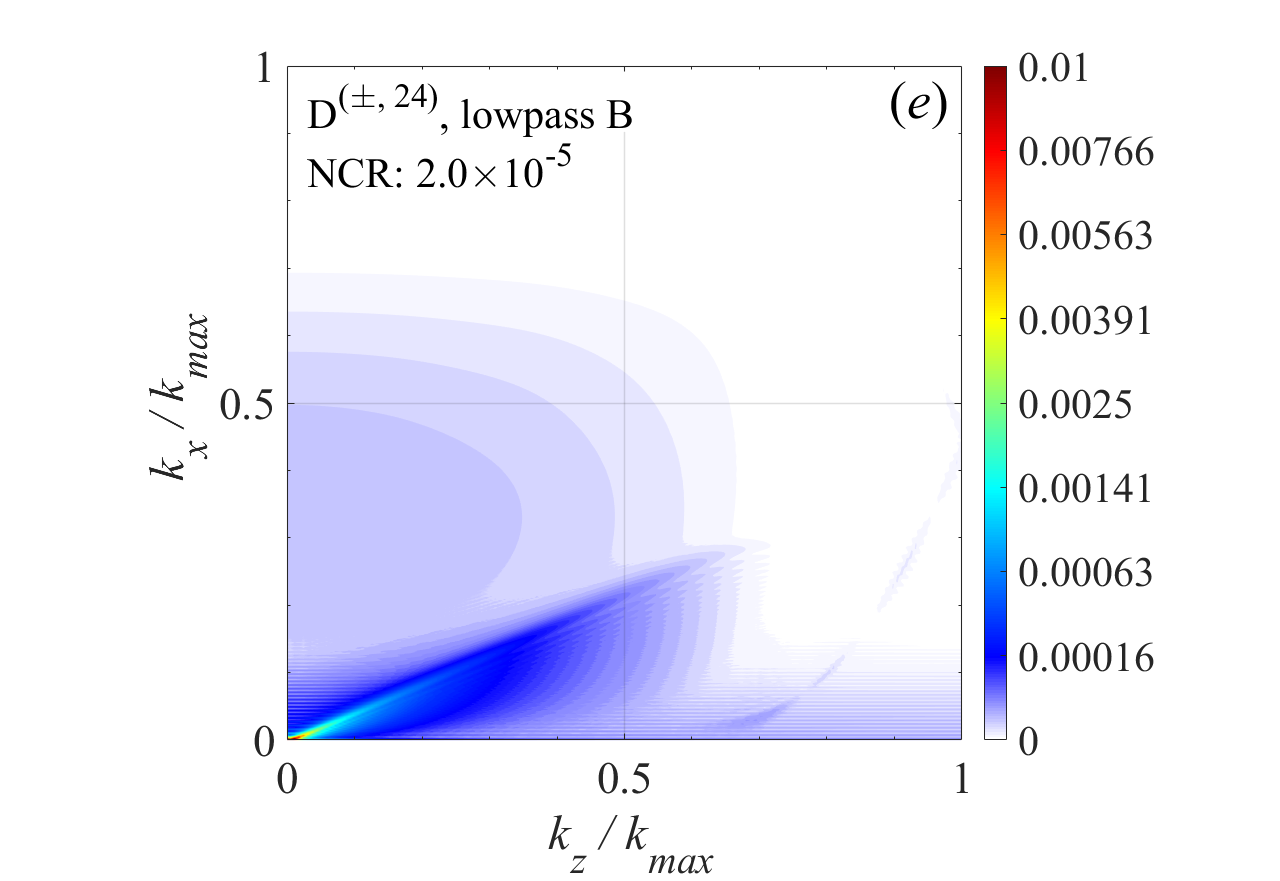}
\includegraphics[trim={198px 0 224px 30},clip,height=4.8cm]{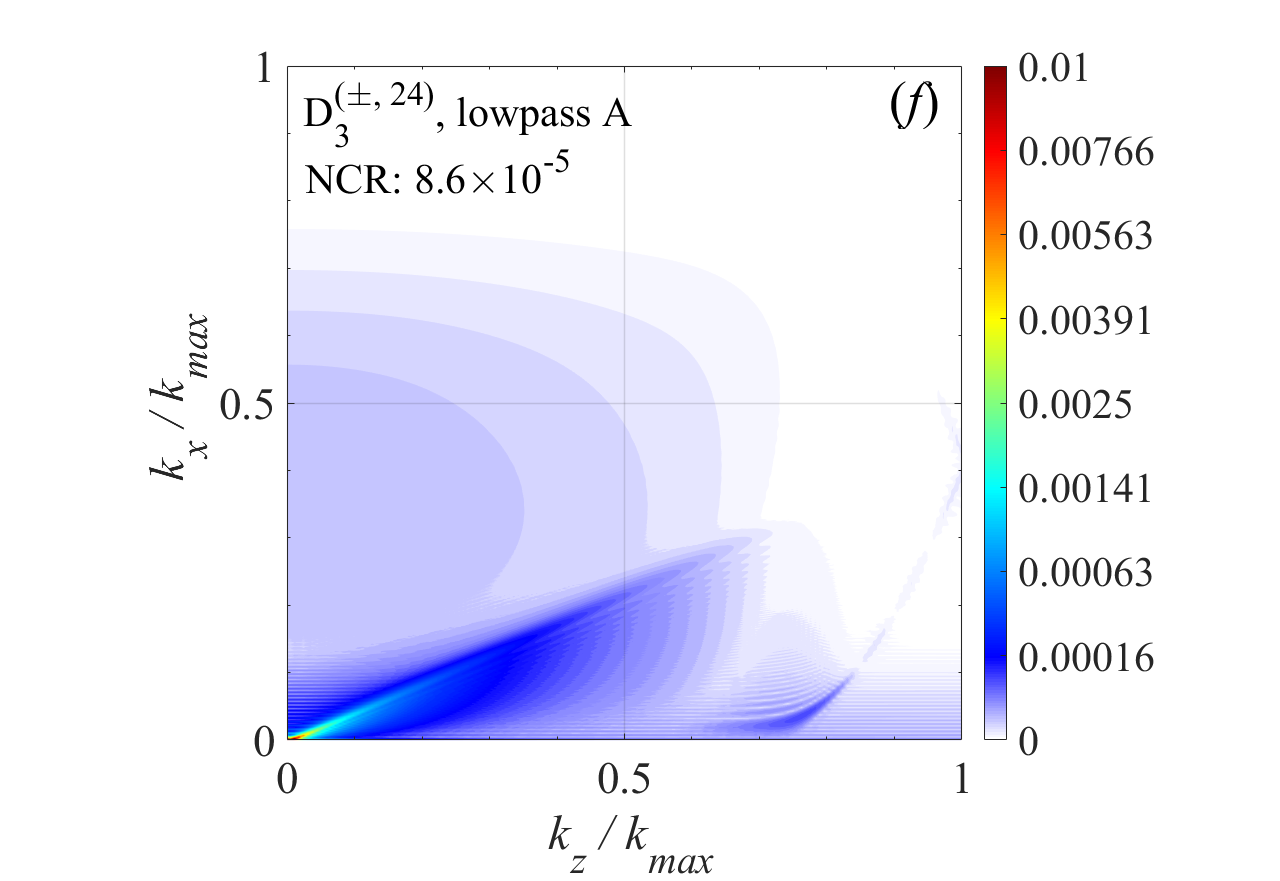} 
\includegraphics[trim={198px 0 224px 30},clip,height=4.8cm]{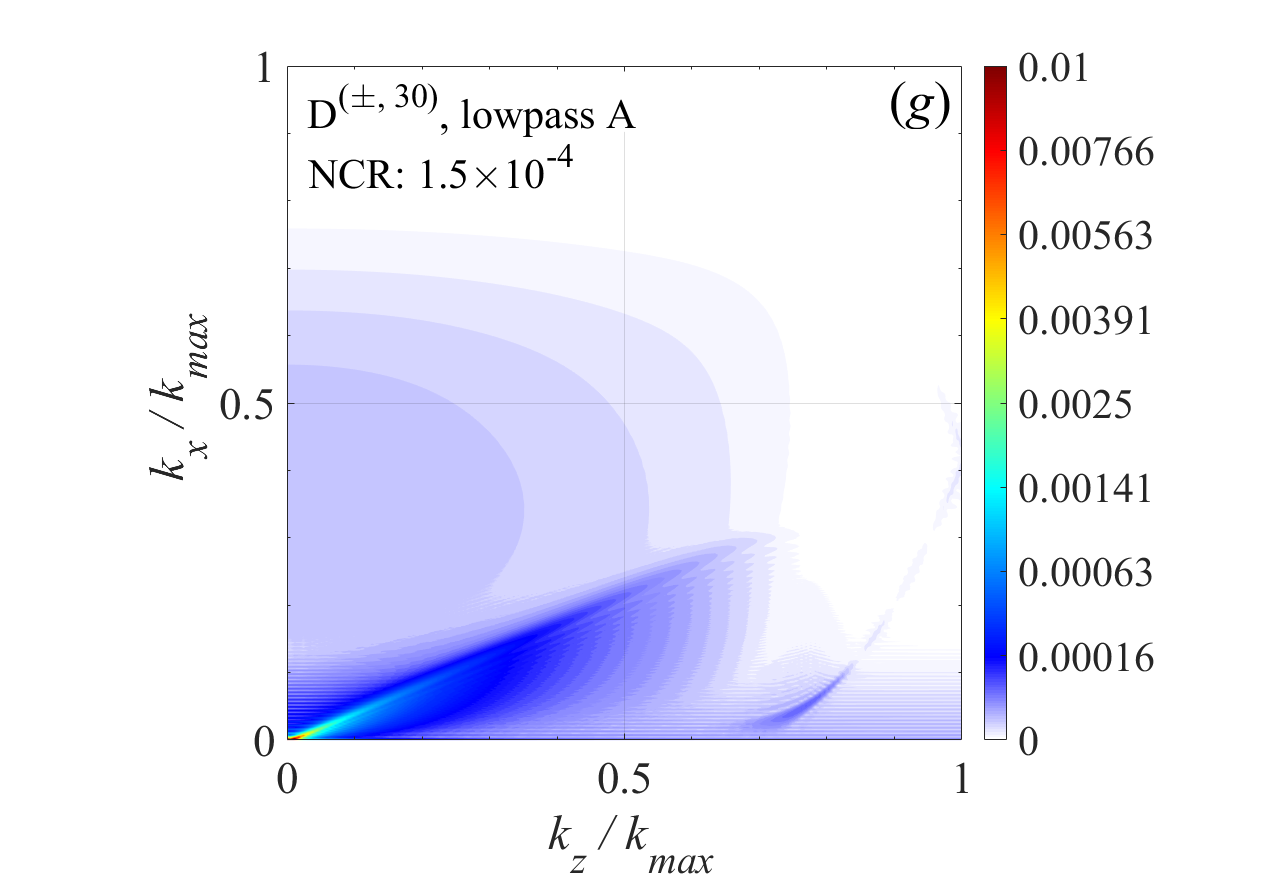}
\includegraphics[trim={198px 0 80px 30},clip,height=4.8cm]{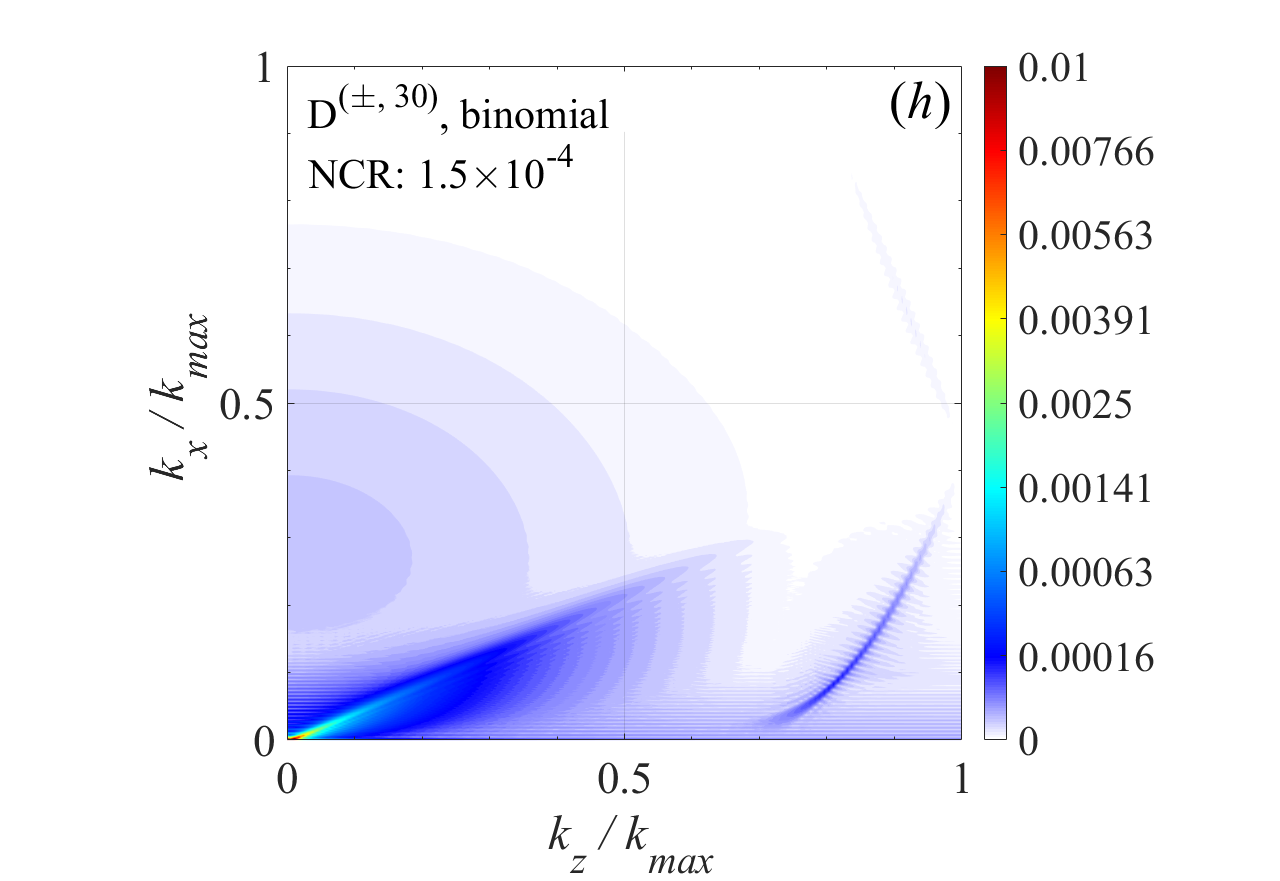}
\caption{
The plot of the \emph{square root} of the normalized spectrum of $E_x-cB_y$ of a relativistically travelling particle bunch after $500\mu m$
propagation in the moving window using different combination of finite differences and current filters with $\Delta t = 0.5\Delta z/c$. The trace of NCR can be seen at high frequencies as parabolas, we also indicated the approximate magnitude of these.
\label{fig:test_ncr_fourier}}
\end{figure*}

\begin{figure*}[tp!]
\centering
\includegraphics[trim={116px 0 100px 35},clip,width=5.5cm]{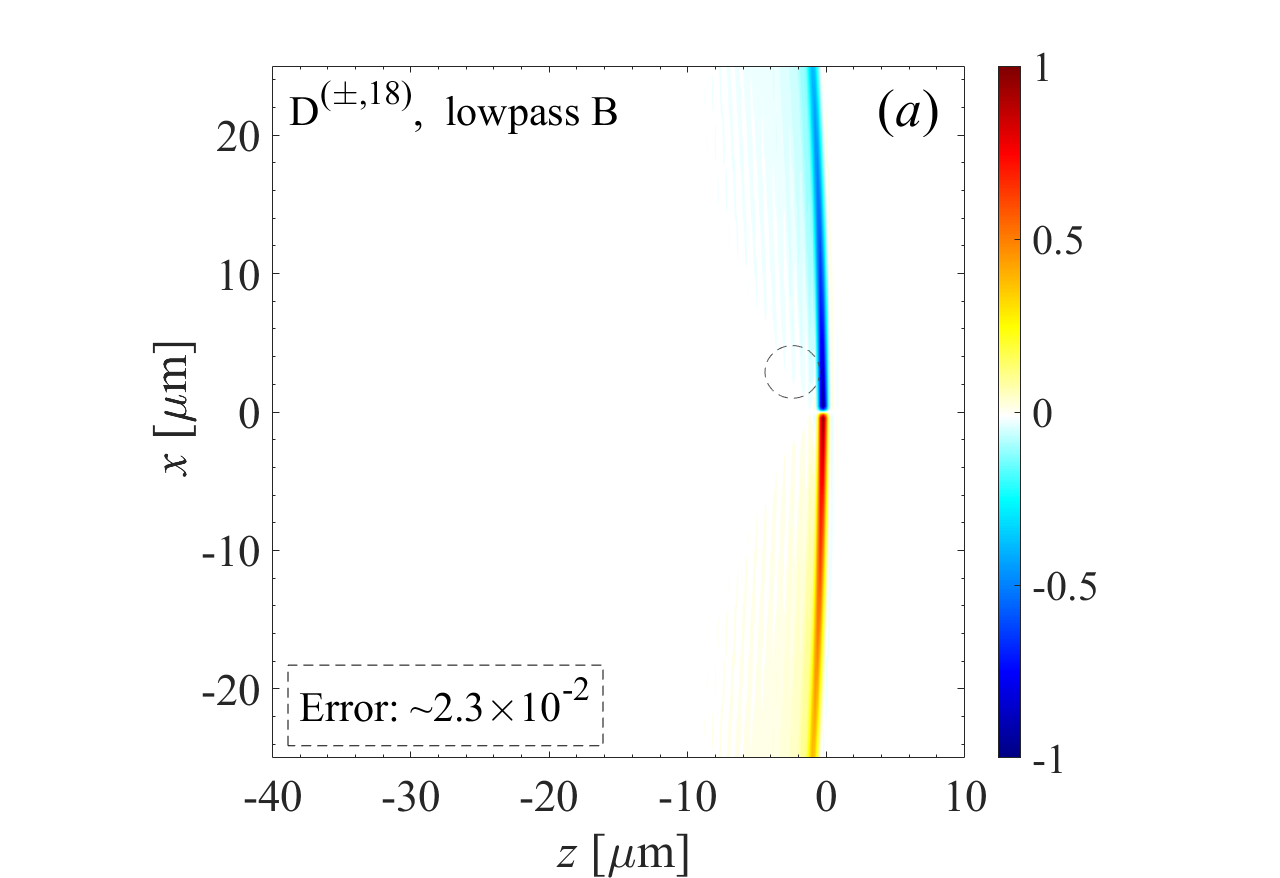}
\includegraphics[trim={116px 0 100px 35},clip,width=5.5cm]{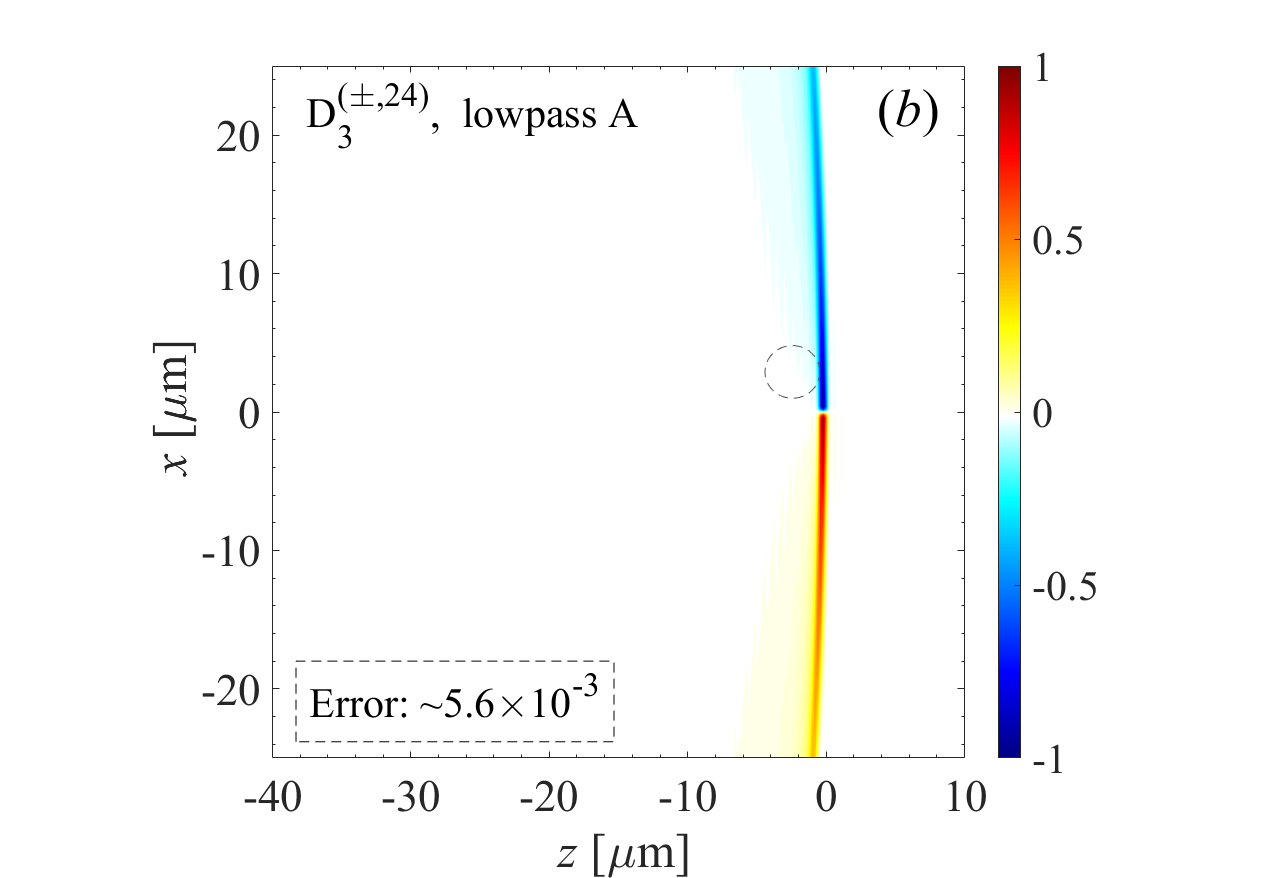}
\includegraphics[trim={116px 0 100px 35},clip,width=5.5cm]{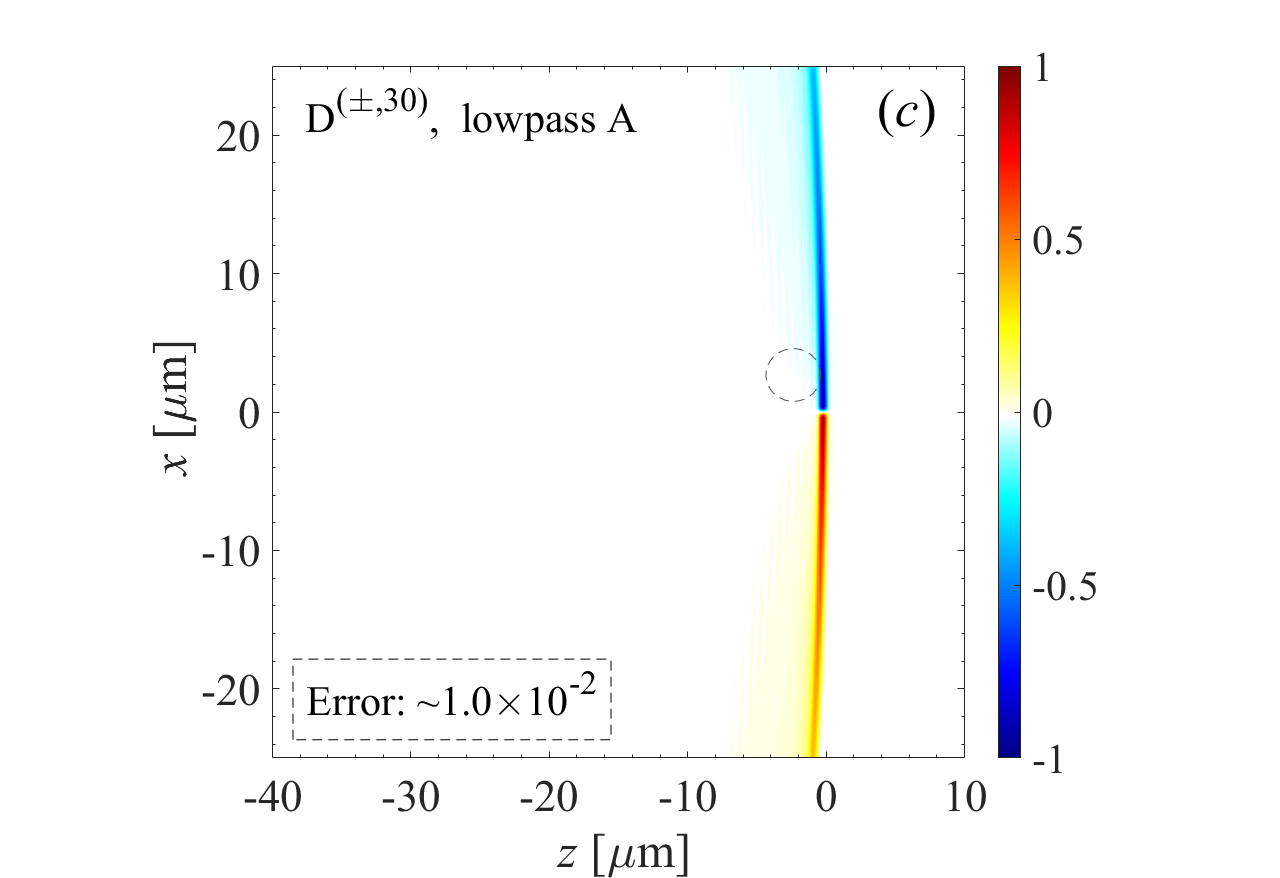}
\caption{The normalized field $E_x+cB_y$ of a relativistically travelling particle bunch after $500\mu m$ propagation with 18th $(a)$, enhanced 24th  $(b)$ and 30th $(c)$ finite differences using current filters and $\Delta t = 0.5\Delta z/c$. We indicated the approximate amount of the oscillations behind the main field, which is due to the dispersion error.
\label{fig:test_ncr_propagation}}
\end{figure*}

In the second test we accelerate  a very small electron bunch along direction $z$ over $50\mum$ distance to $\sim200\gamma$ using an external force, then we propagate it further for $450\mum/c$ duration. The geometry is Cartesian 2D. Initially, this electron bunch overlaps with a proton bunch such that the net charge is zero initially, for simplicity. This electron bunch has a profile of $\exp(-(z^2+x^2)/\Delta z^2)$ with $\Delta x = \Delta z = 0.2\mum$, and the center of the bunch  is located at $z=-500\mum$, $x= 0$ initially. We propagated the fields using 12th order exponentials typically with $\Delta t = 0.5 \Delta z /c$, in a $500\mum\times60\mum$ moving window. This relativistically moving $\sim1$ grid point wide electron spike will emit radiation during the acceleration, but after that its velocity is constant and the \emph{numerical Cherenkov radiation} (NCR) appears, depending on the solver, which propagates forward and becomes visible on Fig. \ref{fig:test_ncr_fields}, also shown in the spectral space in on Fig. \ref{fig:test_ncr_fourier}. In Fig. \ref{fig:test_ncr_propagation} the self-field of the particle bunch is shown, which is the relativistically moving Coulomb field. Because the particle bunch is very narrow, the spectral distribution of the deposited current $\bf J$ will contain very high spatial frequencies (above $0.5\kk_{\max} = 2 \pi /4 \Delta z$) which will contain spurious numerical effects.  For simplicity, we also disabled the self-consistent interaction between the electron bunch and its own radiation. This can be viewed as the worst case electron acceleration scenario numerically, and getting this right is deliberately tough.

In the latter spectral range our high order exponential Maxwell-solver cannot provide high accuracy, our goal here is to show that we can prevent the fields to do anything totally wrong, ideally by removing them, using the combination of high order finite differences (see Section \ref{subsec:differences}) and current filters (see Section \ref{subsec:spatial_current}).  These numerical effects are the \emph{wave dispersion} and the \emph{zeroth-order numerical Cherenkov radiation} (NCR). For analysis purposes we selected two fields of interest: the forward propagating wave $E_x+cB_y$, and the non-forward propagating component $E_x-cB_y$. For the relativistically moving electron bunch the latter also represents the Lorentz-force along the $x$ direction. We normalized the field quantities to the maximal value of  $|E_x+cB_y|$. The field component $E_x-cB_y$ is sensitive to spurious numerical effects, and it is our main subject for our analysis.

The first effect that we found is that there is a spurious field component in $E_x-cB_y$ in the electron bunch, see Fig. \ref{fig:test_ncr_fields} - two dots in the dashed circle. We ran this example with 24th order finite differences and lowpass A current filter. We indicated the magnitude of this erroneous fields on these figures. Surprisingly, we found that this error does not depend on dispersion accuracy of the field solver. Instead, it shows $\sim\Delta t^2$ convergence as it diminishes with smaller $\Delta t$ which indicates that it is created by the $\Delta t^2$ accurate (midpoint) source integral of the deposited current (see Eq. (\ref{eq:exponential_solution2})). Thus this is a self-consistent temporal effect that could limit the particular $\Delta t$ chosen not to be larger than $\Delta z/c$. This effect is dependent on  the bunch charge per its spatial radius. \revisee{We note that this spurious field dot also occurs in the $E_z$ component with the same magnitude - we also found that how it is distributed across these field components is different if we use trapezoid or constant source approximation from  \ref{subsubsec:exponential_source}, even though the total errors are the same.}  We could also see the leftover numerical Cherenkov radiation at $\Delta t = 0.25 \Delta z /c$ after the current filtering on Fig. \ref{fig:test_ncr_fields}$(c)$.

Our next step is to analyze the behaviour of the NCR. For this we did the spatial Fourier-transform of $E_x-cB_y$ fields with various finite difference and current filter combinations, we show these on Fig. \ref{fig:test_ncr_fourier}. The spectral profile of the NCR looks like a parabola at high frequencies \cite{lehe2013numerical,lehe2016fbpic}. Using 4th order finite differences, we can clearly see this on Fig. \ref{fig:test_ncr_fourier} $(a)$ with a relative spectral amplitude of $10^{-2}$ which is consistent with usual Yee-simulations. Using higher order differences pushes the NCR higher in the spectrum, and interestingly even using 12th order differences it is already concentrated above $0.5 \kk_{\max}$. Using even higher order differences we can push the NCR even more up spectrally and we can use less obtrusive current filters. We note that using a single pass binomial filter is not effective at removing the NCR fields (see Fig. \ref{fig:test_ncr_fourier} $(g)$, $(h)$). Overall by carefully selecting the proper current filters we can reduce the NCR by 2 orders of magnitude to the level of $10^{-4}$ (see Fig. \ref{fig:test_ncr_fourier} $(b)$, $(c)$, $(d)$). Using more aggressive filtering the NCR can be removed even further (see Fig. \ref{fig:test_ncr_fourier} $(d)$, $(e)$). This strategy is the same as the reduction of higher order NCR effects in pure spectral solvers \cite{godfrey2014stability_spectral}.

Finally we also show the forward propagating fields $|E_x+cB_y|$ on Fig. \ref{fig:test_ncr_propagation}. Our high order solver is capable of propagating this field structure relatively intact at the speed of light, with current filtering. Problem is that there are still spurious oscillations behind it, which are caused by the wave dispersion error (we indicated the magnitude of these on the panels' lower left corners). Unfortunately this wave dispersion error is still at the order of $10^{-2}$, which is actually two order of magnitude larger than NCR. We note that our 24th order dispersion enhanced differences ${\rm D}_3^{(\pm,24)}$  (see \ref{subsubsec:enhance}) actually result less dispersion error in this case than the staggered 30th order formula (see Fig. \ref{fig:test_ncr_propagation} $(b)$, $(c)$). Curiously, the latter reduces the magnitude of the NCR more (see Fig. \ref{fig:test_ncr_fields} $(f)$, $(g)$). Using stronger lowpass filters also reduce this effect.

Overall, to achieve the best accuracy very high order finite differences or their enhanced versions are required in this example. To suppress the spurious high frequency radiation a properly chosen lowpass current filter is mandatory. Customized derivative and current filter coefficients could significantly enhance the efficiency of the solver, but those may be investigated elsewhere.

\subsection{Accurate force of a laser on a copropagating electron (2D)} \label{subsubsec:test_2d_force}

 \begin{figure*}[tp!]
\centering
\includegraphics[height=5.6cm]{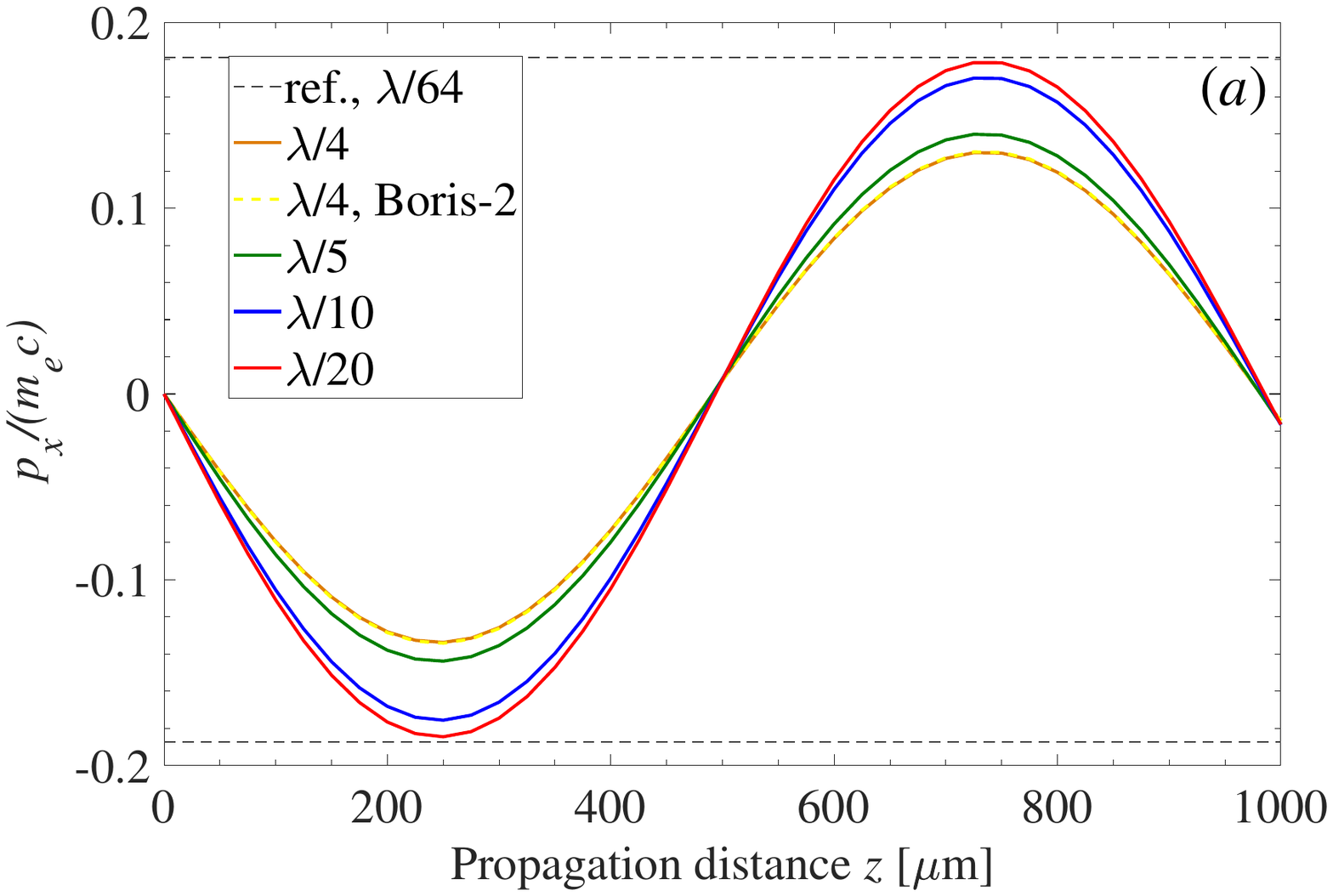}
\includegraphics[height=5.6cm]{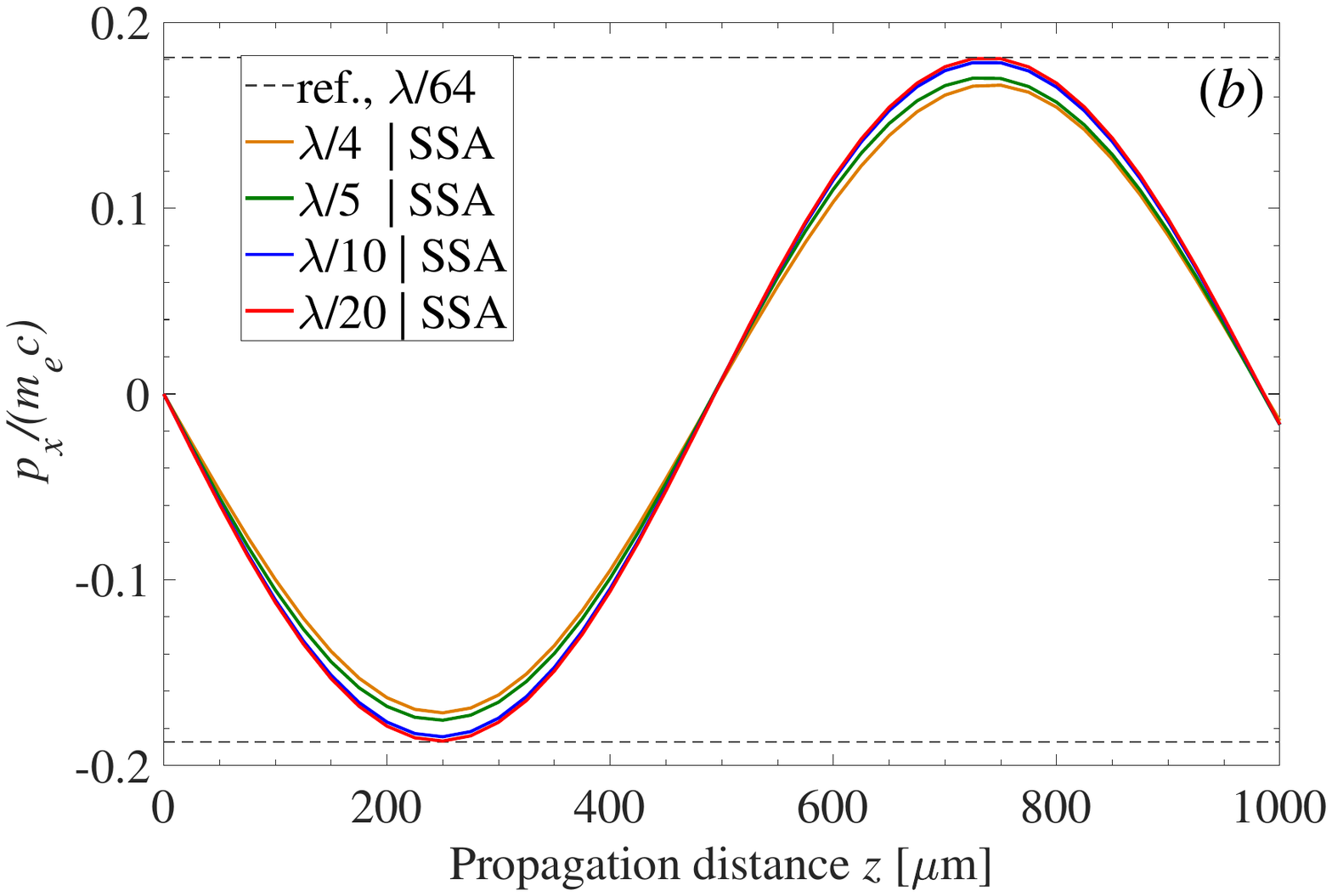}
\caption{ Evolution of the transverse reduced momenta $p_x/(m_e c)$ of a macroparticle, as it copropagates with a laser pulse having $a_0 = 0.2$ and it was placed initially in the center of the pulse with $\gamma = 25$. We plot the results corresponding to  $c \Delta t = \Delta z = \lambda/4$ (orange), $\lambda/5$ (green), $\lambda/10$ (blue), $\lambda/20$ (red) resolutions using the regular interpolation $(a)$, and using $2\times$ supersampling during interpolation $(b)$. We show reference minimum and maximum values with dashed lines acquired from a $\lambda/64$ resolution simulation.
Here, we used the Higuera-Cary pusher for the particle dynamics, but the same curves for the Boris-2 pusher were almost within line thickness (we show this at $\Delta z =\lambda/4$ with a dashed yellow curve on $(a)$). 
\label{fig:test_2d_force}}
\end{figure*}

In this test we copropagate an electron macroparticle of $\gamma = 25$ and a laser pulse of $\lambda = 0.8\mum$ carrier wavelength in vacuum over along direction $z$  over the temporal duration of $1000\mum/c$ in Cartesian 2D geometry. Initially we place them both such that their center is at position $z=0$, $x=0$, which coincides with the laser's focus position. The laser had $w_0=25\mum$ spot size at the focus, and pulse length of $\tau=7\mum/c$ with $a_0 = 0.2$. We performed numerical simulations in $200\mu m\times60\mum$ moving box. We used fixed transverse resolution $\Delta x = 0.2\mum$ with stretched coordinates\footref{foot:trf1}, the presence of the latter did not affect the accuracy. We choose $\Delta t = \Delta z/c$, and we used 24th order differences and 12th order exponentials for the field solver. Aside from the geometry, this test case is identical to that of in FBPIC \cite{lehe2016fbpic}.

In this test we analyze the transverse reduced momenta $u_x$ of the particle, as it copropagates with the laser. In this scenario the canonical momentum of the electron should be approximately conserved as
\begin{equation}
u_x/c-a_x=const.
\end{equation}
as it dephases the laser. It was shown \cite{lehe2016fbpic} that usual Yee PIC codes could spectacularly fail in this a case  due to the insufficient interpolation accuracy mainly caused by the temporal staggering of the $\bf B$ fields. Spectral codes such as FBPIC do not suffer from this issue.

We summarized our results on Fig. \ref{fig:test_2d_force} versus the spatial resolution $\Delta z = \lambda/N$. Using our regular destaggered interpolation (see Fig. \ref{fig:test_2d_force}$(a)$), we can say that our code has sensible behavior even for $\lambda/4$, $\lambda/5$ resolution, but the interpolation somewhat underestimates the Lorentz force felt by the electron. By comparing this with the results found in \cite{lehe2016fbpic} we can say that our method achieves similar accuracy as the spectral codes. We also compared the results of the Higuera-Cary particle pusher and the Boris-2 pusher, their curves were almost within line thickness (the former was a little bit more accurate).

We also tested the the effects of our $2\times$ field supersampling feature which we show on Fig. \ref{fig:test_2d_force} $(b)$. Indeed this procedure resulted interpolation accuracy the same as the double resolution simulations, which made it even usable at $\lambda/4$, $\lambda/5$ resolution. Thus using this method we can achieve interpolation accuracy superior even to the spectral codes but with additional computational cost.

\subsection{Linear propagation in plasma (2D)} \label{subsubsec:test_2d_plasma}
 
 \begin{figure}[ht]
\centering
\includegraphics[height=6.3cm]{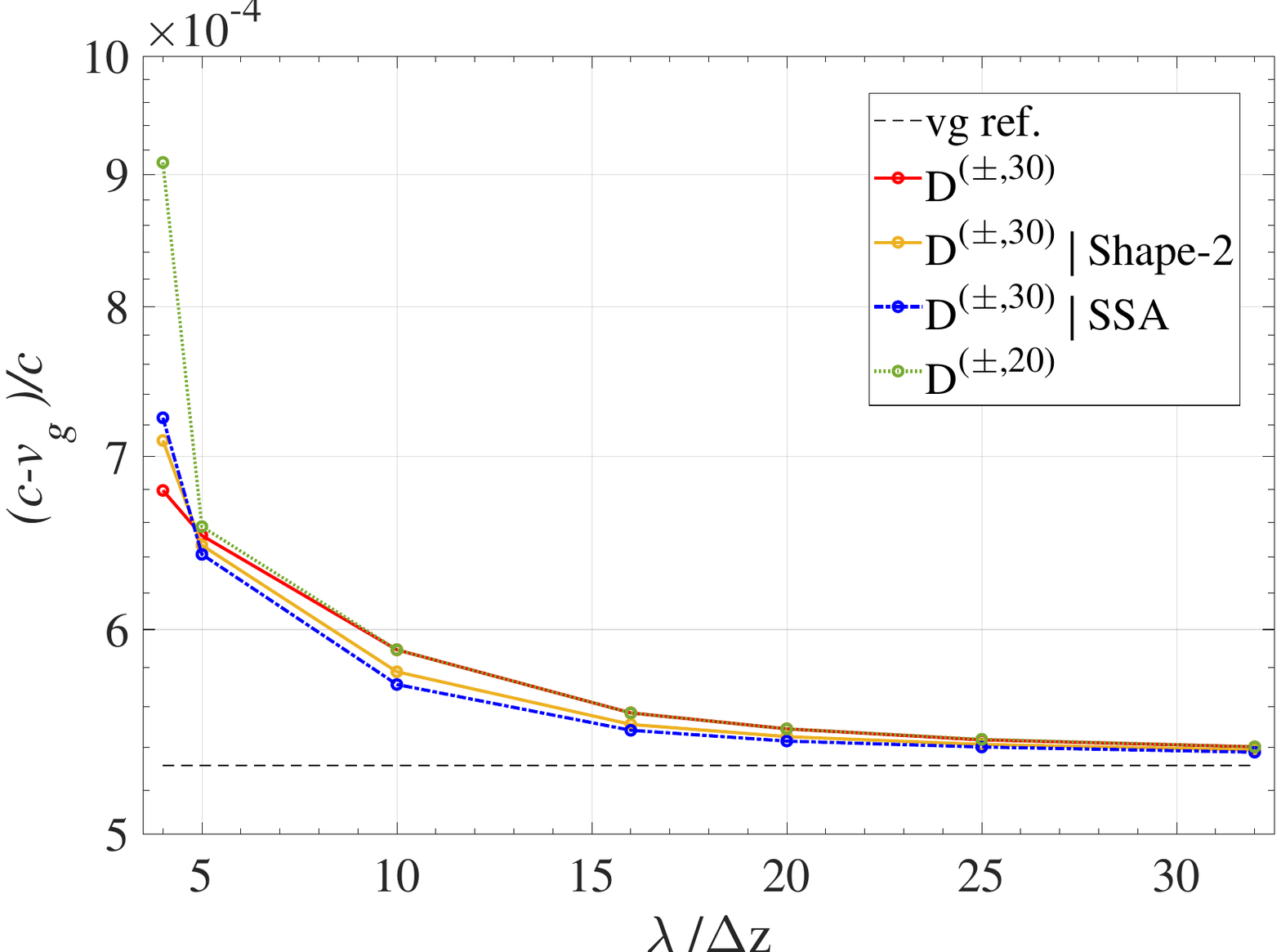}
\caption{ Relative difference between $c$ and the group velocity $v_g$ of the pulse centroid in a plasma of $10^{-3}n_c$ density for different \revise{$N = \lambda/\Delta z$} resolutions in 2D. The dashed line represents the analytical prediction ($5.317\times10^{-4}$) the rest represent PIC simulations with various parameters: with 30th (red, orange, blue) and 20th order (dotted green) finite differences, and using 2nd order particle shape (orange) and using $2\times$ supersampling (dashed blue). We used 3rd order particle shape where it is not indicated.
\label{fig:test_2d_plasma}}
\end{figure}

In this test we propagate a laser pulse of $\lambda=0.8\mum$ carrier wavelength  with $a_0 = 0.01$ in an underdense plasma of $n_0= 1.75\times10^{18} {\rm cm}^{-3} = 10^{-3} n_c$ density, where $n_c$ is the critical density. We perform this test in 2D Cartesian geometry with varying $\Delta z=\lambda/N$ resolution. We do this between $z=0$ and $z=50\mum$, where the focus position of the laser is at $z=0$.  We performed numerical simulations in $140\mu m\times60\mum$ moving box, the laser had $w_0=16\mum$ spot size at the focus, and Gaussian pulse length $\tau=10\mum/c$. We used fixed transverse resolution $\Delta x = 0.2\mum$ with stretched coordinates\footref{foot:trf1}, the presence of the latter did not affect the accuracy. We choose $\Delta t = 0.5\Delta z/c$ unless otherwise stated, and for the field solver we use 30th order finite differences and 12th order exponentials.

There exists an analytical formula for the $v_g$ group velocity for the pulse centroid motion  \cite{esarey1995tightly_focused_pulses}, which is
\begin{equation} \label{eq:test_2d_plasma_vg}
\frac{c-v_g}{c} = \frac{n_0}{2 n_c}+ \frac{1}{2} \left(\frac{\lambda}{2\pi w_0} \right)^2 \quad  \text{in 2D}.
\end{equation}
This yields approximately $-5.32\times10^{-4}$  in this test case. We calculate this using the time derivative of $\left< z \right>$ the pulse centroid, see Eq. (\ref{eq:test_2d_vacuum_Z}). Despite this result is analytical, it is not exact.

We summarize our results on Fig. \ref{fig:test_2d_plasma} versus the resolution \revise{$N = \lambda/\Delta z$}. We can see that due to the high accuracy of the field solver our results are not far from the analytical prediction, and have similar accuracy as those simulated with FBPIC \cite{lehe2016fbpic}. Using 20th order finite differences only seems to impact high frequency ($\Delta z = \lambda/4$) results. However, this does not mean that the result is accurate in the absolute sense, the error converges relatively slowly, which suggest that it comes from the PIC particle representation. Notably, using lower order particle shapes produces better results: when we use 2nd order particle shape (orange curve) the result is more accurate compared to that found using 3rd order shape (red curve). Interestingly, if we turn on $2\times$ supersampling for the 3rd order particle shape the result significantly improves (dashed blue curve). Using the latter, satisfying accuracy can be reached at $\lambda/16$ resolution.
 
Overall, our PIC implementation reproduces the analytical expectation reasonably well. The best accuracy favors the use of lower order particle shapes or our interpolation supersampling scheme.

\subsection{Linear laser-wakefield (3D)} \label{subsubsec:test_3d_linear}

\begin{figure*}[t]
\centering
\includegraphics[trim={140px 0 170px 30},clip,height=5.0cm]{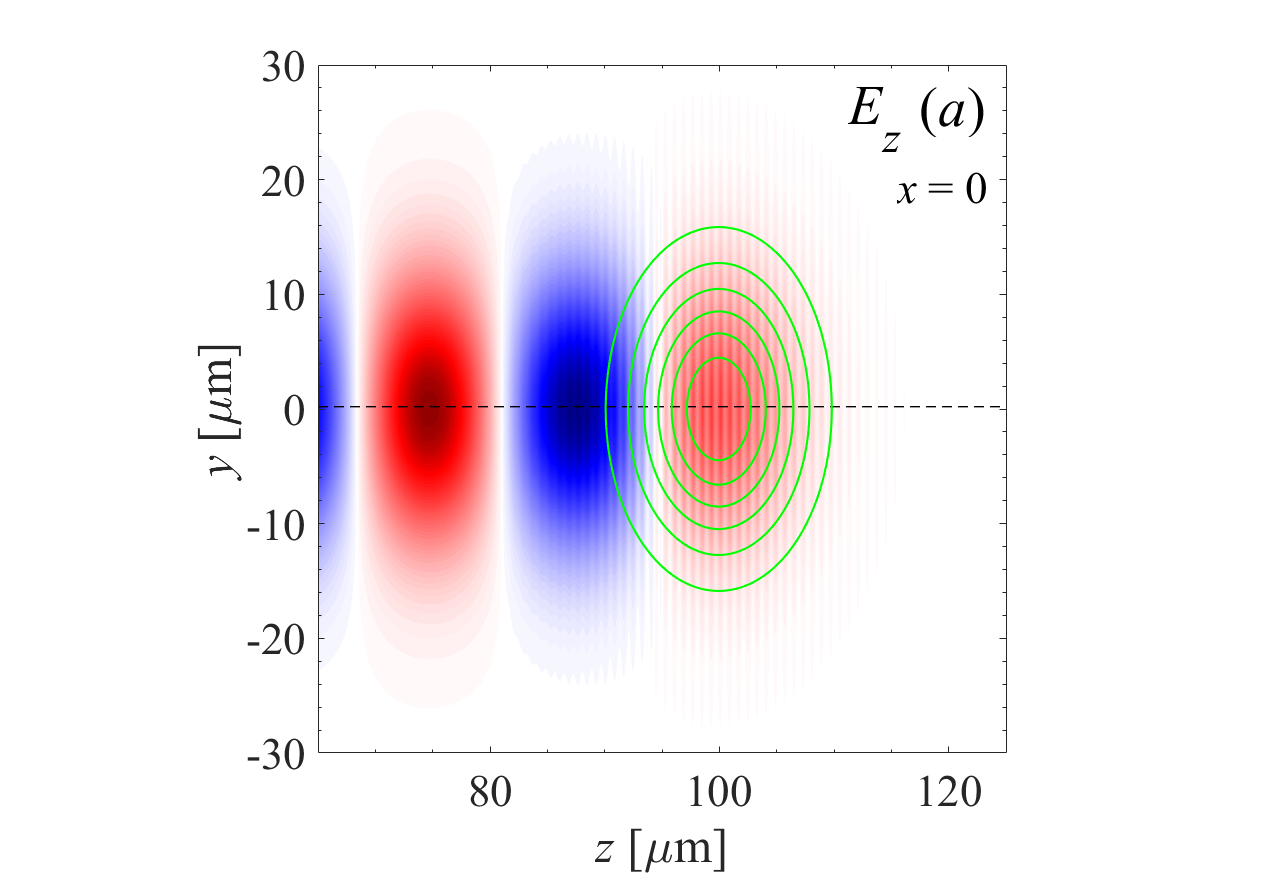}
\includegraphics[trim={0px 0 40px 0},clip,height=5.05cm]{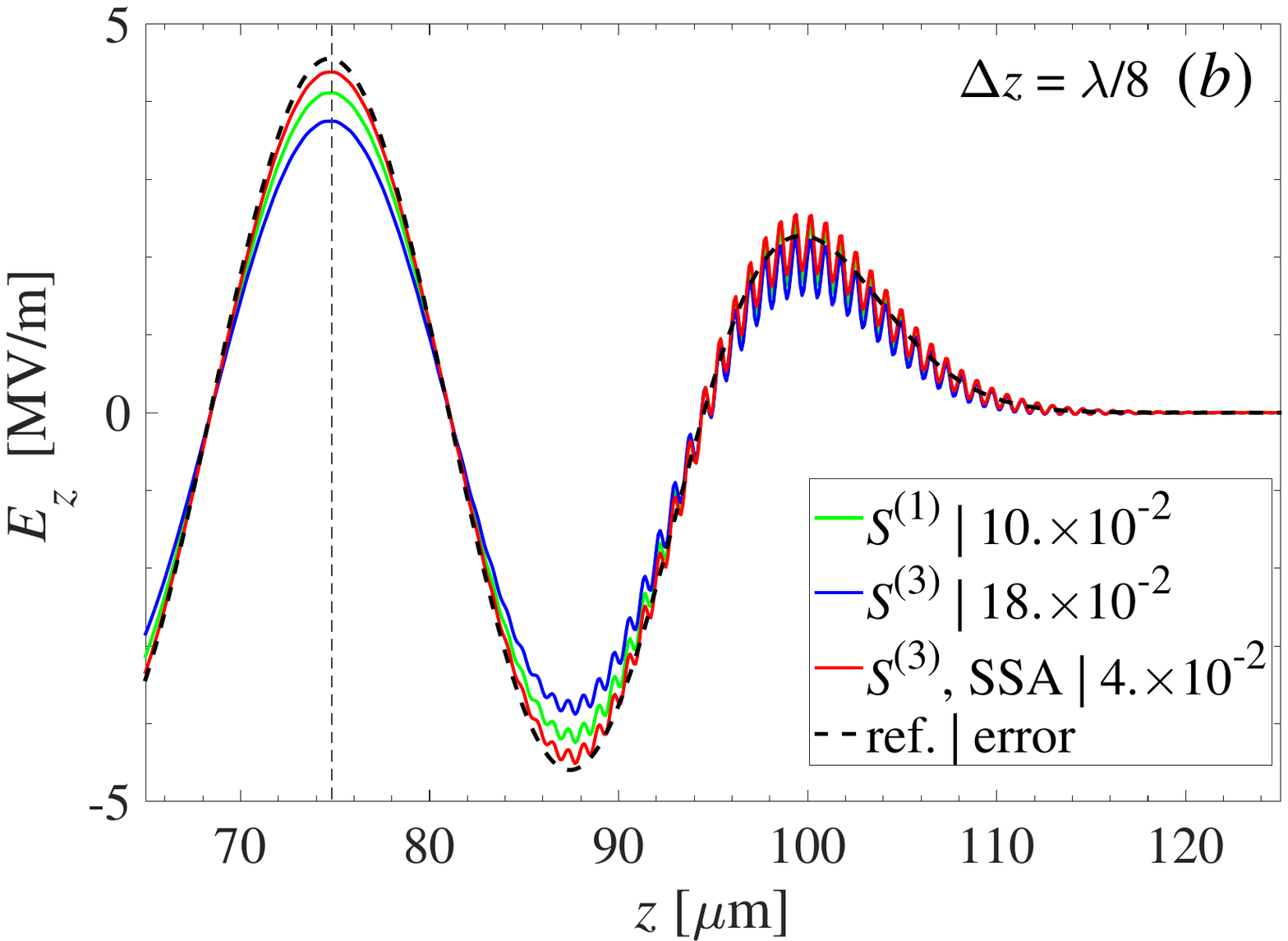}
\includegraphics[trim={70px 0 40px 0},clip,height=5.05cm]{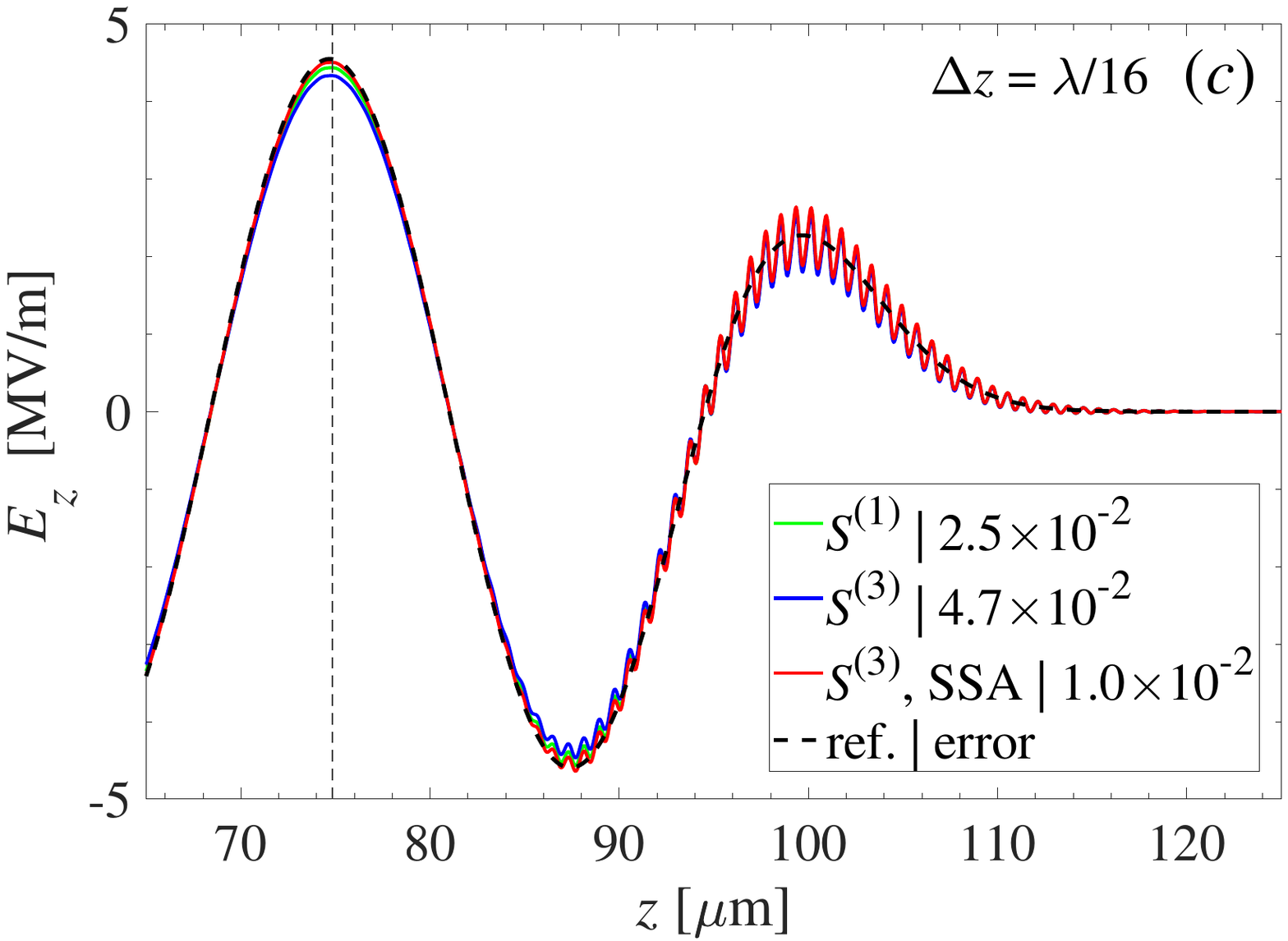}
\includegraphics[trim={140px 0 175px 30},clip,height=5.0cm]{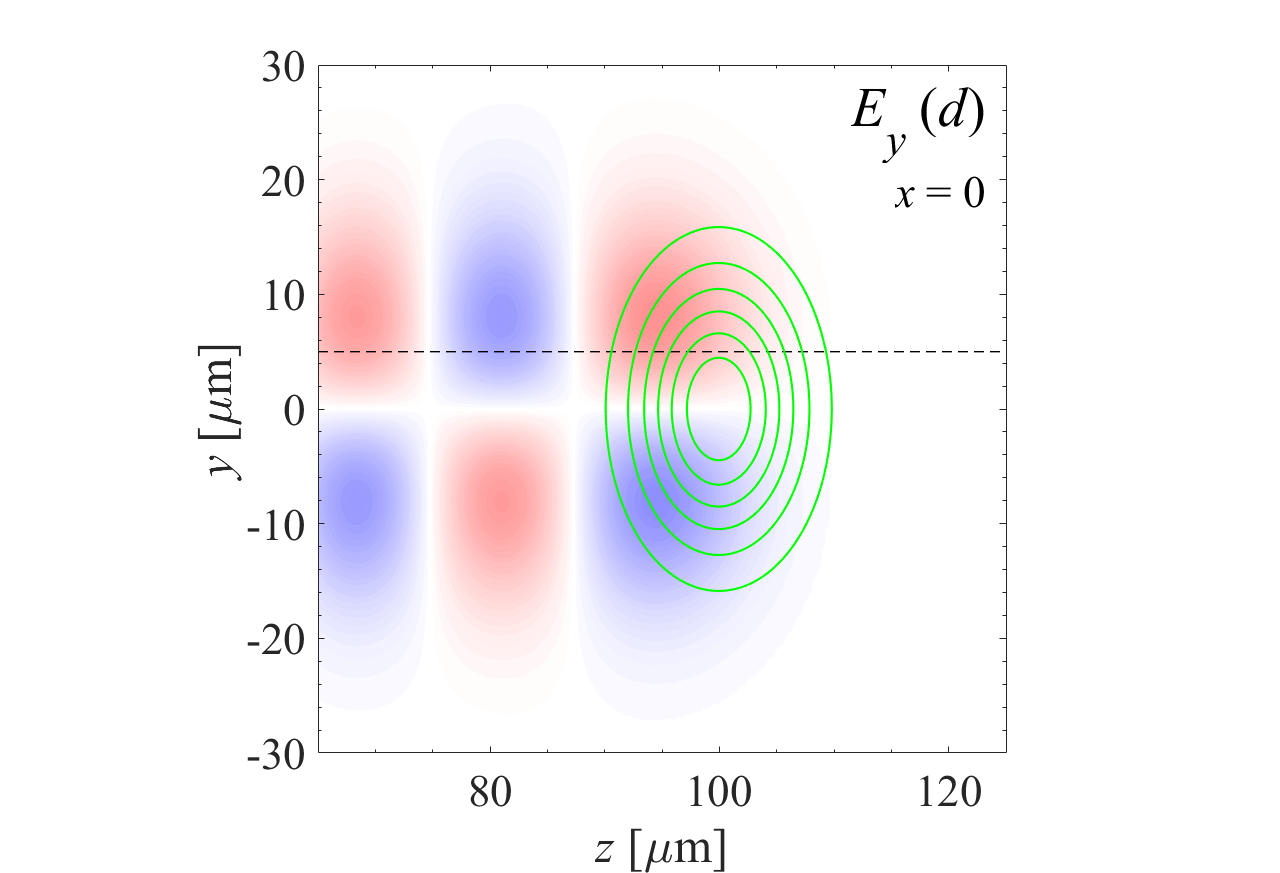}
\includegraphics[trim={0px 0 35px 0},clip,height=5.00cm]{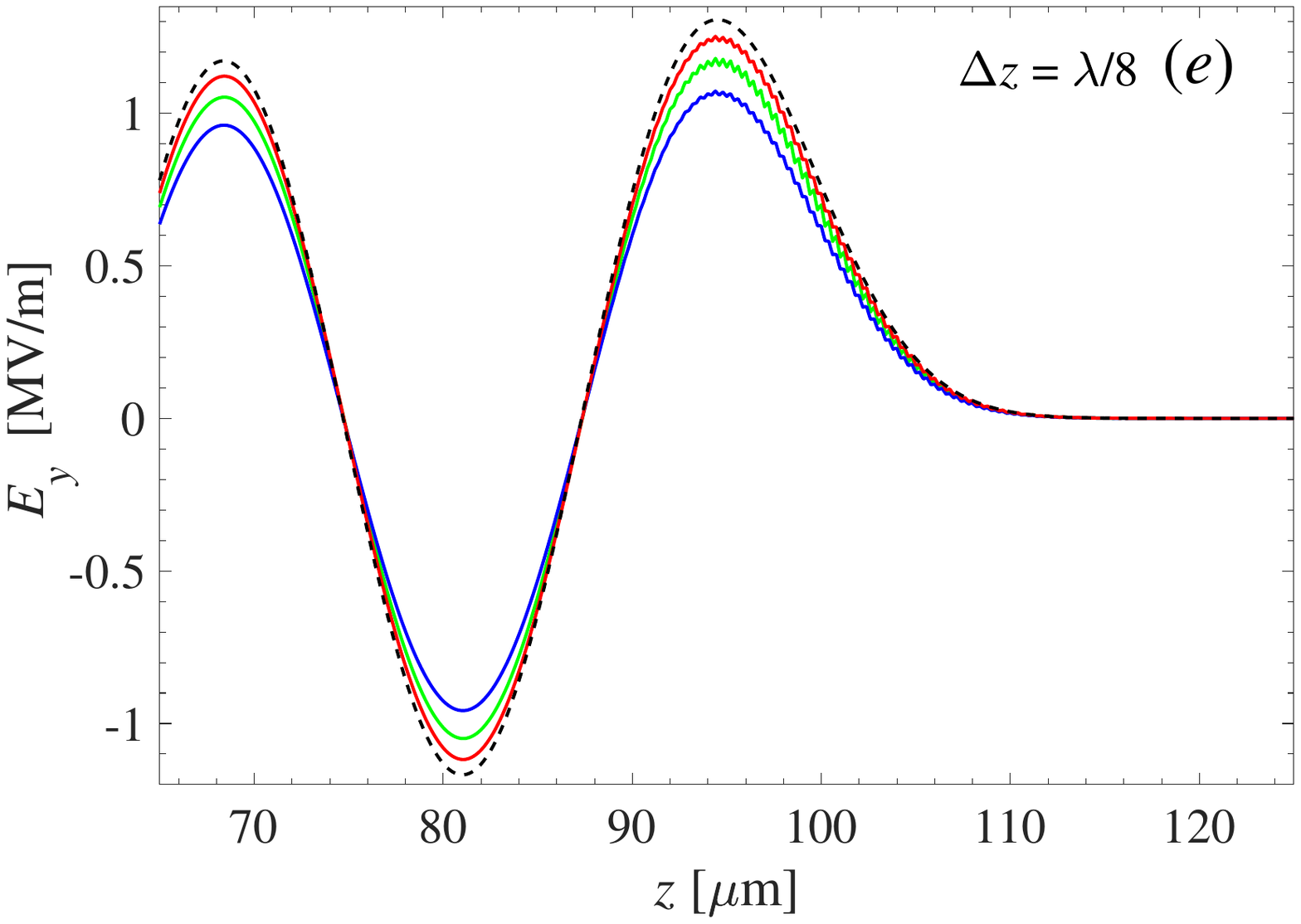}
\includegraphics[trim={55px 0 35px 0},clip,height=5.00cm]{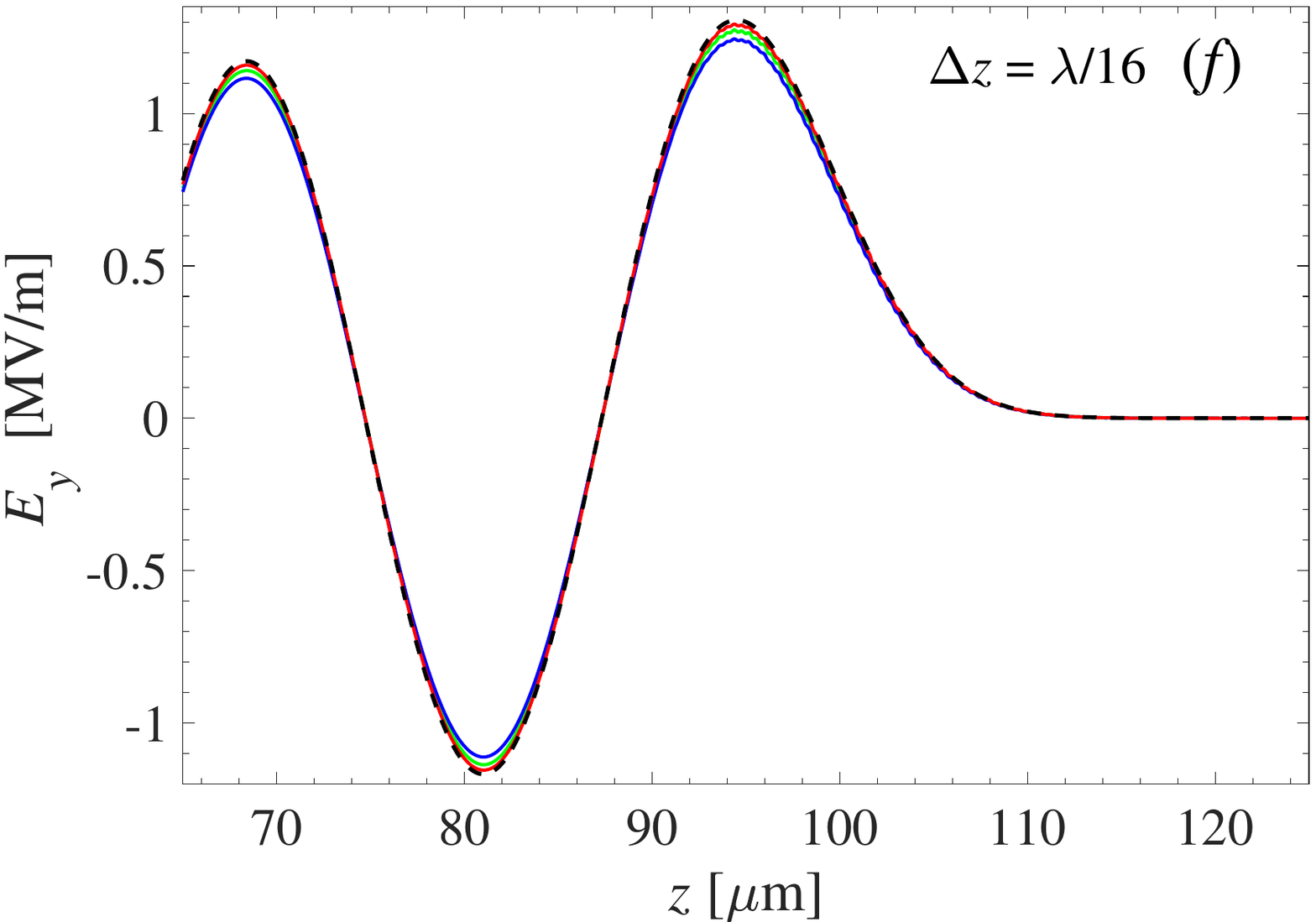}
\caption{Plot of the linear wakefields $E_z$ (top panels), $E_y$ (bottom panels) generated by a laser pulse of $\lambda = 0.8\mu$m  after $100\mu$m propagation in an underdense plasma. We show the colormap of these fields in $(a)$, $(b)$ in the $x=0$ plane where we also indicated the position of the laser envelope and $z$-line cross sections with dashed lines. We show the latter on panels $(b)$, $(e)$ using $\Delta z = \lambda/8 = 0.1\mu$m, and on panels $(c)$, $(f)$ using $\Delta z = \lambda/16 = 0.05\mu$m resolution both at $\Delta t = 0.5\Delta z /c$. We compare at these cross sections the accuracy of the wakefields, where the dashed lines show the expected analytical results, for the green lines we used $S^{(1)}$ particle shape, the blue, red lines show the results using our default $S^{(3)}$ particle shape, where the red lines were simulated with $2\times$ supersampling during field interpolation. We also indicated the approximate relative error of the wakefields at the vertical dashed lines.
\label{fig:test_3d_wake}}
\end{figure*}

Similar to the previous Section, we propagate a laser pulse of $\lambda=0.8\mum$ carrier wavelength with $a_0 = 0.01$ in an underdense plasma of density $n_0= 1.75\times10^{18} {\rm cm}^{-3}$. We perform this test in 3D Cartesian geometry with at $\Delta z = \lambda/8, \lambda/16$ resolutions  and we analyze the resulting wakefields.   We performed numerical simulations in a $140\mu m\times60\mum$ moving box using $w_0=16\mum$ laser spot size at the focus (at $z=0$), and pulse length of $\tau=10\mum/c$. We used fixed transverse resolution $\Delta x = 0.2\mum$ with stretched coordinates\footref{foot:trf1}, the presence of the latter did not affect the accuracy. We choose the temporal step $\Delta t = 0.5\Delta z/c$ such that it provides a temporally convergent result, and for the field solver we use 30th order finite differences and 12th order exponentials.

We propagate the laser pulse in a uniform density between $z=0$ and $z=100\mum$ and at the latter position of the laser pulse we extract the simulated $E_y$, $E_z$ wakefields at the $x = 0$ cross-section. Since this is a low intensity linear propagation, the generated laser wakefields can be given analytically within the laser envelope model, which involves the numerical integration of the ponderomotive potential of the laser ($0.5({e^2}/{m_e^2})|a|^2$) with $\cos \left(k_p(z-z') + \varphi_0 \right)$ plasma oscillations, where $k_p$ is the plasma wave vector. The respective formulas of which can be found in e.g. \cite{esarey2009laser_plasma_accelerators, lehe2016fbpic}. We extracted the approximate longitudinal envelope from the simulations to get the best match with the analytical prediction. Since the PIC scheme is a fully self consistent simulation it knows nothing about effective laser envelope ponderomotive potentials or forces, which makes this case a valuable benchmark.

We summarize our results on Fig. \ref{fig:test_3d_wake}, where we show the planar cross sections on Fig. \ref{fig:test_3d_wake} $(a)$, $(b)$ for the $E_z$, $E_y$ electric fields respectively. We took the axial cutouts corresponding to the horizontal dashed lines of these planar cross sections and show them at $\Delta z = \lambda /8$ (on $(b)$, $(e)$ panels) and $\Delta z = \lambda /16$ resolutions (on $(c)$, $(f)$ panels). We can see that the $E_y$ and $E_z$ wakefields show similar behavior during simulations, so we analyze the latter in the following.

 Unfortunately we can state that our numerical PIC scheme underestimates the ponderomotive force felt by the electrons resulting in  $18\%$ error in the wakefield at $\lambda/8$ resolution using our default 3rd order $S^{(3)}$ particle shape (blue curve on Fig. \ref{fig:test_3d_wake} $(b)$). \revisee{Out of curiosity we performed simulations using the first order $S^{(1)}$ particle shape to show how it might perform in standard PIC codes, the errors for the wakefields are less ($10\%$ vs $18\%$ error at $\lambda/8$ resolution, green curves). Officially we do not support this type of a particle shape because it is subject to substantial aliasing error (see \ref{subsubsec:shapes_alias}), even though the latter does not affect the physics shown in this example due to the low laser intensity.}  %%We observed negative effects of aliasing in the case of higher laser intensities, therefore it should be eliminated. 
 If we double the spatial resolution the wakefield errors decrease by a factor of $1/4$ (see Fig. \ref{fig:test_3d_wake}$(c)$), which suggests that it is caused by an inherent flaw in the PIC representation.
 If we turn on our  $2\times$ field interpolation supersampling feature the wakefield errors decrease by the same $1/4$ factor (red curves), therefore the error is dominated by the field interpolation - we get results similar to performing simulations at double resolution.  
 With supersampling $10^{-2}$ wakefield accuracy can be reached using 3rd order particle shape at $\Delta z= 0.05\mum$ ($=\lambda/16$) resolution which  would otherwise require $\Delta z \approx 0.035\mum$  and $\Delta z \approx 0.025\mum$ using 1st order and 3rd order particle shapes, respectively. This $2\times$ supersampled method is a significant improvement, especially if we take into account that the aliasing errors are also reduced by orders of magnitude.

Overall, the PIC simulations could reproduce the analytical expectation reasonably well but they underestimate the Lorentz force, resulting in weaker ponderomotive wake fields on the order of $10^{-2}$. Using higher order particle shape in the PIC representation instead of the 1st order shape could worsen the accuracy of these wake fields. To achieve the best accuracy at modest resolutions our $2\times$ field supersampling seems to be required. We also experienced the same convergence behavior for nonlinear wake fields.

\subsection{Surface high-harmonic generation (2D)} \label{subsubsec:test_2d_shhg}

\begin{figure*}[t]
\centering
\includegraphics[trim={0px 30 30px 0},clip,height=6.25cm]{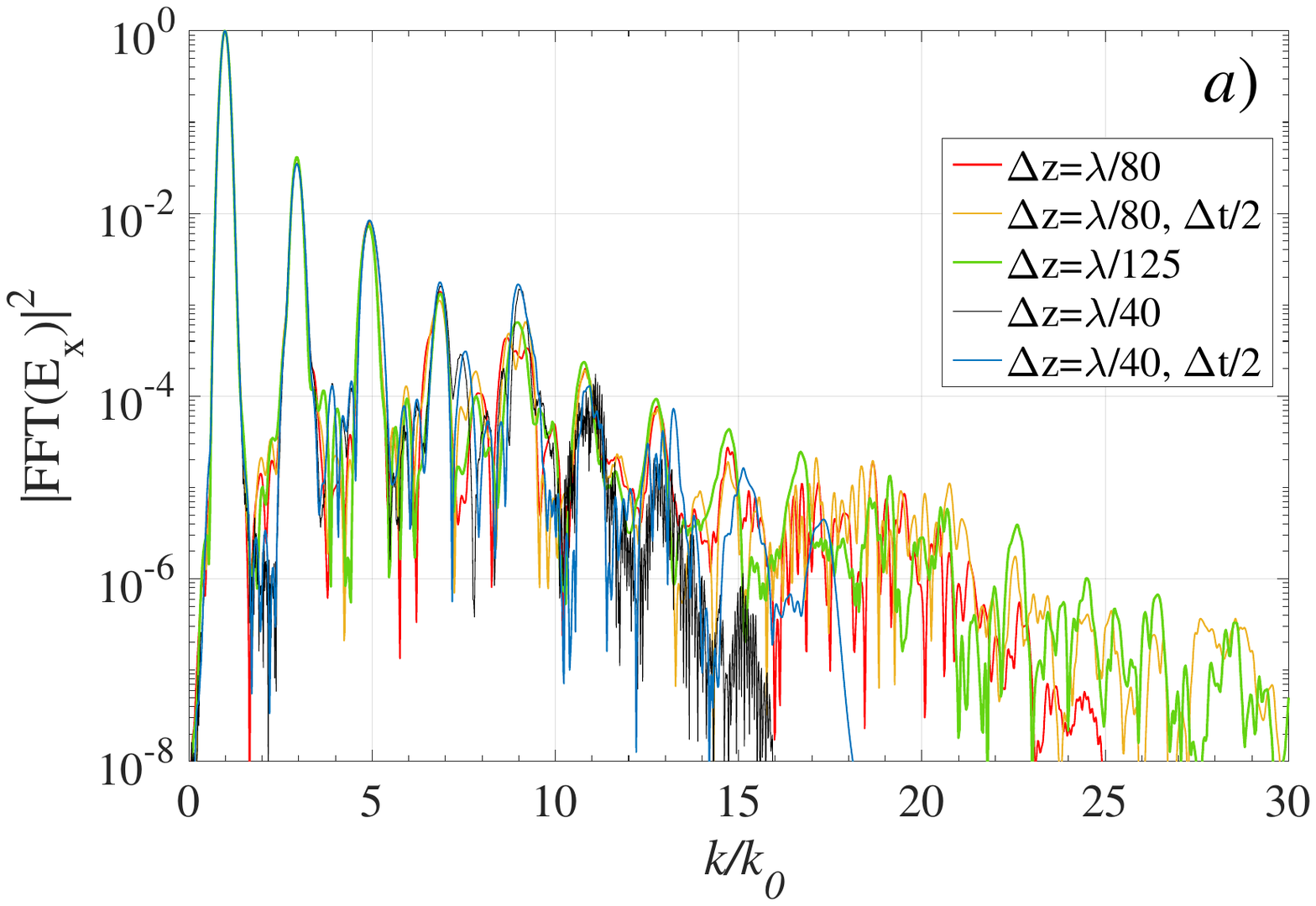}
\includegraphics[trim={0px 30 30px 0},clip,height=6.25cm]{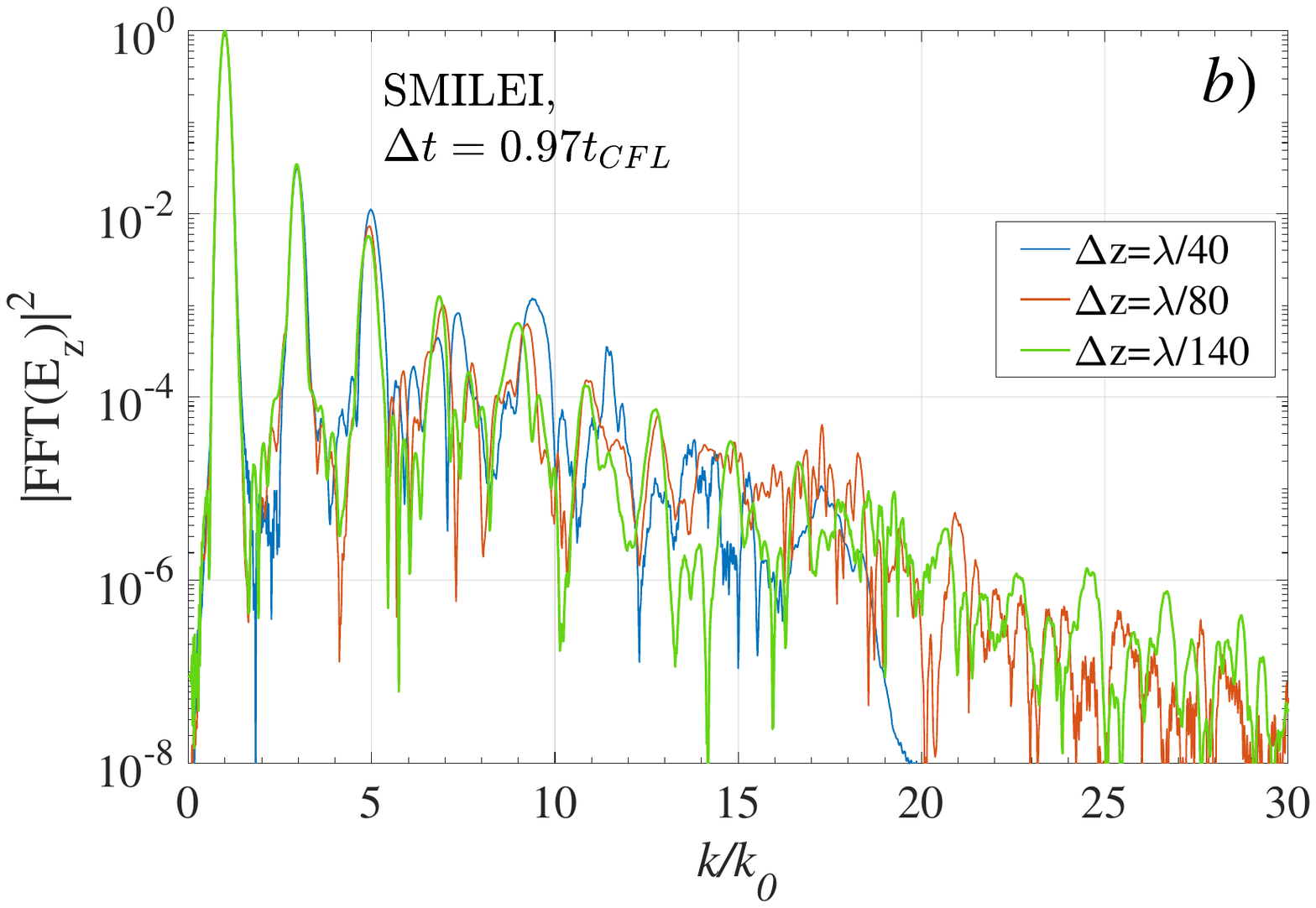}
\includegraphics[trim={0px 0 30px 0},clip,height=6.5cm]{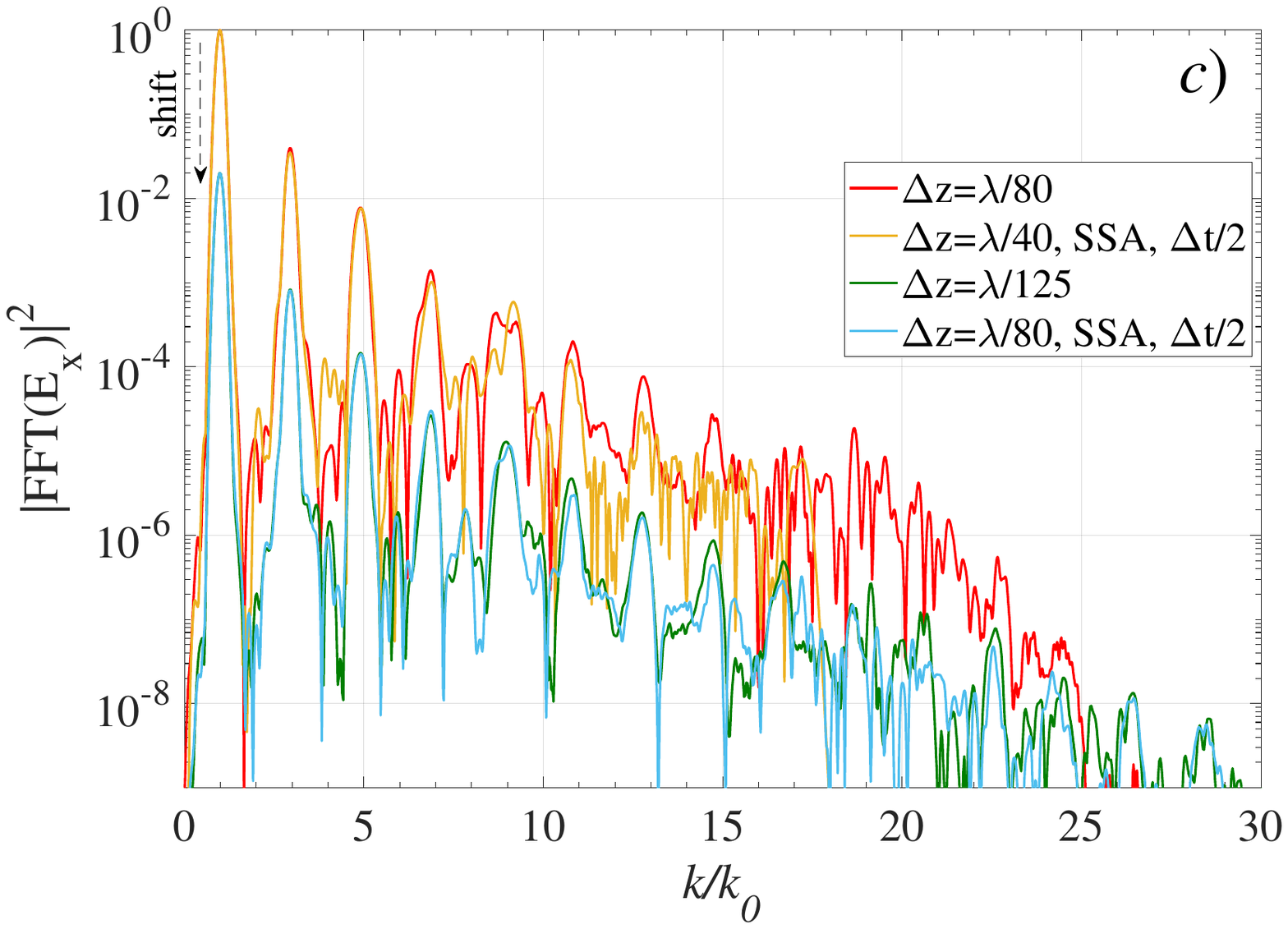}
\includegraphics[trim={80px 20 50px 40},clip,height=6.5cm]{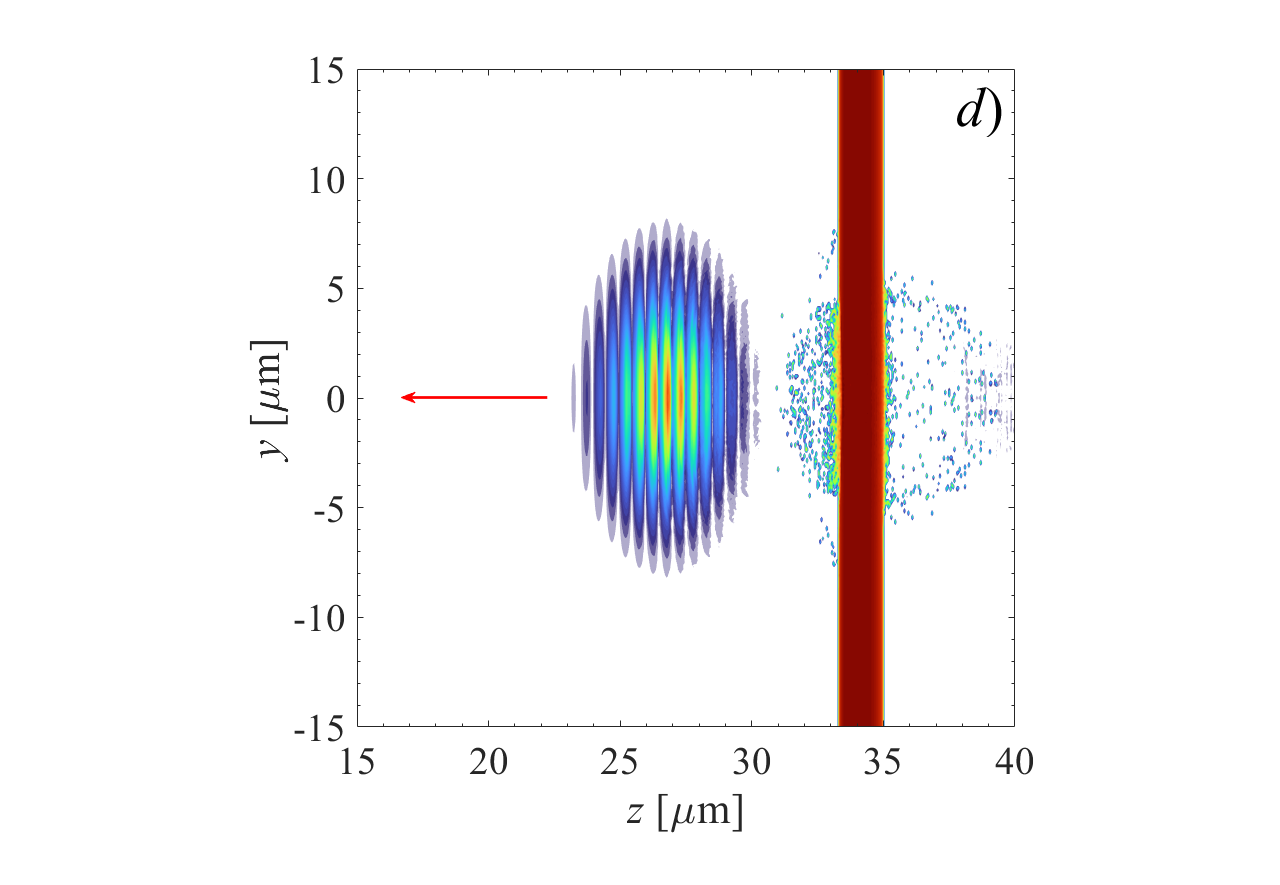}
\caption{ Spectral intensity distribution (square of the Fourier transform) of the reflected laser field along the propagation axis is shown for different numerical parameters in the case of our code $(a)$ and in the case of SMILEI $(b)$. In $(a)$ 8th order exponential solver with D$^{(\pm,24)}$ is used in all cases (without supersampling). In $(c)$ the red and green (divided by 50 for better visibility) curves are the same as in $(a)$, but here they are compared with the corresponding low resolution results (orange, blue) where supersampling and half time step are used. The physical setup and a snapshot of the interaction after reflection is shown in $(d)$.}
\label{fig:test_2d_shhg}
\end{figure*}

In the case of an overdense plasma, where the plasma electron density is much higher than the critical density, the laser pulse gets reflected from the plasma boundary. When an ultra-intense laser pulse impinges on the target surface with sharp density profile a thin electron layer is put into strongly relativistic motion along the direction of the laser propagation axis and the reflected pulse will be strongly modulated. This electron layer can be considered as a relativistic mirror, which oscillates at a frequency twice of the fundamental of the laser pulse. Each time the electron layer moves towards the vacuum, the laser field gets temporally compressed and high frequencies appear in the form of high-harmonics. Because of these characteristics of the process this regime of high-harmonic generation is called relativistic oscillating mirror (ROM) \cite{teuber2009shhg_review, baeva@2006shhg_theory, thaury2010shhg_mirror, quere2008shhg_mirror}. We test the performance of our newly developed code in this extreme regime of laser-plasma interaction, where electron density modulations of very short scale lengths can appear and every tiny detail of the numerical scheme becomes extremely important.

The test case (2D simulation) is visualized in Fig. \ref{fig:test_2d_shhg}$(d)$, where the laser pulse initially propagates in the positive $z$ direction. Its focus position is at the surface of a $1\,\mu$m thick plasma slab with $0.2\,\mu$m linear ramp on both sides, which represent the ever-present preplasma appearing on the target surface in experiments. The plasma density in the plateau region is $n_0=100n_{cr}=1.12\times 10^{29}$ m$^{-3}$ and the normalized peak amplitude of the laser field is $a_0=6$. In each grid cell 30 proton and 60 electron macroparticles are placed. After reflection, the laser pulse propagates in the negative $z$-direction and we analyze the harmonic spectrum by Fourier transforming the one-dimensional field data along the $y=0$ line. 

In principle, the plasma dynamics can be correctly modelled if the spatial grid spacing is smaller than the smallest physical scale length, which is the plasma skin depth in this case: $l_s=\sqrt{\gamma_e}c/\omega_p\approx 30$ nm, where $\gamma_e\approx \sqrt{1+a_0^2/2}$. It is worth noting that the width of the oscillating electron layer, compressed by the Lorentz force, can be even smaller, therefore the coherence of the highest harmonics can be resolved by a grid spacing less than 10 nm. However, the relative intensity of those spectral components is very small ($<10^{-6}$), thus we concentrate on the generation of harmonic number below 30, which are presented in Fig. \ref{fig:test_2d_shhg}. Our aim is to find a convergence in the harmonics spectrum by gradually increasing the resolution along the z-direction. A clear convergence can be seen in Fig. \ref{fig:test_2d_shhg}$(a)$ by comparing the black, red and green lines, while such convergence is not so clear in a code based on the Yee's Maxwell solver (in this case SMILEI \cite{derouillat2018pic_smilei}), shown in Fig. \ref{fig:test_2d_shhg}$(b)$. One of the great advantages of our code is the possibility to use time steps much smaller than $t_{CFL}$ without ruining the accuracy of the field solver. The consequence of using half of the original time step (which is $0.97t_{CFL}$) is that the spectrum is extended to higher frequencies and the spectral noise is reduced. Otherwise, with smaller time step one can reproduce almost the same spectrum, which is seen by comparing the orange-red and cyan-black curves in Fig. \ref{fig:test_2d_shhg}$(a)$.

Another important feature of the code, presented in this paper, is the interpolation supersampling, which, in combination with half time step, is almost equivalent to using two times higher resolution. This is proven in Fig. \ref{fig:test_2d_shhg}$(c)$ showing the results from high-resolution simulations (red, green curves) and from simulations with lower resolution (orange, blue curves) applying the mentioned numerical techniques. We note here that, due to the improved numerical dispersion in our code, it is possible to study the propagation of the reflected pulse over longer distances (hundreds of micrometers), which can lead to filtered harmonics spectrum on the optical axis, thus attosecond pulses appear naturally \cite{pukhov2007shhg_propagation}. In other words, fields with wave number smaller than $0.7\kk_{max}=0.7\pi/\lambda_{min}$, where $\lambda_{min}=2\Delta z$, can propagate without noticeable error if high-order differential operator is used.

\subsection{Laser wakefield acceleration (3D)} \label{subsubsec:test_lwfa}

The phase and group velocities of electromagnetic waves, which were tested in the previous sections, is only one portion of the variables that are required to be precisely modeled when high-amplitude plasma waves are generated by high intensity laser pulses ($a_0>1$). In this regime of laser wakefield acceleration (LWFA) the laser pulse itself can have a nonlinear envelope evolution due to the presence of the wakefield, which generates a non-uniform index of refraction ($\eta$), which depends on the shape of the laser envelope. The large gradients in $\eta$ usually lead to strong self-focusing, if the laser spot size is large enough, and to self-phase modulation, which leads to partial self-compression of the pulse. These two effects have extremely important role in the evolution of the wakefield behind the laser pulse, where electron acceleration takes place, because the evolution of the envelope defines the phase velocity of the plasma wave, which is crucial for controlling the electron's energy gain. 

In the zeroth order approximation the wakefield's phase velocity is equal to the laser's linear group velocity: $v_p\approx v_g=\eta c$, which is basically the velocity of the wake's node, where $E_z=0$. This velocity is decreased by both effects mentioned above, leading to potential wave-breaking, also called as self-injection, which takes place when the electron's velocity in the plasma wave becomes larger than $v_p'<\eta c$. It is evident that a tiny change in the wake potential changes the velocity of the node and the properties of the injected electrons (total charge, energy, transverse emittance, etc.) will also change. Therefore, it is very challenging to correctly simulate the self-injection process, since not only the precision of the Maxwell-solver is highly demanded, but the particle motion and plasma dynamics should also be as exact as possible.

Here, we test our code in such a nonlinear interaction scenario and compare it with EPOCH, which is a standard Cartesian PIC code based on Yee's method, but 4th order spatial derivatives are also implemented, which improves the numerical dispersion at the cost of simulation speed, because the time step should be around one third of that given by the CFL condition. Choosing a larger time step leads to supraluminal group velocity, which is an important issue in advanced Yee schemes \cite{lehe2013numerical}, while it is completely absent in our code. The physical parameters of the simulations are the following. The laser pulse spatial and temporal envelope is Gaussian with $w_0=5\lambda$ focal spot and $\tau=2\lambda/c$, where $\lambda=1\mum$ is the laser wavelength and $a_0=3$. The laser is linearly polarized in the $x$-direction. The target is an underdense plasma which starts with 50$\mum$ long linear up ramp, where the density increases from 0 to $n_0=0.01n_c$ and the density is constant afterwards (for $z>80\mum$). The focus position of the laser pulse is at $z_f=83 \mum$, and the center of the pulse is initially located at $z=20\mum$. The size of the simulation box is $48\mum\times48\mum\times30\mum$ in the $x,y,z$ directions, respectively, and it moves with velocity $c$ in the positive $z$ direction. In all cases the Higuera-Cary (HC) pusher is used, except where it is stated differently. In our code, we used stabilized 9th order exponential expansion (see Section \ref{subsubsec:test_2d_vacuum}) and enhanced 24th finite differences (with $E=3$ enhancement, see \ref{subsubsec:enhance}) and lowpass A current filter (see Section \ref{subsec:spatial_current}). We also tested stretched coordinates transversally in our code such that the uniform inner zone encompassed the entire wake bubble, and doing so had no visible effect on the simulation results we show here.\footnote{Parameters of the stretched coordinates in $x,y$ we tested in the LWFA benchmark: stretching factor of 2, inner zone $10\mum$, transition zone $5\mum$.}

\begin{figure*}[tp!]
\centering
\centering  
\includegraphics[trim={20px 0 118px 35},clip,height=4.55cm]{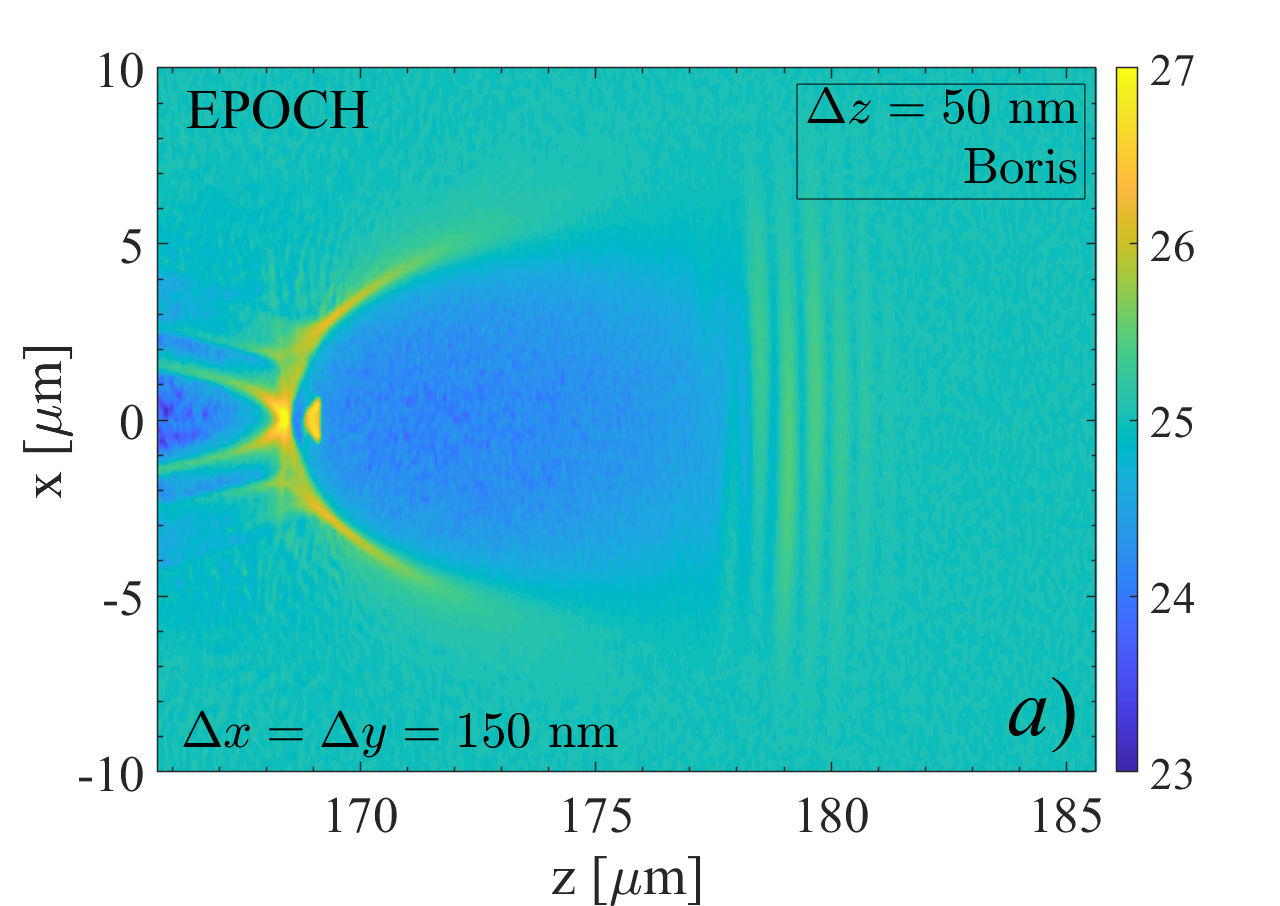}
\includegraphics[trim={15px 0 118px 35},clip,height=4.55cm]{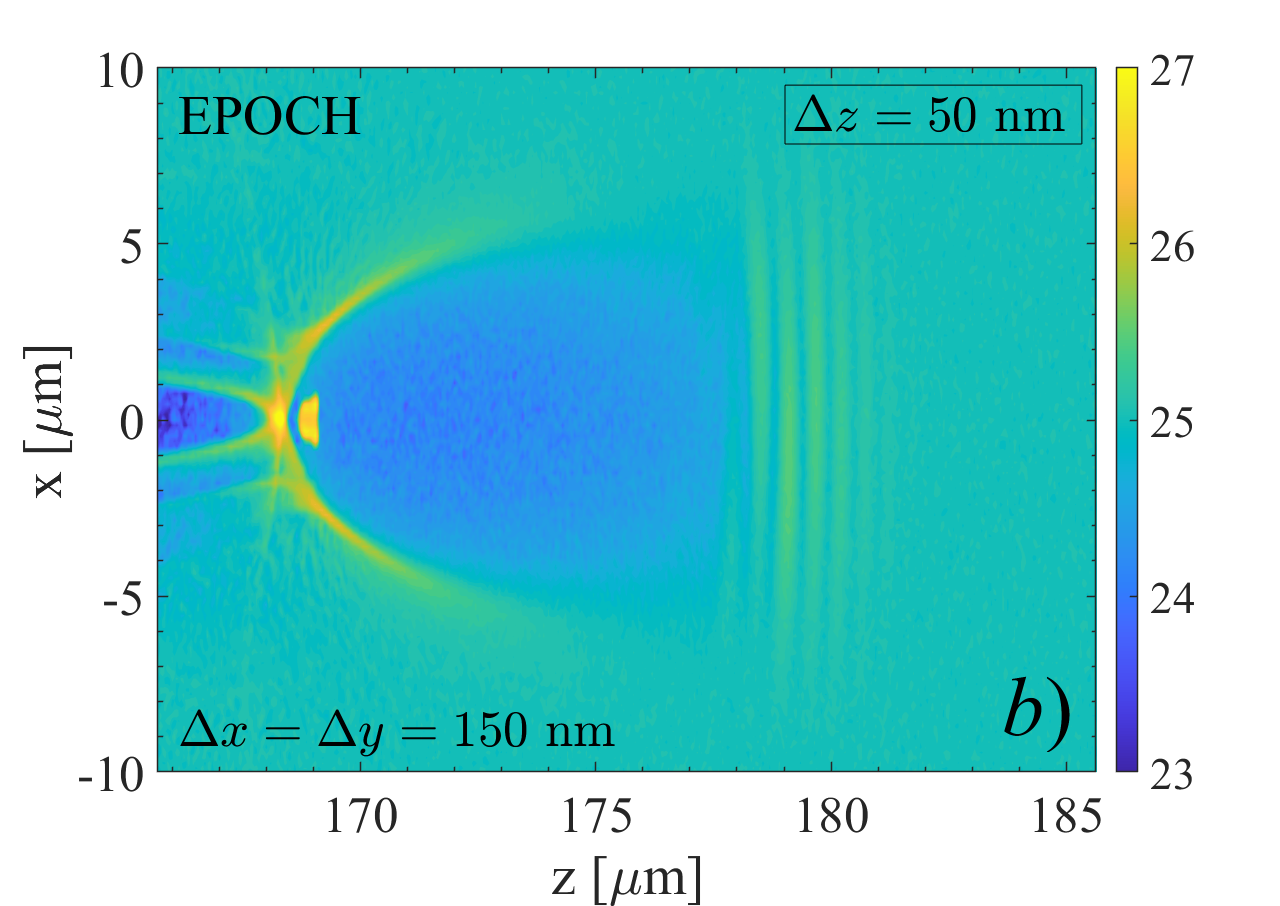}
\includegraphics[trim={15px 0 20px 35},clip,height=4.55cm]{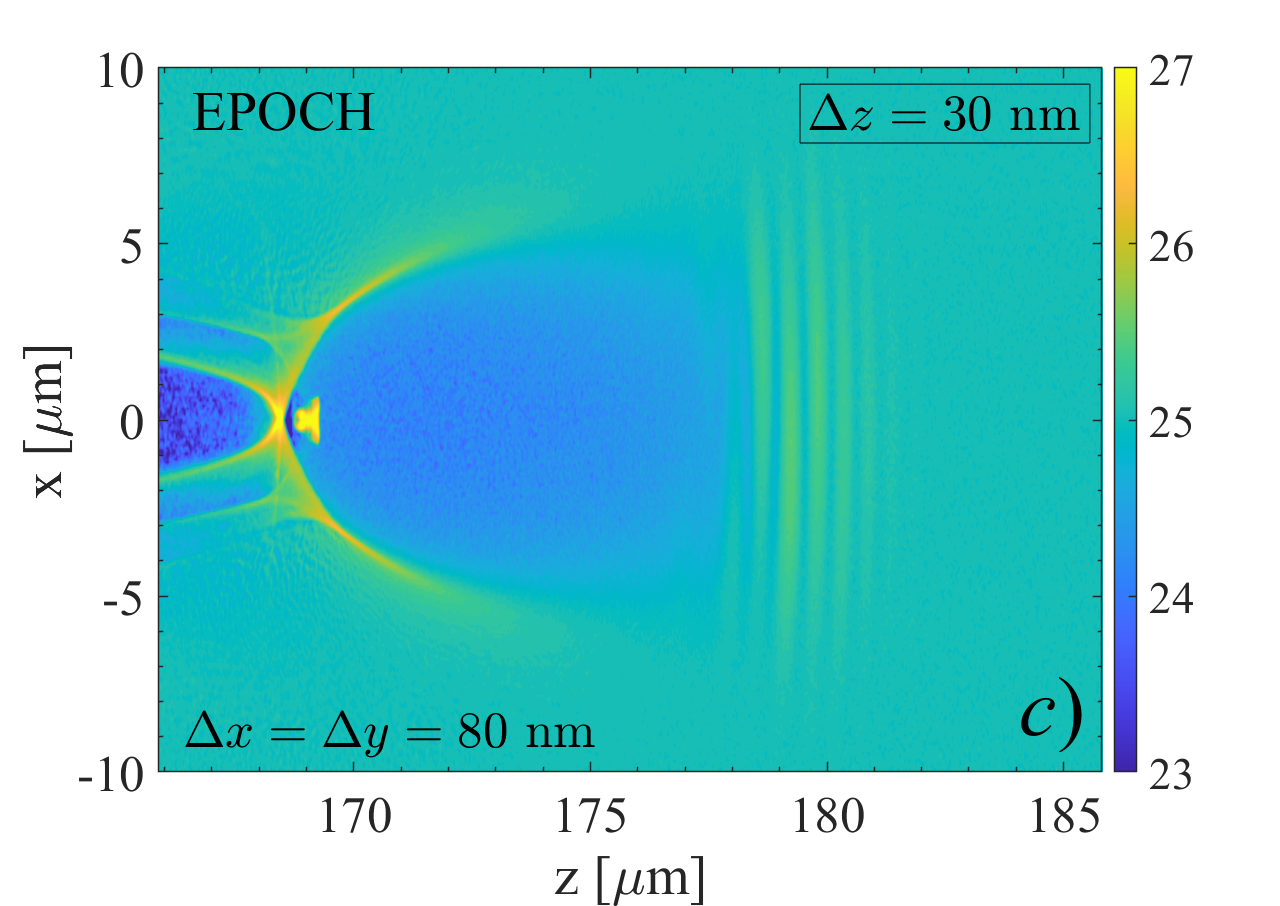}
\includegraphics[trim={20px 0 118px 35},clip,height=4.55cm]{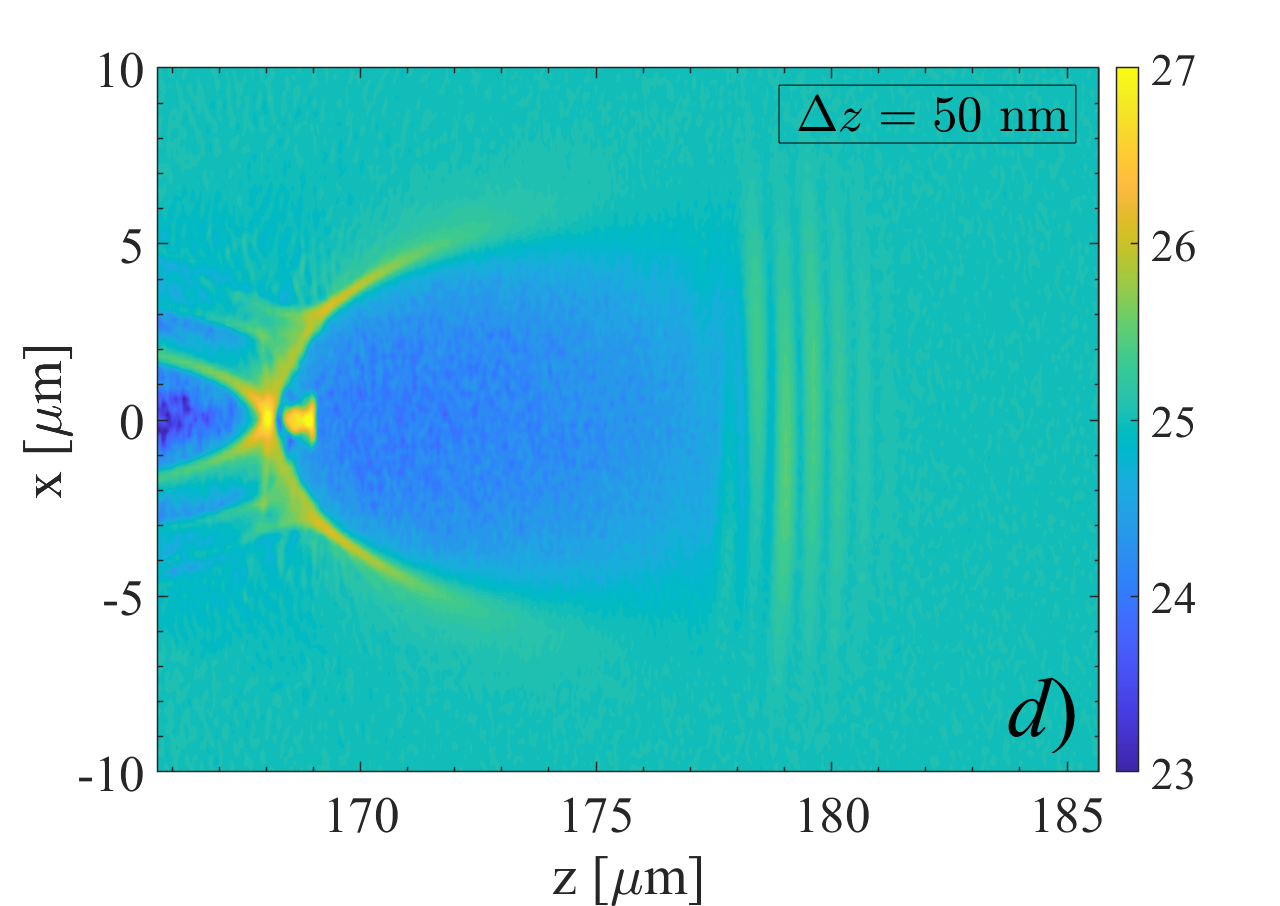}
\includegraphics[trim={15px 0 118px 35},clip,height=4.55cm]{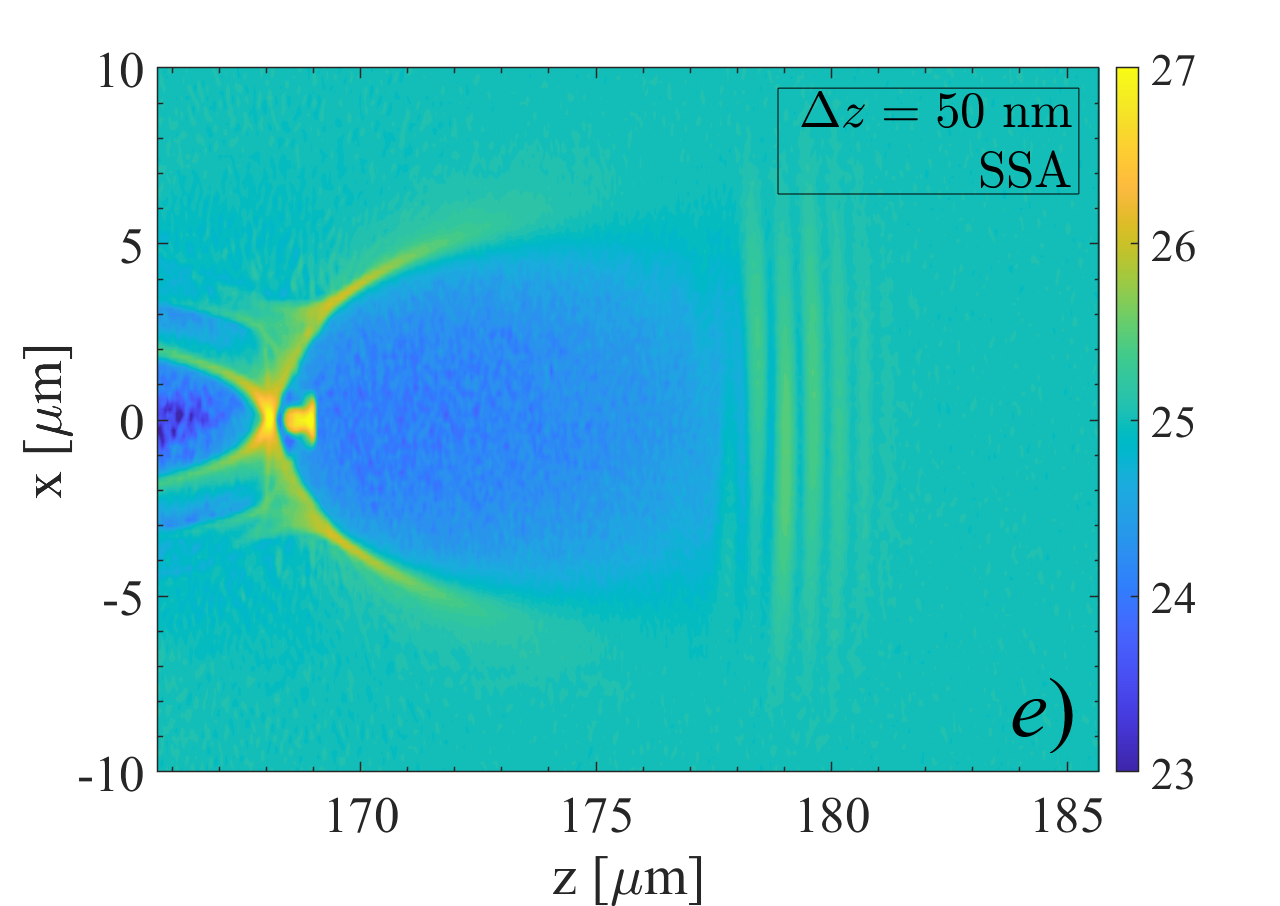}
\includegraphics[trim={15px 0 20px 35},clip,height=4.55cm]{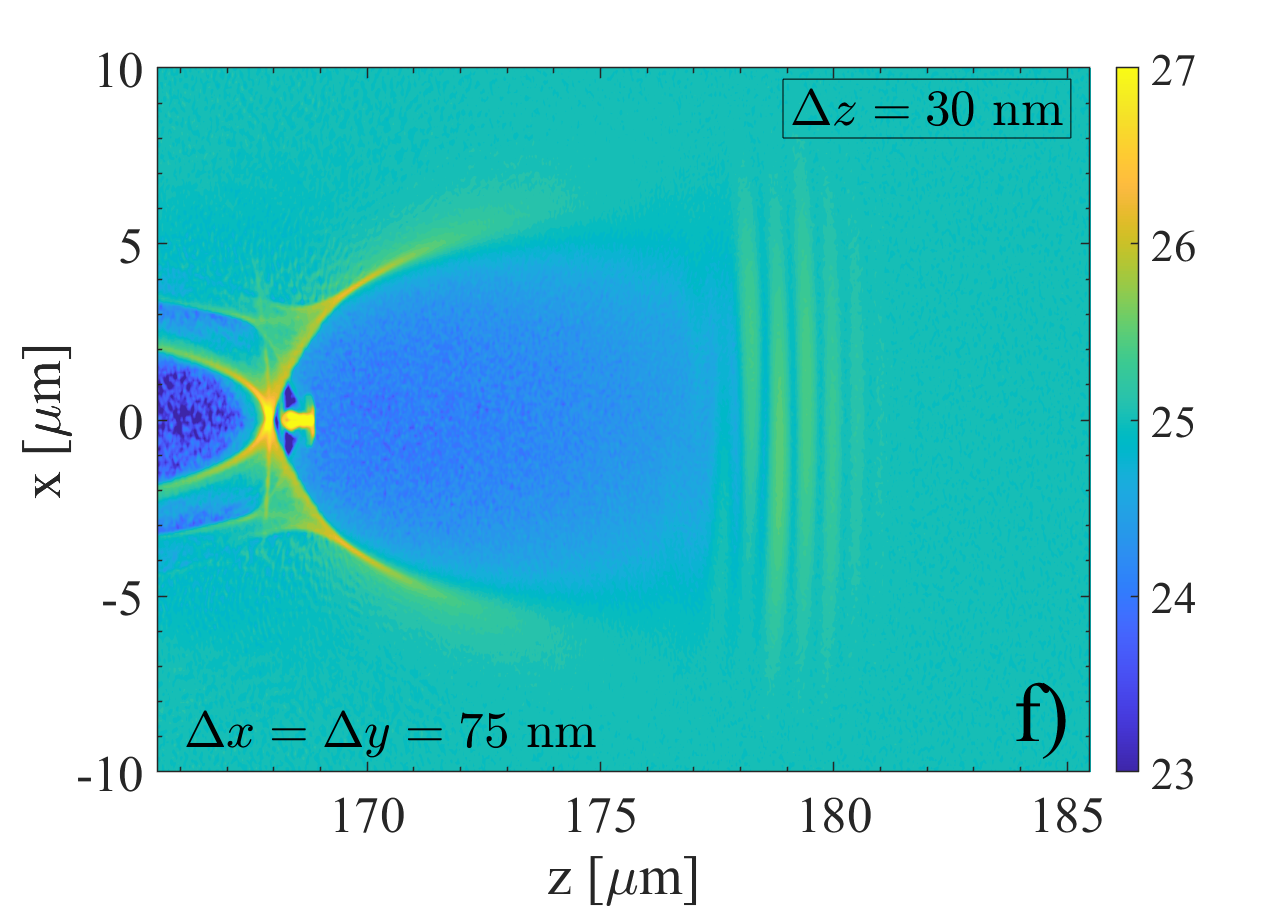}
\caption{Snapshots of electron density distributions on $\log_{10}$ scale (with units $\log_{10}(n_e[m^{-3}])$) obtained with EPOCH $(a)$-$(c)$ and with our code $(d)$-$(f)$. In EPOCH the shape of the injected bunch and its charge, both depend on the resolution and on the particle pusher. } 
\label{fig:test_lwfa_density} 
\end{figure*}

\begin{figure*}[h!]
\centering
\includegraphics[trim={20px 0 118px 35},clip,height=4.55cm]{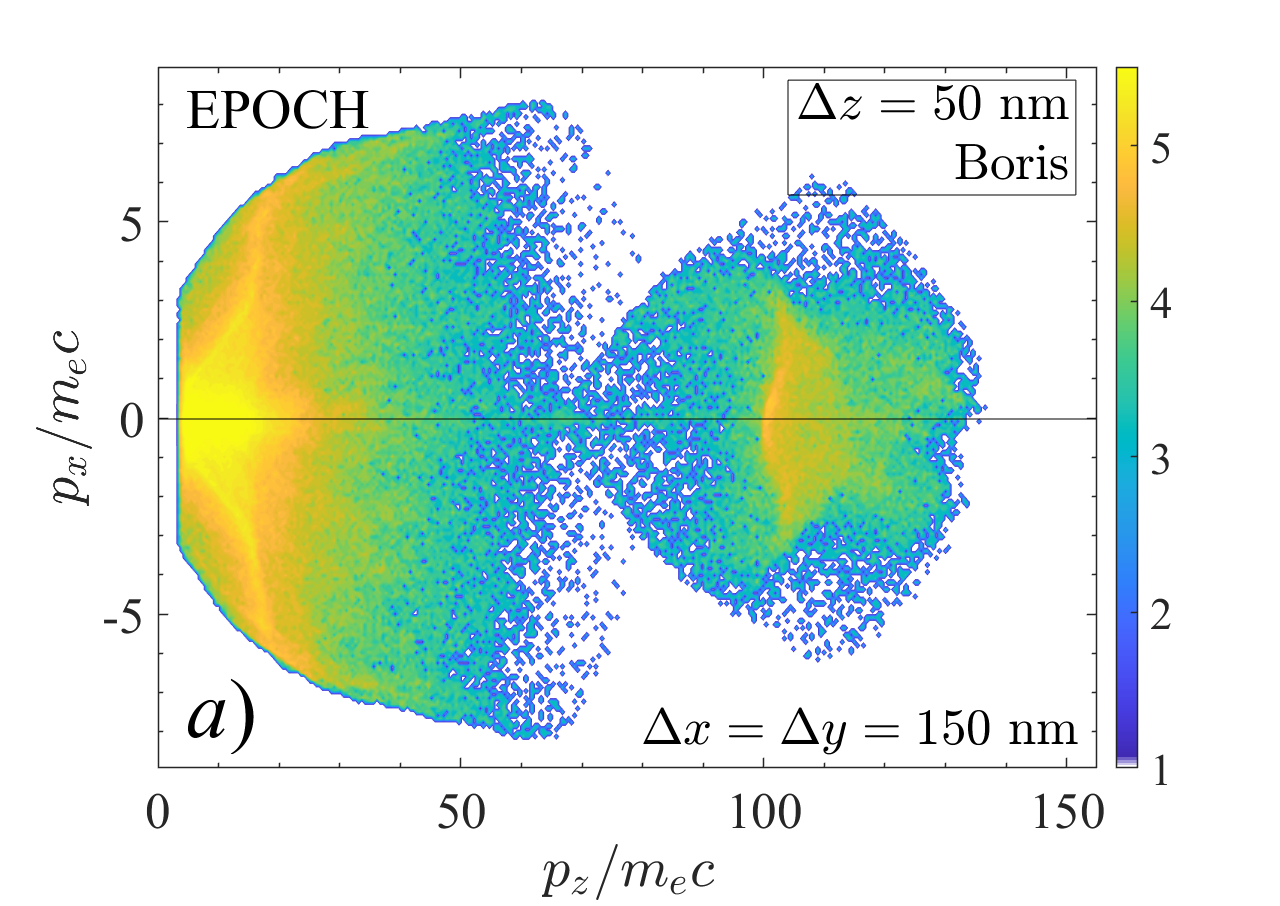}
\includegraphics[trim={15px 0 118px 35},clip,height=4.55cm]{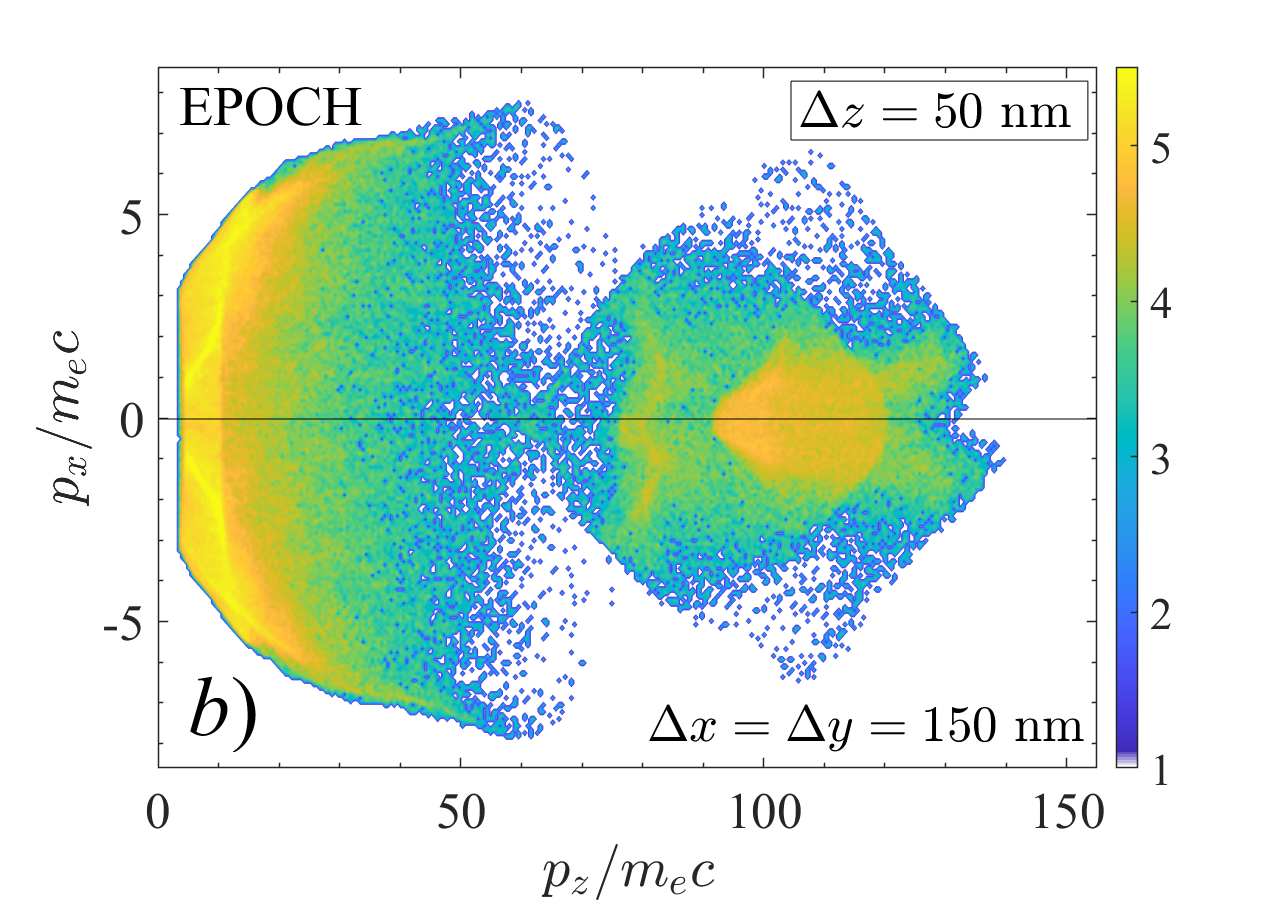}
\includegraphics[trim={15px 0 20px 35},clip,height=4.55cm]{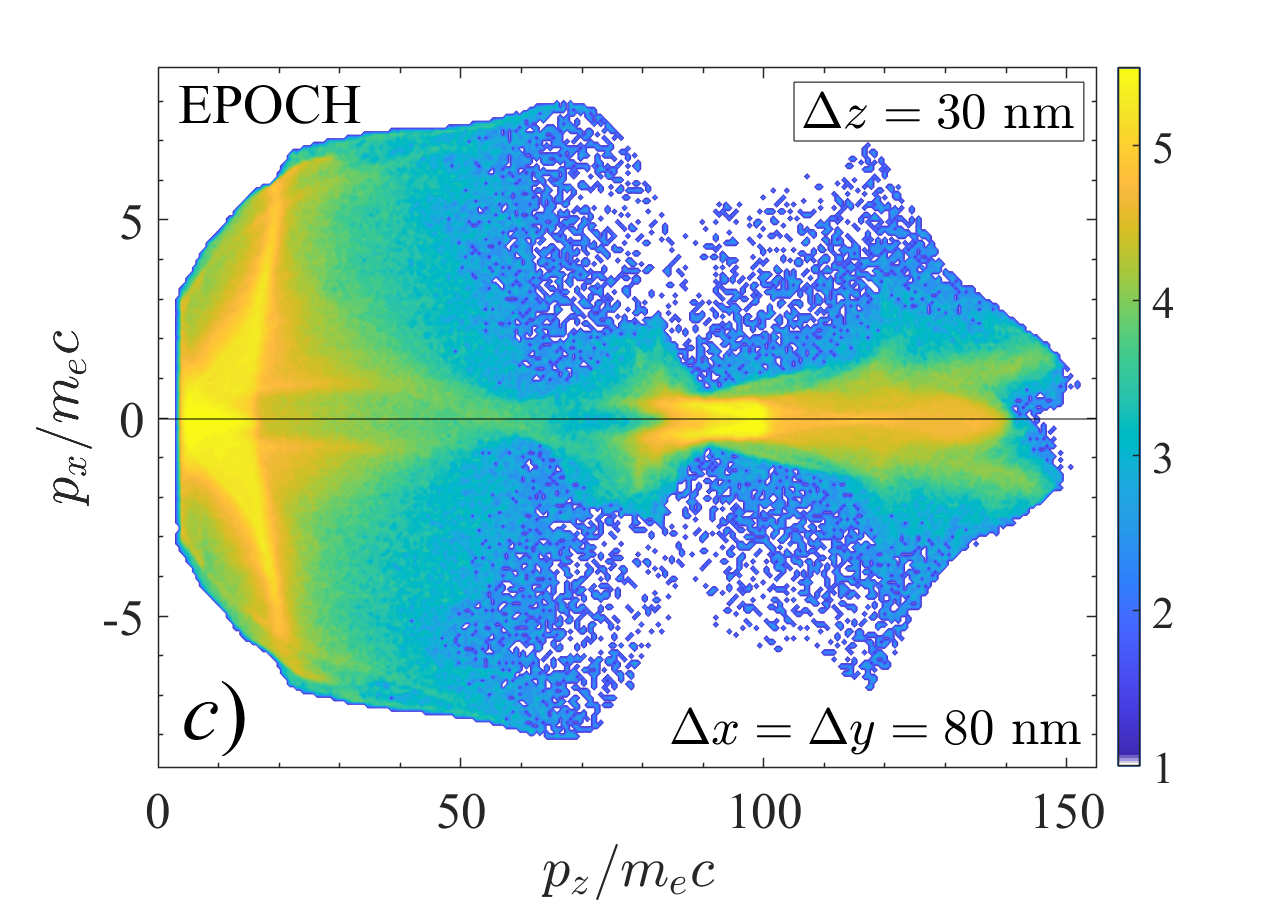}
\includegraphics[trim={20px 0 118px 35},clip,height=4.55cm]{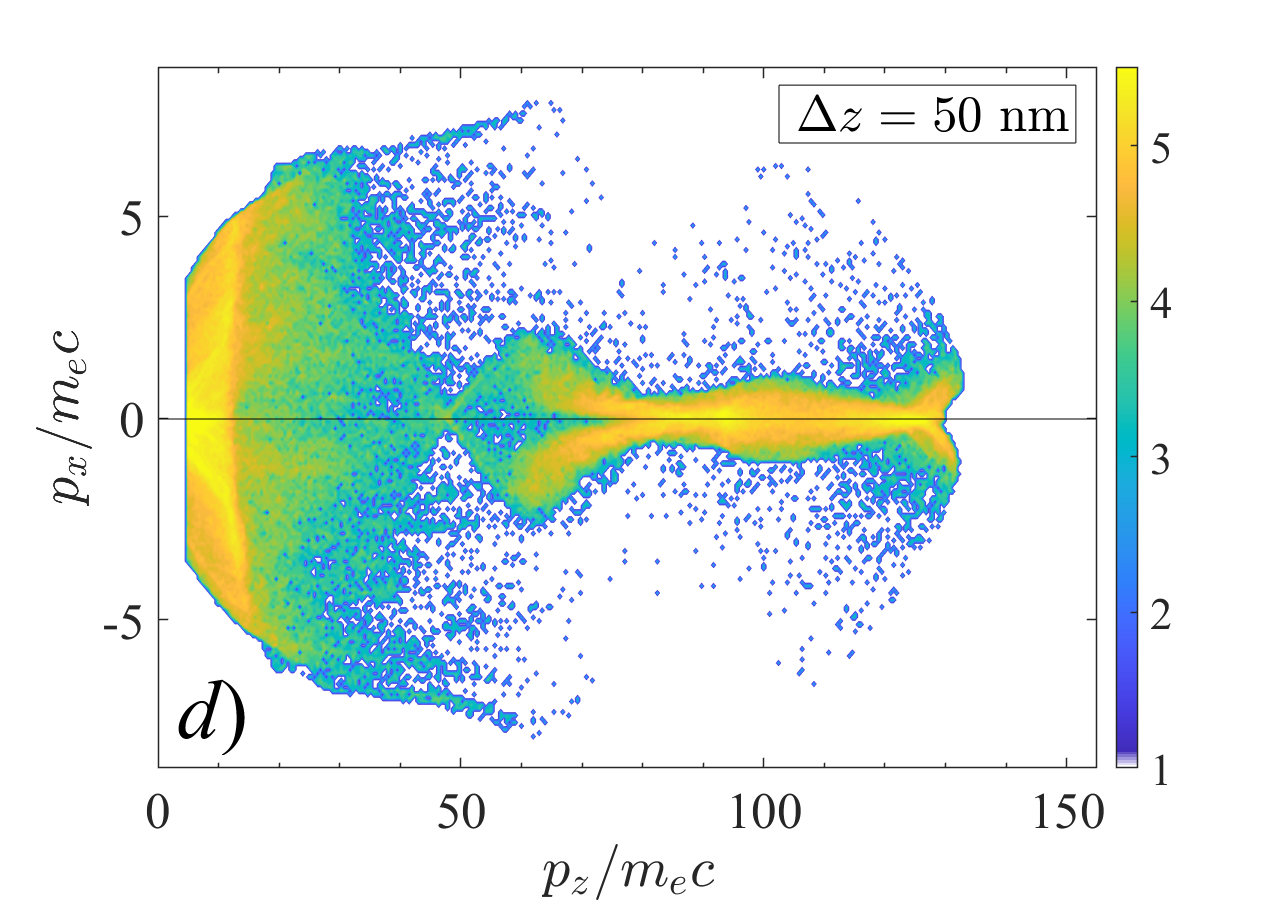}
\includegraphics[trim={15px 0 118px 35},clip,height=4.55cm]{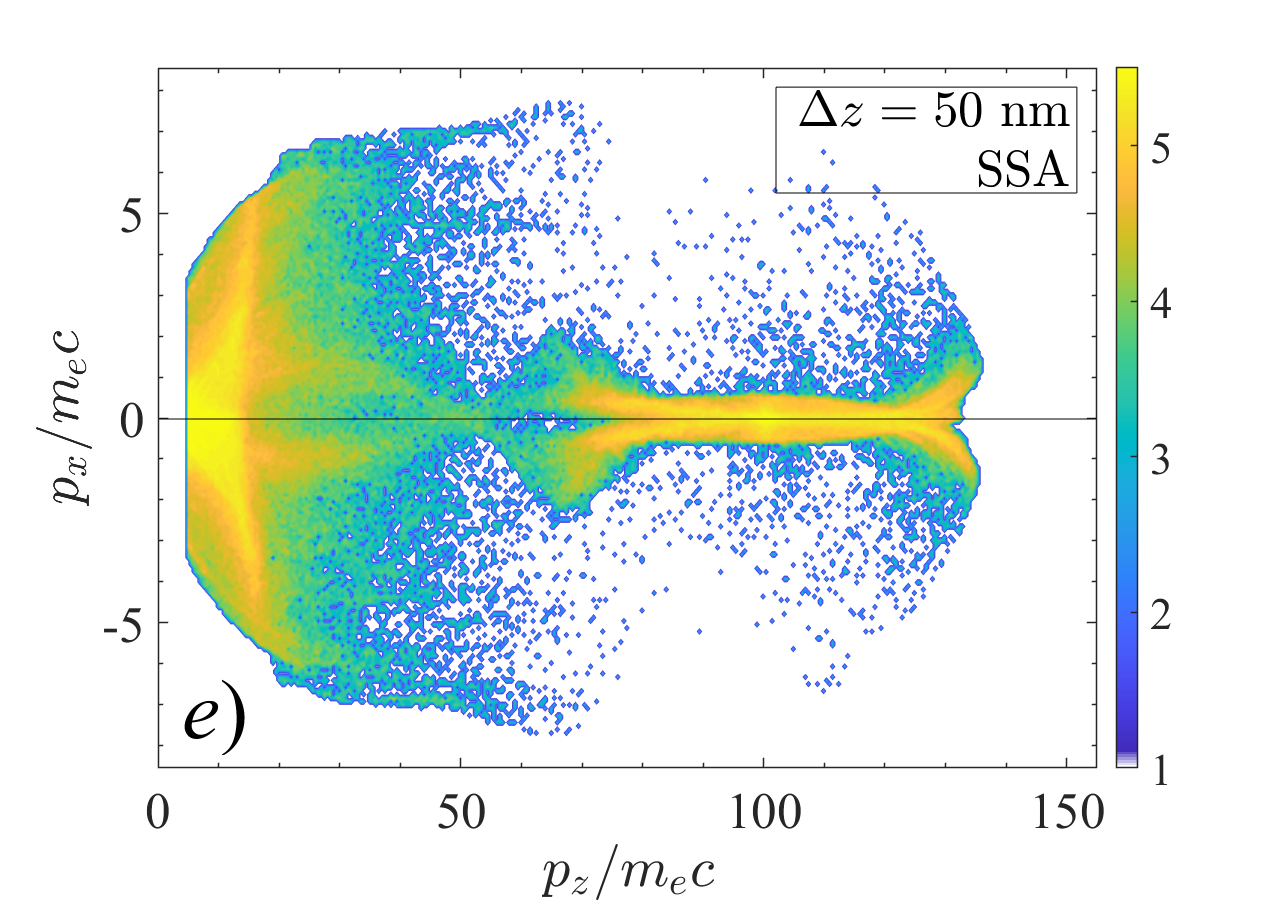}
\includegraphics[trim={15px 0 20px 35},clip,height=4.55cm]{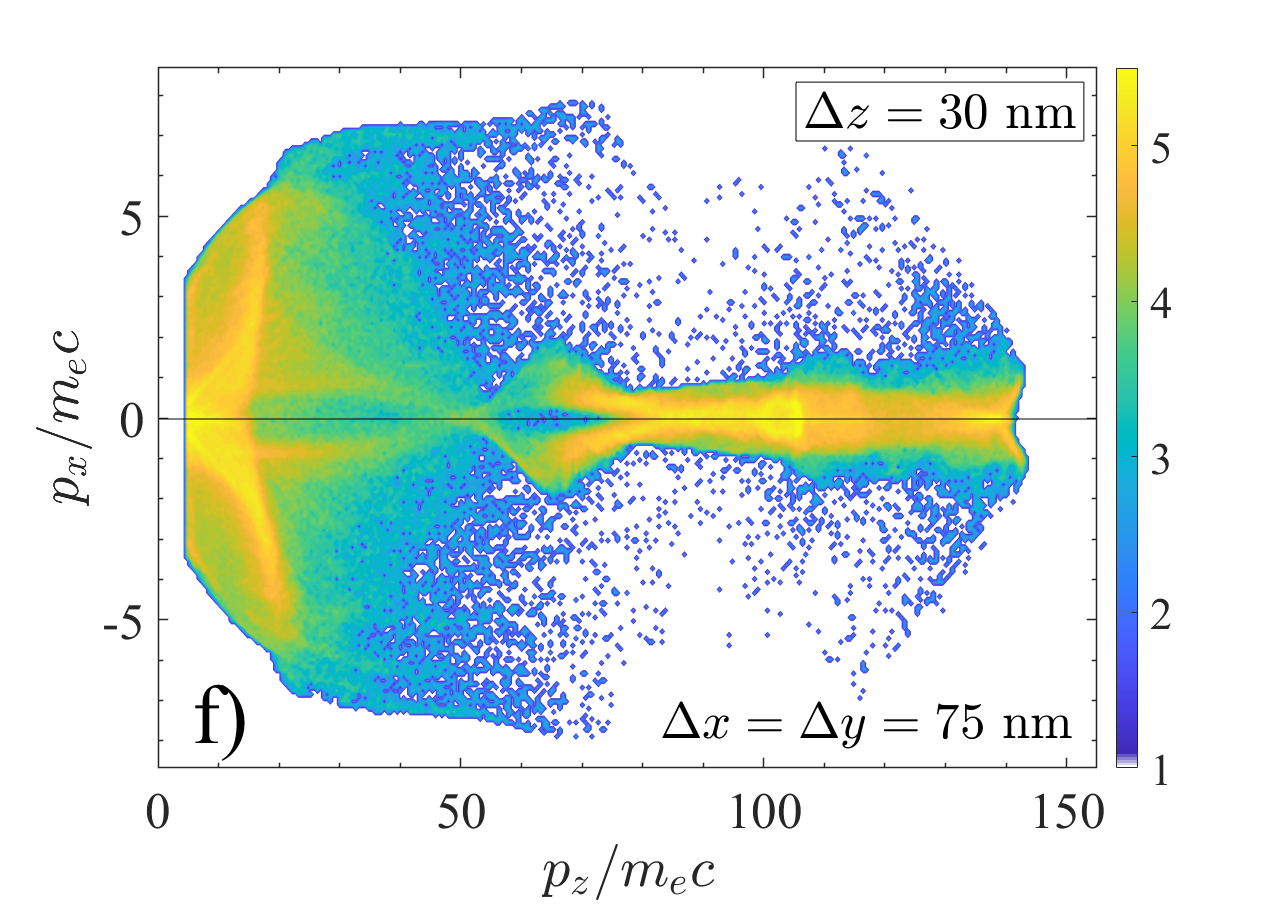}
\caption{Electron distributions on $\log_{10}$ scale in the momentum space at the same time instance as in Fig. \ref{fig:test_lwfa_density}. Our code is less sensitive to the numerical parameters and shows clear convergence in the cut-off energy.}
\label{fig:test_lwfa_momentum}
\end{figure*}

\begin{figure*}[h!]
\centering 
\includegraphics[trim={20px 0 150px 0},clip,height=6.3cm]{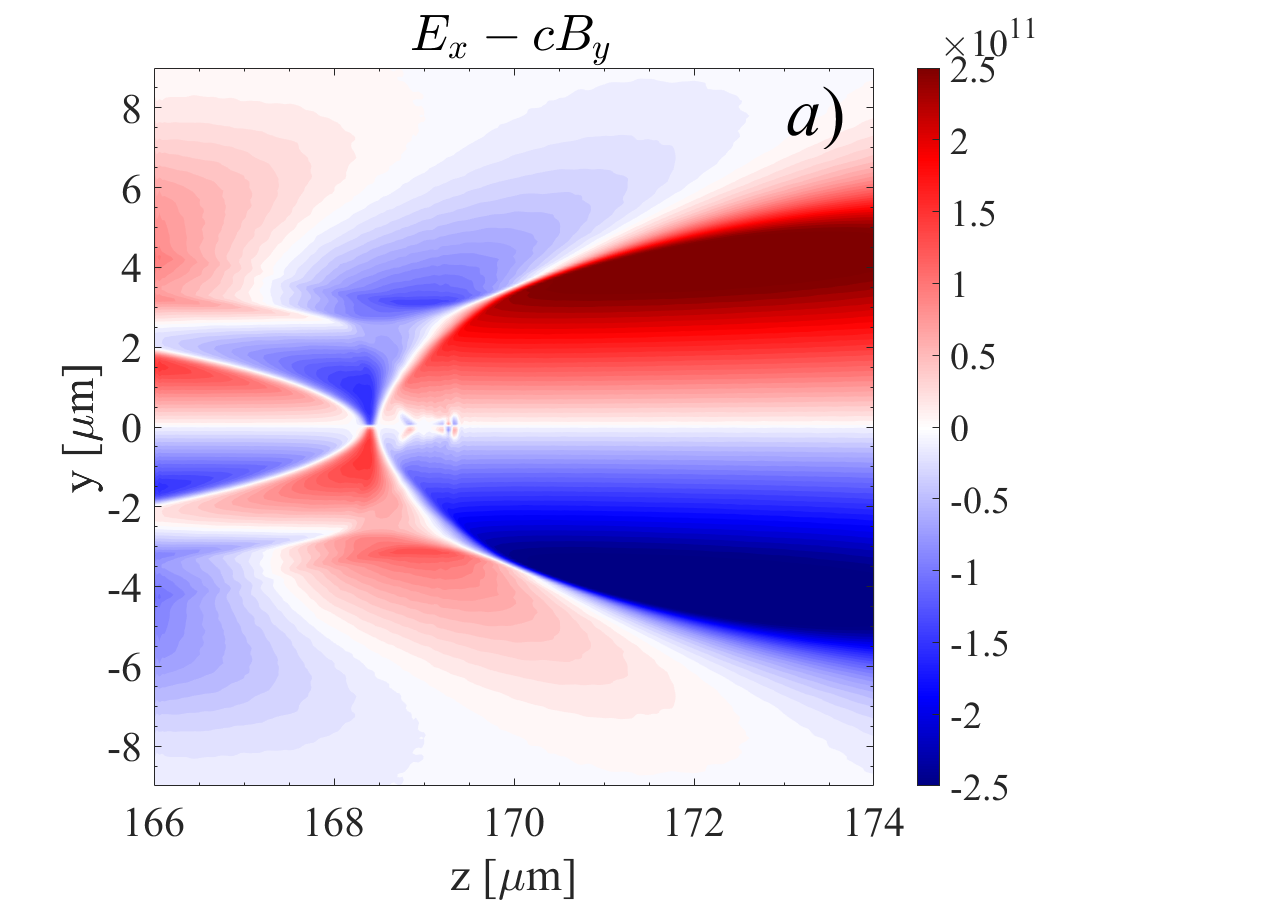}
\includegraphics[trim={20px 0 150px 0},clip,height=6.3cm]{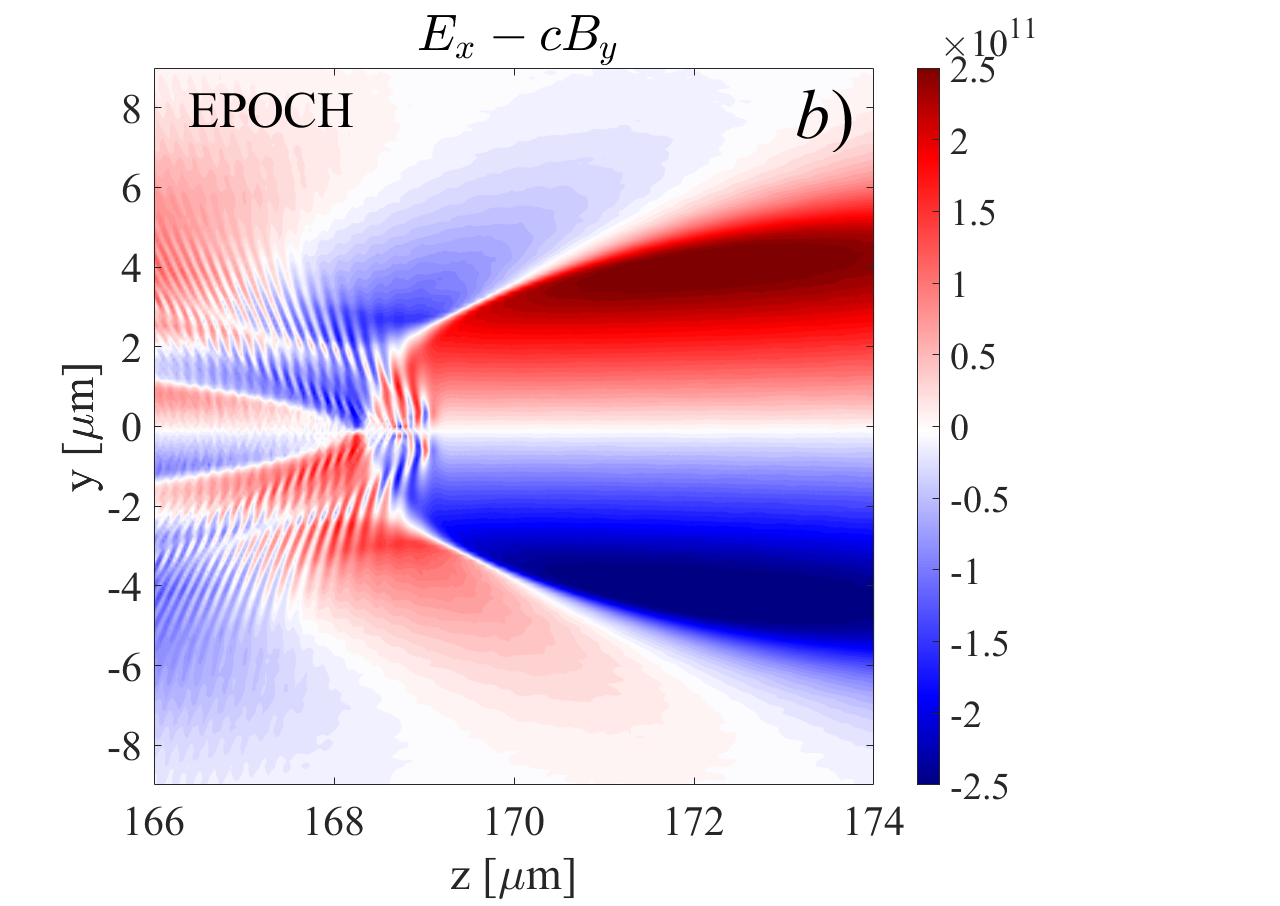}
\caption{Snapshot of the plasma fields $E_x-c B_y$  in the plane perpendicular to the laser polarization (in ${\rm V \cdot m^{-1}}$ units) after electron injection using our code $(a)$ and using EPOCH $(b)$. Numerical Cherenkov radiation can be seen on the EPOCH result. } 
\label{fig:test_lwfa_cherenkov}
\end{figure*}

The electron density distribution, after 160 $\mum$ propagation, is presented in Fig. \ref{fig:test_lwfa_density}. The results of EPOCH simulation are shown in the upper row with $\Delta z=50$ nm, $\Delta x=\Delta y=150$ nm, in Fig. \ref{fig:test_lwfa_density}$(a,b)$ and with $\Delta z=30$ nm, $\Delta x=\Delta y=75$ nm, in Fig. \ref{fig:test_lwfa_density}$(c)$. In the case of the lower resolution the Boris and HC pushers result in quite different spatial distributions and they are also different from the density seen in Fig. \ref{fig:test_lwfa_density}$(c)$. In the case of our code the injected bunch looks very similar for both resolutions. The effect of supersampling (Fig. \ref{fig:test_lwfa_density}$(e)$) becomes visible when the electron distributions in the momentum space are compared, which are presented in Fig. \ref{fig:test_lwfa_momentum}. The upper row shows again the results of EPOCH simulation, where the importance of numerical parameters becomes even clearer. A somewhat surprising result is that the Boris pusher does not reproduce the same electron distribution, as the HC pusher in EPOCH, which is due to its erroneous behaviour in the force-free case, discussed above. We ran our code also with the Boris-2 pusher, which only resulted in minimal differences compared to the results shown here. However, at the back of the bubble the Boris-2 push step might not provide sufficient cancellation of the $E_x-cB_y$ components in the Lorentz-force. Nonetheless, using EPOCH with high resolution ($\Delta z=30$ nm in Fig. \ref{fig:test_lwfa_density}$(c)$) the results converge to the results obtained with our code, which provides almost the same momentum distribution for all configurations. The only quantity that has observable variation is the cut-off energy (or maximum of $p_z$), which is lower in the case of lower resolution. However, applying the supersampling we obtained a maximum energy, which is closer to the cut-off seen at higher resolution, in Fig. \ref{fig:test_lwfa_momentum}$(f)$. The dependence of cut-off energy on the resolution is attributed to the fact that the electric field $E_z$ near the node of a nonlinear wakefield has a very sharp peak with a typical scale length of 10s of nanometers, that has to be resolved by the grid in order simulate the self injection correctly.

The transverse spread in the electron momentum is found to be much larger in EPOCH, which can be explained by the numerical Cherenkov radiation emission, presented in Fig. \ref{fig:test_lwfa_cherenkov}. In our code these fields are filtered out, while in EPOCH they have a strong influence on the electron motion. The emitted radiation in the frequency range between $c/(4\Delta z)$ and $c/(2\Delta z)$ is strongly dispersed, which is visible in Fig. \ref{fig:test_lwfa_cherenkov}(b) as well: the velocity of short wavelength components tends to zero, while the longer wavelengths can co-propagate with the injected electrons and disturb the bunch structure. One can use an improved finite difference solver \cite{lehe2013numerical, pukhov2016pic, bourgeois2020numerical_cherenkov} in order to suppress this artificial field emission, or a current filter can be used in EPOCH to suppress it to some degree (compare with Fig. \ref{fig:test_ncr_fourier}$(a)$).

\section{Conclusion}
 
In this paper we thoroughly explored the possibility of using exponential operators in particle-in-cell simulations in \revise{physics of relativistic laser-plasma interaction}.
%for the first time to the best of our knowledge.
We derived the exponential solution for the Maxwell fields ${\bf E}, {\bf B}$ using high order Taylor-expansions, the results of which converge to the analytical solution, and thus can be made precise for arbitrary accuracy by increasing the computation cost. The advantage of this method is that it is basis and geometry independent, and we used it to give computationally simple solution for the Maxwell equations. This latter means that all important operations are reduced to a lot of matrix-vector multiplications with banded diagonal matrices, using which we provided highly accurate and local solution in real space. Unfortunately, much of the mathematical complexity is due to the problem how to properly form these banded diagonal matrices. We used high-order finite differences for high spatial accuracy and we found out that these had to form staggered system on the Yee-grid, because of the spectral behaviour the first order derivatives. We could specify the latter to very high orders e.g. 12th-32th orders to suppress numerical wave dispersion effects and the effects of the numerical Cherenkov radiation. However, this tactic only worked properly for spatial waves $\leq0.5\kk_{\max}$. We had to filter spatial frequencies of the source current ${\bf J}$ above that threshold to remove all spurious numerical effects in the field solver. Then we adapted the standard quasi particle PIC method to this local field solver, which includes particle push, particle field interpolation, density and current deposition, such that all of the latter are local in nature. We found that using high order spline particle shapes, and  our $2\times$ supersampling method for the interpolated fields are needed for more accurate particle dynamics, especially in nonlinear laser-plasma interactions, like laser wakefield acceleration. In this paper we focused on methods of efficient thread based parallelization, which provides parallelization on multi core CPU-s with minimal overhead. We also designed these to be compatible with GPU-s.

We also verified our results in multiple benchmarks. For our exponential Maxwell solver the most important case was vacuum propagation of a laser pulse, where we completely mapped the accuracy characteristics of our solution. We found  that the exponential Taylor-expansion is not norm conserving, but due to the brute force nature of the solution, the magnitude of the latter error indicated the global high order convergence of the solution reasonably well. We also found that the use of very high order finite differences are also required for accurate wave propagation. Using our code it was possible to reduce the spurious fields of a relativistically travelling particle by orders of magnitude by using the combination of high order finite differences and current filters. We showed that the interpolation accuracy for the particle dynamics are better than conventional Yee codes, similar to the best spectral codes, and even the latter we can surpass using our interpolation supersampling scheme which effectively doubles the resolution of the interpolated fields. The tests we did for linear laser propagation in plasma, and generation of linear wakefield also showed high accuracy, and the benefit using the latter supersampling method. We also benchmarked our solution with more interesting physical phenomena, like surface harmonics generation and laser wakefield acceleration of electrons and compared the results and convergence to that of standard Yee-codes SMILEI and EPOCH. We confirmed better convergence and spectral profile of the generated harmonics using our code and much better and convergent electron momentum distributions and density in the laser wakefield scenario compared to these traditional codes.

Overall, we can say that our PIC solution seems less efficient in raw throughput than codes based on traditional Yee methods, especially enhanced Yee methods \cite{lehe2013numerical, pukhov2016pic, bourgeois2020numerical_cherenkov}, but our solution have the advantage of high degree of global accuracy. On the other hand the fully spectral solutions retain their superior accuracy in the field solver in the expense of their locality. In the future there are many possibilities to extend this work (we gave a glimpse of these in the appendices of this paper) with enhanced methods, physical modules, and parallelization with domain decomposition. We plan to use our code for laser wake field electron acceleration, and we plan to extend these methods to quasi cylindrical geometry.

\section*{Acknowledgments}
The ELI ALPS project (GINOP-2.3.6-15-2015-00001) is supported by the European Union and co-financed by the European Regional Development Fund. We acknowledge KIF\"U for awarding us access to HPC resource based in Debrecen, Hungary.

%%********************************************************************************************* 
 
\appendix
\section{Reference units} \label{subsubsec:units}
We use normalized reference units in the equations and their implementation. We borrow these expressions from SMILEI \cite{derouillat2018pic_smilei}, the definitions of which require a reference frequency $\omega_R$ or a reference length $L_R$ to be specified SI units (see Table \ref{tab:units}).  In our code we provide conversion from SI units to normalized units and vice-versa on the input/output side.  For debugging purposes, we set the reference length to a prefixed SI unit, for example setting $L_R = 10^{-6}$ makes the unit length one $\mu m$, this way the output will be more readable without conversion to SI. We note that volume integrating the number density in these units does not result in the number expected from SI, it has its own unit.
 
\begin{table}[ht]
\begin{tabular*}{\linewidth}{@{\extracolsep{\fill}} rcc }

Name  & Symbol &   Unit  \tabularnewline
\hline 
Velocity       & \bf{v, u}        & $c$\tabularnewline
Charge         & $q$              & $e$\tabularnewline
Mass           & $m$              & $m_{e}$\tabularnewline 
Energy         & $\mathcal{E}$    & $m_{e}c^{2}$\tabularnewline
Time           & $t, \omega_R^{-1}$ &  $\omega_{R}^{-1}$\tabularnewline
Length         & $x,y,z,r, \rho, L_R $  & c/$\omega_{R}$\tabularnewline
Number density & $n_R$          &  $\varepsilon_{0}m_{e}\omega_{R}^{2}/e^{2}$\tabularnewline
Number         & $n {\rm d }V, w$       &  $L_R^3 n_R$\tabularnewline
Current density & $\bf{J}$         &  $e c n_R$\tabularnewline
Electric field & $\bf{E}$   & c $m_{e}\omega_{R}/e$\tabularnewline
Magnetic field & $\bf{B}$   &   $m_{e}\omega_{R}/e$\tabularnewline
\hline
\end{tabular*}
\caption{Expressions for normalized units including the name of the physical quantity, its commonly used symbols and unit values. To complete the expressions the value of the reference length $L_R$ or the reference frequency $\omega _R$ must be provided in SI units.   \label{tab:units}}
\end{table}

\section{Solving Poisson's equation} \label{subsubsec:poisson}
Suppose, we want to solve the divergence equation Eq. (\ref{eq:maxwell_divE}) for $\bf E$  which has 3 unknowns but only one equation - we need to extract the longitudinal component of $\bf E$ using the scalar potential $\phi$:
\begin{equation} \label{eq:potential}
 {\bf E}^{(L)}      = -\nabla \phi.
\end{equation}
Then Eq.(\ref{eq:maxwell_divE}) reduces to Poisson's equation for a single variable  $\phi$:
\begin{equation} \label{eq:poisson_3d}
\partial_x^2 \phi+\partial_y^2 \phi+\partial_z^2\phi = \nabla^2 \phi = -\varrho .
\end{equation}
For solving the Poisson's equation Eq. ({\ref{eq:poisson_3d}}) many methods exist \cite{BOOK_NUMERICAL_RECIPIES}, the most efficient of which are the fast Fourier transformation based methods. In fact, on a Fourier basis the solution of Eq. (\ref{eq:poisson_3d}) is written simply as: 
\begin{equation} \label{eq:poisson_fourier_3d}
\tilde{\phi} = \left({\kk_x^2+\kk_y^2+\kk_z^2}\right)^{-1} \tilde{\varrho}.
\end{equation}
If we have a domain that is periodic in all directions and $\tilde{\varrho}$ is expanded  on the real Fourier-basis, then Eq. (\ref{eq:poisson_fourier_3d}) is the correct solution.

If we assume open boundary at any edge we face complications (like at domain boundaries), because the Poisson equation is completely non-local, i.e. its Green function (in 3D $r^{-1}$, in 2D $\log\rho$) has infinite range. The boundary values strongly affect the solution on the domain's intermediate points. A correct solution method of Eq. (\ref{eq:poisson_3d}) then must involve these steps: ($i$) remove the multipole moments of $\varrho$ so the new $\phi$ will be close to zero at domain boundaries, ($ii$) solve the Poisson equation with the modified $\varrho$ for the domain, ($iii$) add back the multipole moments to $\phi$ (all of them will affect all of the subdomains). This way the infinite range will be contained in the multipole moments.

A flexible numerical representation of Eq. (\ref{eq:poisson_3d}) using high order finite differences can be written as:
\begin{equation} \label{eq:poisson_discrete_3d}
{\rm D}_x^2 \phi+{\rm D}_y^2 \phi+{\rm D}_z^2\phi = -\varrho,
\end{equation}
where ${\rm D}_x^2$,  ${\rm D}_y^2$,  ${\rm D}_z^2$ are the banded diagonal matrix representation of the second order differentials which may include additional boundary conditions (see Eq. (\ref{eq:spatial_derivative2})). This forms a system of linear equations to be solved (after multipole corrections are done). We solve this with the iterative method called biconjugate gradient stabilized (BiCGSTAB) method \cite{vorst1992bicgstab}. It is a so called Krylov subspace method, and it only involves matrix-vector products and vector-vector projections even for non-Hermitian (symmetric) matrices. For multiple subdomains information exchange at the domain interfaces can be incorporated into the iterative procedure, resulting the same explicit style algorithm as the main exponential propagator. This is orders of magnitude less efficient than the Fourier methods, but we only need to do this at the beginning of the simulation if required.

\section{Simplifying the source exponential} \label{subsubsec:exponential_source}
There are a couple of methods that can be used to approximate the integral in Eq. (\ref{eq:maxwell_formal}) instead of Eq. (\ref{eq:exponential_solution2}). 
If $\bf{J}$ is not available forward in time we can approximate the fields to the first order as:
\begin{equation} \label{eq:exponential_source1}
\Psi(t+\Delta t) \approx \exp \left( \Delta t \op{H} \right)
                    \Psi(t)- \Delta t {\bf J}\left(t\right)
\end{equation}
If we already have access to the future value of $\bf J$, we can use the second order trapezoid rule:
\begin{equation} \label{eq:exponential_source2a}
\Psi(t+\Delta t) \approx \exp \left( \Delta t \op{H} \right)
                   \left[ \Psi(t)- \frac{\Delta t}{2} {\bf J}\left(t\right)  \right] -  
                  \frac{\Delta t}{2} {\bf J}\left(t+\Delta t \right) 
\end{equation}
or we can take advantage of the leapfrog method  in the second order midpoint rule with ${\bf J}_{1/2} = {\bf J}\left(t + \Delta t/2 \right)$:
\begin{equation} \label{eq:exponential_source2v}
\Psi(t+\Delta t) \approx \exp \left( \frac{\Delta t}{2} \op{H} \right)
                   \left[  \exp \left( \frac{\Delta t}{2} \op{H} \right) \Psi(t)- \Delta t {\bf J}_{1/2}  \right]
\end{equation}
which shows that it is possible factorize exponentials into substeps. 

However, there is another second order approximation that is used in pseudospectral analytical time domain solution (PSATD).  We assume ${\bf J}$ is constant during the step, and we analytically evaluate the source integral:
\begin{equation} \label{eq:exponential_source2A}
\Psi(t+\Delta t) \approx  \exp \left( \Delta t \op{H} \right)\Psi(t) - \op{H}^{-1} 
                   \left[  \exp \left( \Delta t \op{H} \right)- 1 \right] {\bf J}(t)
\end{equation}
All of the second order methods, like the midpoint Eq. (\ref{eq:exponential_solution2}), trapezoid Eq. (\ref{eq:exponential_source2a})  and analytical Eq. (\ref{eq:exponential_source2A}) have the same Taylor coefficients up to $\Delta t^2$. \revisee{ It is also possible to express the latter in a compact form using Taylor-expansion:
\begin{equation} \label{eq:exponential_sourceT0}
\Psi(t+\Delta t) \approx \Psi(t) + \sum_{n = 1}^{N} \frac{\Delta t ^n}{n!} \op{H}^{n-1} \left[\op{H} \Psi(t) - {\bf J}(t)  \right] .
\end{equation}
For general nonlinear problems, these can be used to form predictor-corrector methods.
}

Finally, we write out a formula in which we merge the two exponentials in the midpoint rule Eq. (\ref{eq:exponential_solution2}) into a single Taylor-expansion provided that we limit the exponential of the source to its lowest orders: 
\begin{align} \label{eq:exponential_sourceT1}
\tilde{\Psi}_{1}(t) = -\frac{3}{2} {\bf J}_{1/2} + \op{H} \left( \Psi + \frac{\Delta t}{4} {\bf J}_{1/2}  \right), \\
\Psi(t+\Delta t) \approx \Psi(t) + \frac{\Delta t}{2} {\bf J}_{1/2} + \sum_{n = 1}^{N} \frac{\Delta t ^n}{n!} \op{H}^{n-1} \tilde{\Psi}_{1}(t). \label{eq:exponential_sourceT}
\end{align}
In fact, the formula Eq. (\ref{eq:exponential_sourceT}) recovers the first 3 Taylor expansion terms of $\exp(\Delta t \op{H}/2){\bf J}_{1/2}$ exactly while not altering the exponential of the state vector $\Psi (t)$.

\section{Absorbing layer} \label{subsubsec:absorbing_layer}

In what follows, we want to adapt the perfectly matched layers (PML) \cite{berenger1996pml, johnson2021PMLnotes} within our exponential method. We want them to form a layer at the domain boundaries which could absorb outgoing electromagnetic waves. Due to the nature of various approximations we make, we simply call the following "absorbing layer".

One standard way to derive PML is done by transforming the spatial coordinates near outer edges to complex domain in each direction as:
\begin{equation} \label{eq:absorbing_layer_deriv}
\partial_p \rightarrow \left( 1+ (i \omega)^{-1} \sigma _p \right)^{-1}\partial_p, \quad { \rm  for } \quad  p = x, y, z,
\end{equation}
where $\omega$ is angular frequency in the temporal Fourier space,  $\sigma_x(x)$, $\sigma_y(y)$, $\sigma_z(z)$ are the layer functions in each direction.  They take nonzero values only near the outer edges. %This expression provides frequency independent absorption.
Next, we rewrite the Maxwell equation (\ref{eq:maxwell_Ex}) using Eq. (\ref{eq:absorbing_layer_deriv}) and $\partial_t \rightarrow i\omega$ without sources:
\begin{multline} \label{eq:absorbing_layer_omegaEz}
\left(i\omega + \sigma _y+ \sigma _z+(i\omega)^{-1} \sigma _y\sigma _z \right) E_x = \\
 - \left( 1+ (i \omega)^{-1} \sigma _y \right)\partial_z B_y +\left( 1+ (i \omega)^{-1} \sigma _z \right)\partial_y B_z,  
\end{multline}
We neglect the terms with multiple $\sigma$, because those only affect the corner values. Then we  transform the similarly modified Eqs. (\ref{eq:maxwell_Ex})-(\ref{eq:maxwell_Ez}) to time domain to get:
\begin{align}
\partial_t E_x &= -\partial_z B_y +\partial_y B_z - (\sigma_z + \sigma_y) E_x  -\partial_z \tilde {B}_y +\partial_y \tilde{B}_z,\label{eq:absorbing_maxwell_Ex}\\
\partial_t E_y &=  \partial_z B_x -\partial_x B_z - (\sigma_z+\sigma_x) E_y + \partial_z \tilde{B}_x -\partial_x \tilde{B}_z,\label{eq:absorbing_maxwell_Ey}\\
\partial_t E_z &= -\partial_y B_x +\partial_x B_y - (\sigma_y+\sigma_x) E_z -\partial_y \tilde{B}_x +\partial_x \tilde{B}_y,\label{eq:absorbing_maxwell_Ez}%\\
%\tilde{B}_x &= \int^{t} \sigma _x B_x {\rm d}t', 
%\tilde{B}_y  = \int^{t} \sigma _y B_y {\rm d}t',
%\tilde{B}_z  = \int^{t} \sigma _z B_z {\rm d}t',%\label{eq:maxwell_Ez}
\end{align}
where $\tilde{B}_x$, $\tilde{B}_y$ and $\tilde{B}_z$ are solved from auxiliary equations as:
\begin{equation}
\tilde{B}_x = \int^{t} \sigma _x B_x {\rm d}t', 
\tilde{B}_y  = \int^{t} \sigma _y B_y {\rm d}t',
\tilde{B}_z  = \int^{t} \sigma _z B_z {\rm d}t'. \label{eq:absorbing_aux_B}
\end{equation}
Then we repeat this process for Eqs. (\ref{eq:maxwell_Bx})-(\ref{eq:maxwell_Bz}) which will introduce  $\tilde{E}_x$, $\tilde{E}_y$ and $\tilde{E}_z$ defined similar as Eq. (\ref{eq:absorbing_aux_B}).

Then, the modified Maxwell-equations will take the form (instead of Eq. (\ref{eq:maxwell_formal})) :
\begin{equation} \label{eq:absorbing_maxwell_formal}
 \partial_t \Psi      = (\op{H}-\Sigma) \Psi - {\bf J} + \op{H}\tilde{\Psi},
\end{equation}
where $\Sigma$ is a diagonal operator which contains $\sigma_x$, $\sigma_y$ and $\sigma_z$ layer functions, and $ \op{H}\tilde{\Psi}$ is a new source term. If we use leapfrog method to calculate Eq. (\ref{eq:absorbing_aux_B}) auxiliary sources, we can merge it with the  staggered current: $\tilde{{\bf J}}_{1/2} = {\bf J}_{1/2}-\op{H}\tilde{\Psi}_{1/2}$. Even though $\Sigma$ can be included in the Taylor-expansion of the main exponential it will adversely affect its stability properties. Instead, we can use the split-operator formula Eq. (\ref{eq:exponential_split2}) to split $\Sigma$ from the main exponential:
\begin{multline} \label{eq:absorbing_maxwell_solution2}
\Psi(t+\Delta t)   \approx \exp (-\Sigma \Delta t/2)  \times \\ 
\left [ \exp (\op{H} \Delta t) \exp (-\Sigma \Delta t/2)  \Psi - \exp (\op{H} \Delta t/2) \tilde{{\bf J}}_{1/2} \Delta t \right] ,
\end{multline}

If we do not include the PML source term $\op{H}\tilde{\Psi}$ in the solution, only the  $\Sigma$ absorbing operator we could still achieve high degree of absorption for outgoing waves in layers such that the absorption scales with the width of the respective layer (which should be wider than the wave length). The PML source terms  $\op{H}\tilde{\Psi}$, however, has significant impact for waves traveling with glazing incidence at the absorbing layers: without them there is a considerable amount of spurious diffracted waves that slowly travel perpendicular to the wave propagation direction.  Accuracy of the source term also depend on $\Delta t$. Overall, PML absorbing layers work with our exponential method, a quantitative analysis may be published elsewhere.

\section{Coordinate transforms} \label{subsubsec:transform}
We directly incorporate into the spatial representation of our numerical solution the possibility of using coordinate transformation on our spatial coordinates in $x$, $y$. Our aim is to achieve real coordinate stretching which results in using less grid points to describe the same physical domain. We limit our self to a single dimension: let $x$ be a coordinate which we discretize on a uniform grid and $\tilde{x}$ coordinate that represent the physical domain. These two are related by $g(x)$ coordinate transform as:
\begin{equation} \label{eq:transform_g}
 \tilde{x} = g(x), \quad  {\rm and } \quad g'(x) = \partial_x g.
\end{equation}
We can express the relevant physical derivatives by applying chain rule to an $\tilde{f}=f(\tilde{x})$ or $f=f(x)$ function:
\begin{align}
\partial_{\tilde{x}}   \tilde{f}  &= (g')^{-1} \partial_x f,\label{eq:transform_deriv}\\
\partial_{\tilde{x}}^2 \tilde{f}  &= (g')^{-2} \partial_x^2 f-g'' (g')^{-3} \partial_x f,\label{eq:transform_deriv2}
\end{align}
Numerically we just replace the analytical derivatives with the respective finite difference versions such as: $g_k \rightarrow g(x_k)$, $\partial_x \rightarrow {\rm D}_x$, $\partial_x^2 \rightarrow {\rm D}_x^2$. It slightly complicates things, when we use staggering. If we stagger Eq. (\ref{eq:transform_deriv}) by $\pm\Delta x/2$ then $g_k^{(\pm)} \rightarrow g(x_k\pm\Delta x/2)$, $\partial_x \rightarrow {\rm D}_x^{(\pm)}$. %We note that the left hand side can be also directly approximated on the non uniform $\tilde{x}$ coordinate using general Lagrange-polynomial method (see \ref{subsubsec:lagrange}).

The integrals will transform as:
\begin{equation}
\int \tilde{f} {\rm d}\tilde{x} = \int f(x) g'(x) {\rm d}x \approx \sum_i f_k \left( g'_k \Delta x \right),\label{eq:transform_int}
\end{equation}
This means that the grid samples $\Delta x_k$ will be replaced by $g'_k \Delta x$ on the uniform grid. Using Eqs. (\ref{eq:transform_g})-(\ref{eq:transform_int}) is all it takes to transform the electromagnetic field solvers to stretched coordinates.

In case of PIC particles additional steps needed to be done, since the field interpolation and deposition must be done on the physical coordinate $\tilde{x}$. We have done this by transforming the particle properties to the uniform grid $\tilde{x} \rightarrow x$, then we proceed by doing everything as we would be doing on a uniform grid - just replacing the local cell volume with  $g'_k \Delta x$ (like as in cylindrical 3D PIC codes). The extra difficulty here is that we must evaluate the inverse $g^{-1}(\tilde{x})$, which is not known analytically (except in simple cases). We circumvent this by using Halley's root finding method \cite{BOOK_NUMERICAL_RECIPIES} to iterate the inverse value as:
\begin{equation}
g^{-1}(\tilde{x}) \rightarrow x_{n+1} = x_{n}-\frac{2 \left[ g(x_n)-\tilde{x} \right] g'(x_n) }{
2 \left[ g'(x_n)\right]^2 - \left[ g(x_n)-\tilde{x} \right]  g''(x_n)},\label{eq:transform_inverse2}
\end{equation}
for a chosen reasonable start value of $x_0$ not far from the ideal $x$. In absence of an analytical inverse, this is ideal in the sense that it only requires fast evaluation of $g(x), g'(x)$  and $g''(x)$, which are needed for the transformation of the field solver anyway. It also provides faster convergence than Newton's method.
    
\begin{figure}[ht]
\centering
\includegraphics[width=6.5cm]{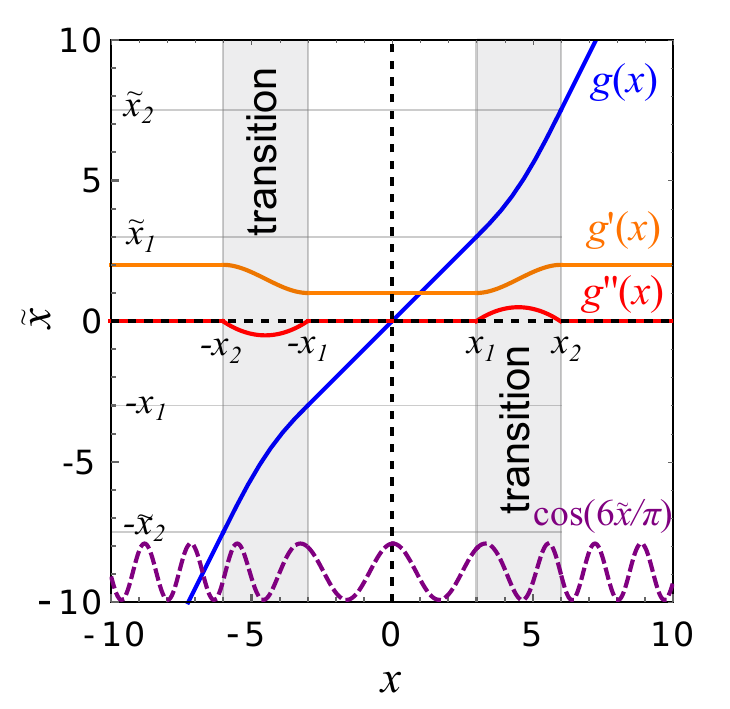}
\caption{ Real coordinate transform function Eq. (\ref{eq:transform_gpoly}) (blue) and its  derivatives $g'(x)$ (orange) and $g''(x)$ (red) with $c_1 = 2, x_1 = 3, x_2 = 6$. The resulting polynomial in the transition zone is $g_\text{m}(x) = -\frac{9}{2}+6 x -2 x^2 +\frac{1}{3}x^3-\frac{1}{54}x^4$ with $c_0 = -\frac{9}{2}$. The end of the transition zone in the physical coordinate is $\tilde{x}_2=7.5$. We also show a transformed physical wave $\cos(6\tilde{x}/\pi)$ at the bottom with purple dashed lines.
\label{fig:transform_g}}
\end{figure}

All that is left to prescribe $g(x)$ that can achieve the coordinate stretching and local enough so that the calculation of the inverse Eq. (\ref{eq:transform_inverse2}) is reasonably fast. We assume the following piecewise polynomial form (for $x > 0$):
\begin{equation} \label{eq:transform_gpoly}
g(x) = 
\begin{cases}
g_\text{in}(x)  = x,  & \text{if } x \leq x_1 \\
g_\text{m}(x) = b_0 + b_1 x + b_2 x^2 + b_3 x^3 + b_4 x^4,  & \text{between} \\
g_\text{out}(x) = c_0+c_1 x,  & \text{if } x_2 \leq x \\
\end{cases}
\end{equation}
The coordinate domain is composed of an inner zone, outer zone with uniform grids, and a transition zone between them with grid scaling $c_1$. We specify the transformation with $c_1$, $x_1 = \tilde{x}_1$ and $x_2$. The rest of the parameters we have to calculate analytically by requiring that $g(x)$ be smooth at the zone boundaries $x_1, x_2$ as much as possible: 
\begin{align} 
g_\text{in}(x_1) &= g_\text{m}(x_1), \quad g'_\text{in}(x_1) = g'_\text{m}(x_1), \quad g''_\text{m}(x_1) = 0,  \label{eq:transform_gpoly1} \\
g_\text{out}(x_2) &= g_\text{m}(x_2), \quad g'_\text{out}(x_2) = g'_\text{m}(x_2), \quad g''_\text{m}(x_2) = 0, \label{eq:transform_gpoly2}
\end{align}
from these $b_0, b_1, b_2, b_3, b_4$ and $c_0$ can be solved in closed form dependent on only $x_1, x_2$ and $c_1$ such that the resulting $g(x)$ will be continuously differentiable twice (see Table \ref{tab:transform_poly}). It is possible to generalize this to higher order polynomials to increase the smoothness. It worth noting that the value of $c_0$ will not depend on the polynomial order yielding  $c_0 = (1-c_1)(x_1+x_2)/2$. This polynomial expression for $g(x)$ will allow us to evaluate the inverse function Eq. (\ref{eq:transform_inverse2}) efficiently. We show an example for the transformation function on Fig. \ref{fig:transform_g}.

\begin{table}[ht]
\begin{tabular*}{\linewidth}{@{\extracolsep{\fill}} rcc }

Coefficient  &  Value  \tabularnewline
\hline 
$b_0$   & $ {(1-c_1) x_1^3 (x_1-2 x_2)}/{(2 (x_1-x_2)^3)}$        \\
$b_1$   & $ ({c_1} x_1^2 (x_1-3 x_2)+x_2^2 (3 x_1-x_2))/(x_1-x_2)^3$              \\
$b_2$   & $ -{3 (1-c_1) x_1 x_2}/{(x_1-x_2)^3}$              \\ 
$b_3$   & $ {(1-c_1) (x_1+x_2)}/{(x_1-x_2)^3}$    \tabularnewline
$b_4$   & $-(1-c_1)/( 2 (x_1 - x_2)^3)$   \tabularnewline
$c_0$   & $(1-c_1)(x_1+x_2)/2$ \tabularnewline
\hline
\end{tabular*}
\caption{Table of the polynomial coefficients of Eq. (\ref{eq:transform_gpoly}) expressed with parameters $x_1, x_2$ and $c_1$ using Eqs. (\ref{eq:transform_gpoly1}) and (\ref{eq:transform_gpoly2}).   \label{tab:transform_poly}}
\end{table}

The accuracy field solver using these coordinate transformations will be limited by the accuracy of  finite difference order used (considering the inner and outer zones) and the smoothness of the transition layer between them. The smoothness can be increased by increasing the polynomial order or by increasing the transition with $x_2-x_1$. Even though this can be implemented in usual Yee-solvers and many have done similar \cite{belayev2015PICsar}, the low order accuracy of the derivatives used there will cripple the field solver accuracy in the outer zones. On the other hand doing this on spectral basis is prohibitive because the fast Fourier transform algorithm cannot be used on the physical grid. 
\revise{We note that there exist a wide variety of structure preserving methods that are based on the theory of mimetic differences \cite{palha2014physics_discretization}, and nonrelativistic PIC methods that employ these \cite{kraus2017gempic,campos2024variational_particle} which are well suited to handle complex geometries and general curvilinear coordinates.}

Even though this coordinate transformation is effective for stretching the physical domain, for example to prevent distortion of a laser pulse, it may introduce smoothness artifacts: the highest frequency physical spatial  waves that are in the inner zone and that cannot be represented in the outer zone (due to the larger grid spacing) are reflected in the transition zone if they travel outwards. Despite this, we found it to be important to natively support coordinate transformations in our code.

\section{Dispersion enhanced finite differences} \label{subsubsec:enhance}

There is an interesting numerical technique that makes it possible to further improve the accuracy of the high-order finite difference representation, which is also fully compatible with our banded diagonal matrix formulation (see Section \ref{subsec:bandedmatrices}) without modifications. It has been already used in laser-plasma PIC codes to reduce the effects of \revise{anisothropic field propagation \cite{sekido2024accuracy_difference}}, the numerical Cherenkov radiation (NCR) \cite{blinne2018maxwell_dispersion, lu2020cherenkov_pic} at the higher spatial frequencies, and it is based on customizing the coefficients of the high order differences in spectral space. Xu et. al. proposed to do this kind of dispersion improvement with incorporating temporal information into widened difference formulas, and improving the fields surrounding relativistically drifting particles \cite{xu2020relativistically_moving}. Though this latter is not applicable directly to our exponential solution, but dispersion enhancement is possible by improving the spatial accuracy of the differences without any performance cost. We want to illustrate the key concepts here, detailed investigation may be published elsewhere.

\begin{figure}[ht]
\centering
\includegraphics[height=4.5cm]{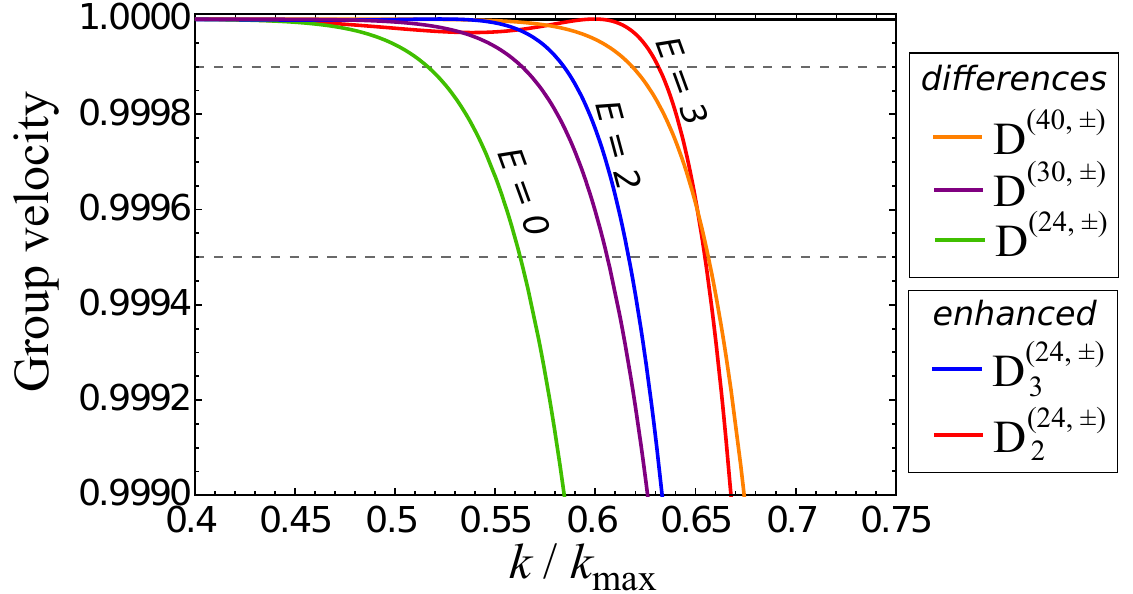}
\caption{ Spectral functions of our customized dispersion enhanced  24th order staggered finite differences using $E=2$ (blue) and $E=3$ (red) number of independent coefficients in Eq. (\ref{eq:enhance_difference}). We also show the group velocity curves for the staggered 24th (green), 30th (purple) and 40th (orange curves) order differences.
\label{fig:enhance24}}
\end{figure}

\begin{table}[ht]
\begin{tabular*}{\linewidth}{@{\extracolsep{\fill}} c|cc }

Coefficient  &  $E = 2$   &  $E = 3$ \tabularnewline
\hline
$C^{(24)}_{0}$   & $\ \  4.23126860472023$ & $\  10.7340070467821$  \\
$C^{(24)}_{1}$   & $-4.33978633826126$ & $-16.8700581115584$  \\
$C^{(24)}_{2}$   & $ $ & $\quad  8.4538046813569$   \tabularnewline
\hline
\end{tabular*}
\caption{Table of the coefficients in Eq. (\ref{eq:enhance_difference}) for our customized  dispersion enhanced 24th order staggered finite differences corresponding to Fig. (\ref{fig:enhance24}).    \label{tab:enhance24}}
\end{table}

We outline the technique as the following. We assume that the customized finite difference formula ${\rm D}^{(\pm, n)}_{E}$ is the linear combination of the $E+1$ the highest order finite differences of the staggered variants Eq. (\ref{eq:spatial_derivativeSt}) such that
\begin{equation} \label{eq:enhance_difference}
{\rm D}^{(\pm, n)}_{E} = \sum_{\epsilon = 0}^{E-1} C^{(n)}_{\epsilon} {\rm D}^{(\pm, n-2\epsilon)}+ \left(1-\sum_{\epsilon = 0}^{E-1} C^{(n)}_{\epsilon} \right) {\rm D}^{(\pm, n-2E)},
\end{equation}
where  $n$ is the largest difference order selected and $C^{(n)}_{\epsilon}$ are the coefficients. Note that the ansatz Eq. (\ref{eq:enhance_difference}) is always a finite difference operator itself, and it is at least $n-E$ order accurate. Let us denote the spectral function of ${\rm D}^{(\pm, n)}_{E}$ as $g^{(\pm, n)}_{E}(\kk)$ and its derivative as $G^{(\pm, n)}_{E}(\kk)$. The exponential method inherits the real part of its dispersion curve from  $G^{(\pm, n)}_{E}(\kk)$. Then we have to search for $C^{(n)}_{\epsilon}$ coefficients such that they minimize the wave dispersion error, like using the following $L^2$ error functional:
\begin{equation} \label{eq:enhance_minimize}
h\left( C^{(n)}_{\epsilon} \right) = \int_{0}^{\kk_{max}} w(\kk) \left[1 - G^{(\pm, n)}_{E}\left(\kk, C^{(n)}_{\epsilon} \right)\right]^2  {\rm d}\kk,
\end{equation}
where $w(\kk)$ is a weight function of choice. Its clear that after minimization ${\rm D}^{(\pm, n)}_{E}$ will be more accurate on the average than any of its constituents. One downside of this is that the group velocity could overshoot the speed of light (= 1) in vacuum. The more coefficients one uses the higher dimensional the parameter space gets, and there are more parameters that can be constrained to met one's design goals.

We looked for coefficients that directly enhances the $n$th order derivatives such that their enhanced versions at most have $2\times 10^{-7}$ overshoot or undershoot in dispersion curve compared to the $n$th order in the ansatz. We did this for the 24th order finite differences with $E=2$ and $E = 3$ which we show in Fig.  \ref{fig:enhance24} and list their coefficients in Table \ref{tab:enhance24}. We can see that what we can achieve with this optimization technique: $E=2$ extended the group velocity plateau region to 30th order finite difference curve and $E=3$ near to the 40th order finite difference curve, significantly improving the dispersion properties without any computational cost. Regarding the effects on simulations, these are ultimately just customized finite difference formulas without any temporal component, they behave exactly as can be predicted from Fig. \ref{fig:enhance24}, further enhancing our exponential method. In the future we want to make this an integral part of our code, but a systematic analysis of this method, however, is out of scope of this paper.

\section{Pseudospectral exponential solution} \label{subsubsec:spectral_maxwell}

\revise{In the following we summarize how the exponential method can be used in spectral space to propagate the electromagnetic fields and how it is related to the pseudospectral analytical time domain (PSATD) algorithm.}

The goal of (pseudo)spectral methods is to transform the fields into a  basis set that diagonalizes or greatly simplifies the representation of the Maxwell-equations.  We suggest to transform the fields using the combination of $\mathcal{C}$ cosine transforms and $\mathcal{S}$ sine transforms in the following way: if the sampling along a spatial dimension would be non-staggered in Yee-grid representation we do the one dimensional $\mathcal{S}$ sine transform, if it would be staggered we do $\mathcal{C}$ cosine transform as:
\begin{align} \label{eq:spectral_transform_EBx}
\tilde{E}_x = \mathcal{S}_z \mathcal{S}_y \mathcal{C}_x \left[ E_x \right] (\kk_x, \kk_y, \kk_z), \ 
\tilde{B}_x = \mathcal{C}_z \mathcal{C}_y \mathcal{S}_x \left[ B_x \right] (\kk_x, \kk_y, \kk_z)  \\
\tilde{E}_y = \mathcal{S}_z \mathcal{C}_y \mathcal{S}_x \left[ E_y \right] (\kk_x, \kk_y, \kk_z), \ 
\tilde{B}_y = \mathcal{C}_z \mathcal{S}_y \mathcal{C}_x \left[ B_y \right] (\kk_x, \kk_y, \kk_z)  \\
\tilde{E}_z = \mathcal{C}_z \mathcal{S}_y \mathcal{S}_x \left[ E_z \right] (\kk_x, \kk_y, \kk_z), \ 
\tilde{B}_z = \mathcal{S}_z \mathcal{C}_y \mathcal{C}_x \left[ B_z \right] (\kk_x, \kk_y, \kk_z) 
\end{align}
Doing $\mathcal{S}$ transform poses antisymmetric, and doing $\mathcal{C}$ poses symmetric boundary conditions at the edges of the domain. The bulk of the computational load is concentrated to the latter 3D basis transforms. To keep actual computational complexity manageable, one should use the discrete fast cosine transform and the discrete fast sine transform algorithms.  

We suggested this type of Fourier-basis because then the Maxwell-equations (\ref{eq:maxwell_Ex})-(\ref{eq:maxwell_Bz}) remain completely real after the spectral transform  according to:
\begin{align}
\partial_t \tilde{E}_x &= -\kk'_z \tilde{B}_y + \kk'_y \tilde{B}_z - \tilde{J}_x , \quad
\partial_t \tilde{B}_x =  -\kk'_z \tilde{E}_y + \kk'_y \tilde{E}_z \label{eq:spectral_maxwell_EBx}\\
\partial_t \tilde{E}_y &= \kk'_z \tilde{B}_x - \kk'_x \tilde{B}_z - \tilde{J}_y , \quad
\partial_t \tilde{B}_y =  \kk'_z \tilde{E}_x -\kk'_x \tilde{E}_z ,\label{eq:spectral_maxwell_EBy}\\
\partial_t \tilde{E}_z &= -\kk'_y  \tilde{B}_x + \kk'_x \tilde{B}_y - \tilde{J}_z, \quad
\partial_t \tilde{B}_z =  -\kk'_y \tilde{E}_x +\kk'_x \tilde{E}_y ,\label{eq:spectral_maxwell_EBz}
\end{align}
where $\kk'_x = g(\kk_x), \kk'_y=g(\kk_y), \kk'_z=g(\kk_z)$ denotes the spectral function of the derivative \revise{(see Section \ref{subsec:differences})}. For the exact representation choosing the \revise{analytical} $\kk'_x=\kk_x, \kk'_y=\kk_y, \kk'_z=\kk_z$ is possible. Notice that ${ \bf J }$ current density also has to be transformed the same way as the ${\bf E}$ electric field. 
\revise{It is possible to map the real space solution of Section \ref{subsec:spatial_maxwell} into this  spectral basis exactly using the same boundary conditions if we take into account the grid staggering. This process needs specialized forms of the above transformations which are called DST-I and DCT-II in the FFTW  library \cite{FFTW_ODD}.} 

In this spectral basis the Maxwell-equations become almost diagonal.  We note that for each wave number $\kk_x,\kk_y$ and $\kk_z$ the propagation is described by simple 6x6 matrix: 
\begin{equation}  \label{eq:spectral_maxwell_H}
\mathdutchcal{ H} =
\left( \begin{array}{cccccc}
0            & 0         & 0       &      0      &-\kk'_z &  \kk'_y \\
0            & 0         & 0       &  \kk'_z & 0           &-\kk'_x \\
0            & 0         & 0       &  -\kk'_y & \kk'_x  & 0           \\
0            &  -\kk'_z & \kk'_y & 0          & 0         & 0          \\
 \kk'_z & 0           &  -\kk'_x & 0          & 0         & 0         \\
-\kk'_y & \kk'_x  & 0           & 0          & 0         & 0         \\
\end{array} \right) .
\end{equation}

Now, using the midpoint solution Eq. (\ref{eq:exponential_solution2}) with matrix Eq. (\ref{eq:spectral_maxwell_H}) one can propagate the Maxwell-fields in this spectral basis with the formula:
\begin{equation} \label{eq:spectral_solution2}
\tilde{\Psi}(t+\Delta t) \approx \exp \left( \Delta t \mathdutchcal{ H} \right)\tilde{\Psi}(t)- 
                \exp \left(\Delta t \mathdutchcal{ H} /2 \right)\tilde{\bf J}\left(t+\Delta t / 2\right) \Delta t
\end{equation}
For the expansion of exponential operator we propose two options: use high order explicit Taylor-expansion Eq. (\ref{eq:exponential_taylor}), or use a Pad\'e-approximant, such as Eq. (\ref{eq:exponential_CN2}), this latter will yield unconditionally stable algorithm. Both of these are easy to do numerically in this  representation. If we use the spectral function of (staggered) finite differences for $\kk'_x, \kk'_y, \kk'_z$, and use Taylor-expansion of the exponential we get an equivalent numerical solution compared to our main finite difference algorithm (same accuracy, same stability, same parallelization and locality properties). \revise{This forms a \emph{pseudospectral exponential solution}, but we prefer our main algorithm which provides more flexibility regarding the boundary conditions and spatial representation.}
%While this  spectral method could provide higher spatial-temporal accuracy effectively, our main algorithm provide more flexibility for the boundary conditions and spatial representation.
    
The analytical evaluation of Eq. (\ref{eq:spectral_solution2}) is also possible: we need to transform that expression into the eigenbasis of $\mathdutchcal{H}$, which will become a diagonal matrix containing its doubly degenerate eigenvalues of 0, and $\pm \kk' = \sqrt{\kk_x^{'2}+\kk_y^{'2}+ \kk_z^{'2}}$. Then, $ \exp \left( \Delta t \mathdutchcal{ H} \right) \tilde{\Psi}$ can be evaluated in this basis, and finally one transforms the expression back. The \emph{analytic expression for the exponential} becomes: 
\begin{multline} \label{eq:spectral_solutionA_E}
\tilde{\bf E} (t+\Delta t) = \cos(\kk' \Delta t) \tilde{\bf E} + \sin ( \kk' \Delta t) \kk'^{-1} ({\bf k'} \times \tilde{\bf B}) + \\ \left[ 1-\cos (\kk' \Delta t) \right] \kk'^{-2} {\bf k'} ({\bf k'} \cdot \tilde{\bf E})  
\end{multline}
\begin{multline} \label{eq:spectral_solutionA_B}
\tilde{\bf B} (t+\Delta t) = \cos(\kk' \Delta t) \tilde{\bf B} + \sin ( \kk' \Delta t) \kk'^{-1} ({\bf k'} \times \tilde{\bf E}) + \\ \left[ 1-\cos (\kk' \Delta t) \right] \kk'^{-2} {\bf k'} ({\bf k'} \cdot \tilde{\bf B}),
\end{multline}
where ${\bf k}' = (\kk'_x, \kk'_y, \kk'_z)$. To complete Eq. (\ref{eq:spectral_solution2}), we also need to subtract the current component $\exp \left(\Delta t \mathdutchcal{H} /2 \right)\tilde{\bf J}$ which follows from Eqs. (\ref{eq:spectral_solutionA_E})-(\ref{eq:spectral_solutionA_B}) straightforwardly. These results are the PSATD formulas containing $\cos(\kk' \Delta t )$ and $\sin(\kk' \Delta t )$ terms \cite{vay2013decomposition_spectral}. The Taylor expansion of the Maxwell exponential is equivalent to the Taylor-expansion of the PSATD results, without the divergence equations.\footnote{Oddly enough $\kk'$ does not have a reasonable operator equivalent in real space, because it is $\sqrt{-\nabla^2}$.} 
\revise{For a true pseudospetral analytical solution like the one in \cite{godfrey2014stability_spectral} using $\kk'=\kk$, the basis transformations have to be done on the global system even if domain decomposition is used across many processors.}

\section{Exponential Boris pusher} \label{subsubsec:pusherexp}

In this Section we show that it possible to solve the particle equation Eq. (\ref{eq:particles_motionu}) with exponential method, and what it yields the exponential form of the usual Boris pusher \cite{BOOK_PLASMA_SIMULATION, hairer2023boris_integrator}. We also outline an improved Boris pusher using a better expansion of the exponential.

We derive the formal solution of Eq. (\ref{eq:particles_motionu}) using the operator notation $\overline{\rm R}_n = - \overline{\bf B}_n \times $ of Eq. (\ref{eq:pusher_matrixR}). 
Using the formal exponential solution Eq. (\ref{eq:maxwell_solution}) on Eq. (\ref{eq:particles_motionu}) yields:
\begin{multline} \label{eq:pusherexp_solutionu}
  {\bf u}_{n+1} =
 \exp \left(2 \overline{\gamma}_{n}^{-1} \overline{{\rm R}}_{n} \right) {\bf u}_{n} +\\
 \frac{2}{\Delta t}
 \int_{s=0}^{\Delta t} \exp \left( 2 \left( {\Delta t}-s \right) \overline{\gamma}_{n}^{-1} \overline{{\rm R}}_{n} \right)
 \overline{\bf E}_{n} {\rm d}s,
\end{multline}
where we already included $r_\sss \Delta t /2$ factors in the definition of the average field values $\overline{\bf E}_n$ and $\overline{\bf B}_n$. The above formula also require
 $\overline{\gamma}^{-1}$, which is the average of the inverse of the particle's $\gamma$ factor:
\begin{equation} \label{eq:pusherexp_solutiongamma}
\overline{\gamma}^{-1} = {\Delta t}^{-1} \int_{s = 0}^{\Delta t} \gamma^{-1} (t_n +s) {\rm d}s
\end{equation}
These equations form an exact solution in laboratory frame. Main problem is how to self consistently approximate the average inverse gamma Eq. (\ref{eq:pusherexp_solutiongamma}). Complete exact solutions with closed formulas can be derived in boosted frame \cite{petri2020relativisic_pusher}.

To derive the exponential form of the Boris pusher, we substitute trapezoid rule into Eq. (\ref{eq:pusherexp_solutionu}) and multiply both sides with $\exp \left(- \overline{\gamma}_{n}^{-1} \overline{\rm R}_{n} \right)$:
\begin{equation} \label{eq:pusherexp_boris}
 \exp \left(- \overline{\gamma}_{n}^{-1} \overline{\rm R}_{n} \right) ( {\bf u}_{n+1} - \overline{\bf E}_{n}) =
 \exp \left(\overline{\gamma}_{n}^{-1} \overline{{\rm R}}_{n} \right) ( {\bf u}_{n} +\\
 \overline{\bf E}_{n} ).
\end{equation}
Using first order Taylor expansion on the left and right hand sides results the Boris formula Eq. (\ref{eq:pusher_solutionB}). Then, when calculating ${\bf u}_{n+1}$ on the right hand side the diagonal Pad\'e approximant Eq. (\ref{eq:exponential_CN2}) of $\exp \left(2\overline{\gamma}_{n}^{-1} \overline{{\rm R}}_{n} \right)$ appears which is norm (gamma) conserving. The average gamma is chosen as:
\begin{equation} \label{eq:pusherexp_borisgamma}
 \overline{\gamma}^{-1}_n \approx  {\gamma}^{-1}_{-} = {\gamma}^{-1}_{+} \quad \text{with} \quad  {\gamma}^{-1}_{\pm} = \left( {1+ ( {\bf u}_{\pm}^2}) \right)^{-1/2},
\end{equation}
where ${\bf u}_{-} = {\bf u}_n+\overline{\bf E}_n$ and ${\bf u}_{+} = {\bf u}_{n+1}-\overline{\bf E}_n$.

Using high order exponential expansions in Eq. (\ref{eq:pusherexp_boris})  the rotation term can be evaluated to arbitrary order. We propose the following choice: we use Pad\'e approximants individually on each side as: 
\begin{multline} \label{eq:pusherexp_pade1}
\exp \left(\overline{\gamma}_{n}^{-1} \overline{{\rm R}}_{n} \right) =
\left[ \exp \left(- \overline{\gamma}_{n}^{-1} \overline{\rm R}_{n} \right) \right]^{-1} \approx \\
\left[\II - \frac{1}{2} \overline{\gamma}_{n}^{-1} \overline{\rm R}_{n} \right]^{-1}
\left[\II + \frac{1}{2} \overline{\gamma}_{n}^{-1} \overline{\rm R}_{n} \right],
\end{multline}
If we do this, then it yields our proposed scheme Eq. (\ref{eq:pusher_solutionB2}), which has the property that each individual approximant preserves the vector norm such that each side of Eq. (\ref{eq:pusherexp_boris}) have the gamma:
\begin{equation}
\gamma_{n+1/2} = \sqrt{1+({\bf u}_n + {\bf E}_n)^2} = \sqrt{1+({\bf u}_{n+1} - {\bf E}_n)^2},
\end{equation}
which is unlike the usual Boris algorithm. By taking two exponential substeps, it directly means that the accuracy of the rotation improves by a factor of $4$.

Similar to Eq. (\ref{eq:pusher_inverseR}) it is also possible write the Pad\'e approximant in explicit form by taking advantage of the definition of $\overline{\rm R}_n = -{\bf B}_n \times$ from Eq. (\ref{eq:pusher_matrixR}) as:
\revise{
\begin{multline} \label{eq:pusherexp_pade2}
 \left[\II - \frac{1}{2}\overline{\gamma}_{n}^{-1} \overline{\rm R}_{n} \right]^{-2}
 \left[\II + \frac{1}{2}\overline{\gamma}_{n}^{-1} \overline{\rm R}_{n} \right]^2 = \\
 \II + 4
  \left( 1  + \overline{\gamma}^{-2}_{n} \overline{\bf B}_n^2  \right)^{-2}
 \left[ \overline{\gamma}_{n}^{-1} \left( 1 - \overline{\gamma}_{n}^{-2}\overline{\bf B}_n^2 \right)  \overline{\rm R}_n +  2\overline{\gamma}_{n}^{-2}  \overline{\rm R}_n^2  \right].
\end{multline}
}
\revisee{The evaluation cost of this is almost the same as the usual Boris-pusher. We note that the analytical expression of the exponential (\ref{eq:pusherexp_pade1}) is also possible using Rodrigues' rotation formula \cite{cheng1989historical_rotations} with $\theta = {\gamma}_{n}^{-1} |\overline{\bf B}|$ angle - this is the convergence limit of substepping the exponential more than twice.} 

The particular point of failure for the Boris pushers are the ${\bf E}\times{\bf B}$ conservation \cite{vay2008pusher}, which is most notable in the force free case where $\gamma^{-1}\overline{\rm R}_{n} {\bf u}_n = -\overline{\bf E}_n$. This is because that the average exponential rotators will not cancel the net ${\bf E}_n$ generally. Unfortunately, this is caused by the low order approximations in the integrals of Eqs. (\ref{eq:pusherexp_solutionu}) and (\ref{eq:pusherexp_solutiongamma}). This seems to suggest that all exponential pushers using low order quadratures suffer from this issue.

Nevertheless, we wrote out two simple test cases to our new Boris pusher. First we choose a weakly relativistic parameters
\begin{equation} \label{eq:pushexp_test1}
 {\bf E}_0 = {\bf i}, \quad {\bf B}_0 = \sqrt{2} {\bf j}, \quad  {\bf u}_0 =  {\bf k}. 
\end{equation}
This is a stationary case, so no reduced momentum component is expected in the $x$ direction.  We calculated a single push which is enough to illustrate the point, and it showed us interesting results (summarized in Table \ref{tab:pushexp_test1}).
The accuracy of the usual Boris pusher scales with $\sim \Delta t^2$ as expected,  but the Boris 2 scheme scales with $\sim \Delta t^4$ per $\Delta t$ time step, which means two orders of magnitude error suppression in this edge case. Evaluating the exponential more accurately (Boris 2N, 8 substeps) yields less than factor of 4 improvement. Higuera-Cary pusher \cite{higuera2017pusher} is not subject to this error. 

However, a second test case with more strongly relativistic parameters (which are more relevant for LWFA schemes)
\begin{equation} \label{eq:pushexp_test2}
 {\bf E}_0 = 10 {\bf i}, \quad {\bf B}_0 = 1.00005 {\bf j}, \quad  {\bf u}_0 =  100{\bf k},
\end{equation}
turned out to be less interesting. In this case all Boris schemes showed us $\sim \Delta t^2$ convergence. The error values were $1.25\times10^{-5}$ for Boris, $6.22\times10^{-6}$ for Boris 2, $4.68\times10^{-6}$ for Boris 2N doing a single $\Delta t= 0.1$ step. This means approximately a factor of 2 and a factor of 3 improvements for the Boris 2 and Boris 2N pushers. Selecting other strongly relativistic parameters yields the same relative improvements.

Overall, we can conclude that the Boris 2 scheme is a direct incremental upgrade over the regular Boris pusher, especially with weakly relativistic parameters.

\begin{table}[ht]
\begin{tabular*}{\linewidth}{@{\extracolsep{\fill}} lcccc }

$\Delta t$  &  $0.2$ &  $0.1$ & $0.05$  \tabularnewline
\hline 
Boris   & $4.93\times10^{-4}$ &  $6.23\times10^{-5}$ & $7.70\times10^{-6}$       \\
Boris 2 & $6.16\times10^{-7}$ &  $1.95\times10^{-8}$ & $6.10\times10^{-11}$           \\
Boris 2N  & $1.50\times10^{-4}$ & $1.55\times10^{-5}$ & $1.95\times10^{-6}$           \\ 
\hline
\end{tabular*}
\caption{Amplitude of the spurious reduced momenta component $u_x$ in the simple force free test case Eq. (\ref{eq:pushexp_test1}) in a single $\Delta t$ time step. We list data for the usual Boris pusher, our improved Boris 2 pusher, Boris 2N containing more accurate exponential evaluation (8 substeps).   \label{tab:pushexp_test1}}
\end{table}

\section{Quasi particle shape functions} \label{subsubsec:shapes}

We adopt the standard form of the normalized quasi particle shape functions \cite{BOOK_PLASMA_SIMULATION}. In this section we briefly summarize the relevant formulas. First, we assume that the shape functions are separable in each dimension:
\begin{equation} \label{eq:shapes_3D}
\hat{\mathcal{S}} ({\bf r}) = \hat{S}_x (x) \hat{S}_y (y) \hat{S}_z (z) \quad \text{with} \quad \int \hat{\mathcal{S}} {\rm d} V = 1,
\end{equation}
where $\hat{S}_x(x)$ is one dimensional normalized shape function. For simplicity, we assume they are the same in each direction and drop the subscript $x$.

The shape functions can be defined with the following convolution \cite{derouillat2018pic_smilei}:
\begin{equation} \label{eq:shapes_def}
\hat{S}^{(n)}(x) =  \int_{-\infty}^{\infty} \hat{S}^{(0)}(x'-x) \hat{S}^{(n-1)}(x') {\rm d} x',
\end{equation}
where the superscript denotes the order of the particle shape. Here $\hat{S}^{(0)}$ denotes zeroth order particle shape (``top-hat''):
\begin{equation} \label{eq:shapes_S0}
\hat{S}^{(0)} (x) = 
\begin{cases}
1 \quad &\text{if} \quad  |x| \leq \frac{1}{2} \Delta x, \\
0 \quad &\text{otherwise},
\end{cases}
\end{equation}
which is in a form of a crenel function. This function represents the local cell volume that the particles belong to.

For reference, the next three lowest order few shape functions are written as (``triangle'', quadratic and cubic spline):
\begin{align} \label{eq:shapes_S1}
&\hat{S}^{(1)} (x) =
\begin{cases}
1 - \left| \frac{x}{\Delta x}  \right| \quad &\text{if} \quad  |x| \leq \Delta x, \\
0 \quad &\text{otherwise},
\end{cases}  \\
\label{eq:shapes_S2}
&\hat{S}^{(2)} (x) =
\begin{cases}
\frac{3}{4} \left[1-\frac{4}{3} \left(\frac{x}{\Delta x} \right)^2 \right] \quad &\text{if} \quad  |x| \leq \frac{1}{2}\Delta x, \\
\frac{9}{8} \left(1-\frac{2}{3} \left(\frac{x}{\Delta x} \right)^2 \right) \quad &\text{if} \quad    \frac{1}{2}\Delta x \leq |x|\leq \frac{3}{2}\Delta x, \\
0 \quad & \text{otherwise} ,
\end{cases}  \\
\label{eq:shapes_S3}
&\hat{S}^{(3)} (x) =
\begin{cases}
\frac{2}{3} \left[1-\frac{3}{2} \left(\frac{x}{\Delta x} \right)^2 + \frac{3}{4} \left|\frac{x}{\Delta x} \right|^3 \right]  \   &\text{if} \  |x| \leq \Delta x, \\
\frac{4}{3} \left(1-\frac{1}{2} \left|\frac{x}{\Delta x} \right| \right)^3  \ \    &\text{if} \ \     \Delta x \leq |x|\leq 2\Delta x, \\
0  \   & \text{otherwise}.
\end{cases}
\end{align}
These higher order shapes also called splines, because they are piecewise continuous polynomials of order $n$.
The discretization of the above 1st, 2nd and 3rd order shapes require 2, 3, 4 grid points in the vicinity of the grid point closest to $x=0$. This means increasing computation cost especially in higher dimensions.

Assuming one of the above normalized particle shapes, the one dimensional interpolation formula for particle $\pp$  of field $\phi$ in Cartesian coordinates reads as:
\begin{equation} \label{eq:shapes_interpolate1}
\phi_\pp =  \Delta x \sum_{k} \hat{S}^{(n)} (x_k - x_\pp) \phi(x_k) .
\end{equation}

The direct deposition of density of a particle property $\tau_\pp$ at grid point $k$ reads as:
\begin{equation} \label{eq:shapes_deposit1}
{\varrho}_k^{(\tau)} =\sum_{\pp}  \tau_\pp   w_\pp \hat{S}^{(n)} (x_k - x_\pp).
\end{equation}

This yields the $\varrho_k$ charge density if we substitute $\tau_\pp$ with the particle charge $q_\pp$. Note that while the particle interpolation Eq. (\ref{eq:shapes_interpolate1}) can be independently done for each particle, this is not the case for density deposition Eq. (\ref{eq:shapes_deposit1}), because the particle functions might overlap.

If we use coordinate transformations (see \ref{subsubsec:transform}), these formulas apply to the uniform grid coordinates. For the interpolation formula Eq. (\ref{eq:shapes_interpolate1}) we ignore variations in the local volume completely. In the deposition formula Eq. (\ref{eq:shapes_deposit1}) these variations must be taken into account, such that $\sum_{k} \hat{S}(x_k-x_\pp)g_k \Delta x = 1$, where $g_k$ is the discretized transformation function.  We also need in both cases to get the particle position $x_\pp$ in the uniform grid coordinates from the iteration Eq. (\ref{eq:transform_inverse2}).

In Cartesian 3D, the above formulas will contain three summations corresponding to each dimension, assuming separable particle shape Eq. (\ref{eq:shapes_3D}). To the question which particle shape to use we the answer generally is that the lowest order that provides sufficient accuracy. To explain this in detail we analyze the interpolation accuracy in \ref{subsubsec:shapes_alias}.

\section{Interpolation accuracy analysis} \label{subsubsec:shapes_alias}
After initial testing with laser wakefield simulations, we found that the choosing which particle shape to use (see \ref{subsubsec:shapes}) for PIC field interpolation substantially affect the artifacts introduced and the accuracy of the simulations. (We used the same particle shapes in the density deposition).

To model the interpolation errors introduced by the discretized particle shapes $\hat{S}^{(n)}$ we propose the following  test scheme. Let the particle position coincide with a peak of a cosine of wavelength $\lambda = N \Delta x$, then the interpolated value reads as: 
\begin{equation} \label{eq:shapes_alias}
h(\tilde{x}) = \Delta x ^{-1} \sum_{k} \hat{S}^{(n)} (x_k - \tilde{x}) \cos \left( \frac{2 \pi}{\lambda} (x_k-\tilde{x}) \right), 
\end{equation}
which we consider as the function of $\tilde{x}$. The variable $\tilde{x}$ is an offset compared to the grid points $x_k$.  (Also $h(\tilde{x})$ is a periodic function with period $\Delta x$). We know that the value of Eq. (\ref{eq:shapes_alias}) should not depend on the variable $\tilde{x}$, but this is actually not the case: this is a consequence of the \emph{aliasing error} introduced by the discretization of the particle shape on the grid \cite{BOOK_PLASMA_SIMULATION}. 

The (average) interpolated value of $h(\tilde{x})$ depends on the particle width. As the reference value we consider the averaged value of the cosine in the grid cell corresponding to its peak (this is the average value of $h(\tilde{x})$ using the zeroth order shape). We call the offset from this value the \emph{average interpolation error}. To get accurate interpolation we must suppress both type of errors at once.

\begin{figure}[ht]
\centering
\includegraphics[width=7.5cm]{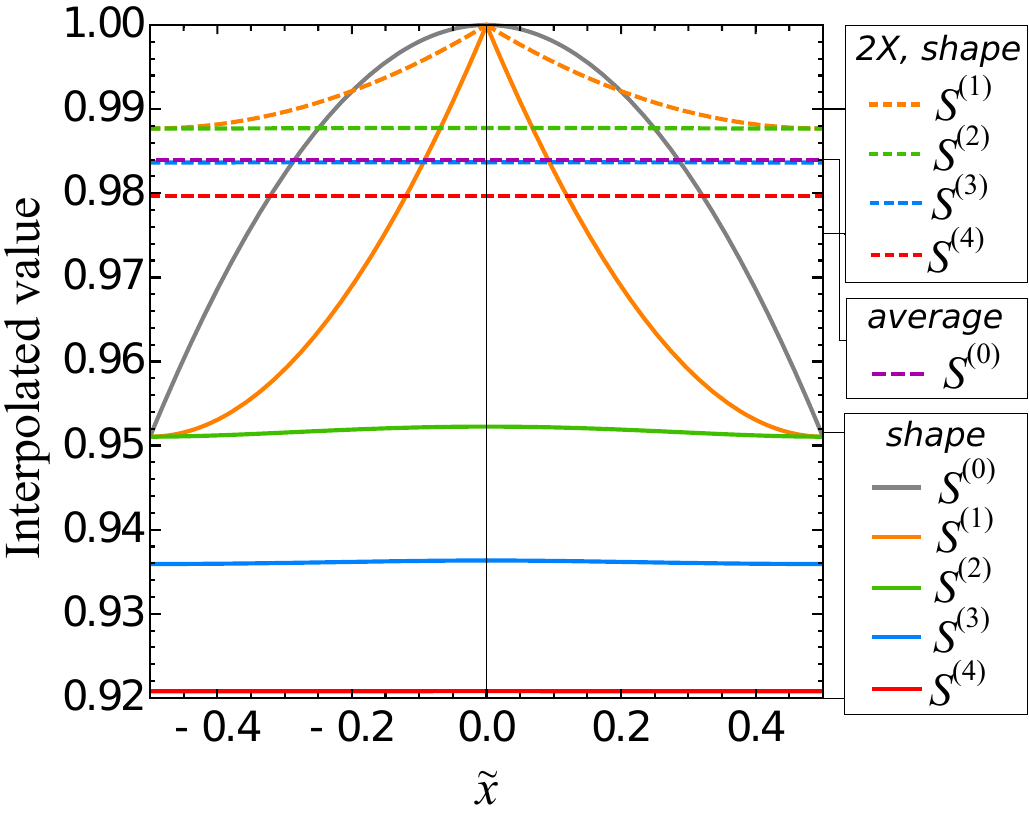}
\caption{The interpolated field value $h$ as the function of the offset $\tilde{x}$ from Eq. (\ref{eq:shapes_alias}) with $\lambda = 10 \Delta x$ (solid lines) and $\Delta x = 1$. We show the interpolated values using  the zeroth  ($\hat{S}^{(0)}$, gray), 1st ($\hat{S}^{(1)}$, orange), 2nd ($\hat{S}^{(2)}$, green), 3rd ($\hat{S}^{(3)}$, blue) and 4th ($\hat{S}^{(4)}$, red) order particle shapes. We also show the offset averaged field value corresponding $\hat{S}^{(0)}$ with dashed purple lines. We also plot the same interpolated values using double resolution with dashed lines ($2\times$ supersampling, which means $\lambda = 20 \Delta x$). 
\label{fig:shapes_alias}}
\end{figure}

\begin{table}[ht]
\begin{tabular*}{\linewidth}{@{\extracolsep{\fill}} cccc }

   &  Aliasing error & Average error  \tabularnewline
\hline 
$\hat{S}^{(1)}$  & $4.89\times10^{-2}$ &  $-1.89\times10^{-2}$      \\
$\hat{S}^{(2)}$  & $1.20\times10^{-3}$ &  $-3.23\times10^{-2}$      \\
$\hat{S}^{(3)}$  & $4.19\times10^{-4}$ & $-4.78\times10^{-2}$       \\ 
$\hat{S}^{(4)}$  & $1.97\times10^{-5}$ & $-6.32\times10^{-2}$       \\ 
${2\times} \ ,\ \hat{S}^{(1)}$  & $1.23\times10^{-2}$ &  $9.84\times10^{-3}$      \\
${2\times} \ ,\ \hat{S}^{(2)}$  & $7.57\times10^{-5}$ &  $3.72\times10^{-3}$      \\
${2\times} \ ,\ \hat{S}^{(3)}$  & $2.55\times10^{-5}$ & $3.27\times10^{-4}$       \\ 
${2\times} \ ,\ \hat{S}^{(4)}$  & $3.12\times10^{-7}$ & $4.36\times10^{-3}$       \\ 
\hline
\end{tabular*}
\caption{Table of numerical errors of field interpolation $h(\tilde{x})$ of Eq. (\ref{eq:shapes_alias}) with $\lambda = 10\Delta x$ and $\Delta x=1$ for 1st to 4th order particle shapes $\hat{S}^{(n)}$ corresponding to Fig. \ref{fig:shapes_alias}. The bottom four rows show  the same errors with double resolution ($2\times$ supersampling, which means $\lambda = 20 \Delta x$). The reference value for third column was single cell field average ($\hat{S}^{(0)}$) near the cosine peak ($\approx 0.984$). The aliasing error refers to the difference between the smallest and largest interpolated values from the function $h(\tilde{x})$.   \label{tab:shapes_alias}}
\end{table}

We plotted the interpolated values of  $h(\tilde{x})$ for zeroth order ($\hat{S}^{(0)}$) to 4th order particle shapes ($\hat{S}^{(4)}$) in Fig. \ref{fig:shapes_alias} (solid lines) with $\lambda = 10 \Delta x$ with $\Delta x = 1$. We can see that the interpolated value of $h(\tilde{x})$ substantially ($\sim 0.05$) varies for the zeroth and first order particle shapes.\footnote{Zeroth order particle shape shows serious amount of aliasing even if the particle position do not coincide with the cosine peak, it should not be used in PIC simulations.}
This aliasing problem has serious implications since it results in an nonphysical particle momentum jitter. In practice this could manifest itself with the so-called numerical self-heating \cite{BOOK_PLASMA_SIMULATION, arber2015pic_epoch} which could also strongly affect LWFA simulations \cite{cormier2008unphysical_kinetic_lwfa}. 
Using high order particle shapes reduces the aliasing effects by orders of magnitude - but the interpolated values are substantially downshifted.

We quantify the aliasing error using the difference between the largest and smallest interpolated values of $h(\tilde{x})$, which we show in the first 5 rows of Table \ref{tab:shapes_alias} (2nd column). We also show the averaged errors in reference to $\hat{S}^{(0)}$ case, which has the value of $\approx 0.984$ (3rd column). First we can see that using second order shape $\hat{S}^{(2)}$ suppresses the aliasing error by a factor of 40. Using $\hat{S}^{(3)}$ and $\hat{S}^{(4)}$ suppresses this further by factor of 3 and 60, respectively, suggesting uneven convergence properties. The average error worsens approximately by $0.016$ per shape order which is substantial. Surprisingly, using these 1st-4th order particle shapes for field interpolation suffer \emph{almost equal amount of error}, which approximately 5\%. So choosing a particle shape we can only balance between aliasing (artifacts) and average errors (slightly erroneous motion). We prefer the 3rd order particle shape.

There is, however, one thing that can be done which can reduce interpolation errors: increase the resolution of the fields (supersample) with respect to the particle shape (this effectively means $\lambda = 20 \Delta x$). In Fig. \ref{fig:shapes_alias} (dashed lines) and in Table \ref{tab:shapes_alias} we show these results with "$2\times$" mark. Now, averaged errors of the high order shapes are up shifted to be in the vicinity of the reference value (average $\hat{S}^{(0)}$).  The aliasing errors are also reduced further, but for $\hat{S}^{(1)}$ is only means factor of 4 - which is a very weak error suppression. So, the only way to accurately interpolate the fields is to simultaneously supersample and antialias them (\emph{supersampling antialiasing}), the latter of which involves using higher order ($\geq 2$) particle shapes. As a rule of thumb interpolating with third order shape at $2\times$ resolution is roughly equivalent to interpolating with $\hat{S}^{(0)}$ without aliasing error.

In practice, one increases the resolution in which the laser (wave) is sampled in PIC simulations, which is the usual way to handle these errors. However, during nonlinear laser plasma interaction these short wavelength field peaks could self-consistently appear, which still suffer from the above phenomena. This means that the only way to eliminate these errors is to supersample the fields during PIC field interpolation step, which inspired us write our interpolation method in Section \ref{subsec:particles_interpolate}.

\section{On charge conserving current depositions} \label{subsubsec:deposit_conserve}

In this section we discuss the concept of charge conserving current deposition methods in the context of our exponential field solver method. In the standard Yee PIC codes it was found that using such a scheme is mandatory to avoid self-consistently arising errors in the plasma motion \cite{BOOK_PLASMA_SIMULATION, pritchett2003pic_tutorial, boris1970relativistic}. In practice practice it only means that during or after deposition of the current ${\bf J}$ one enforces the continuity Eq. (\ref{eq:maxwell_divJ}) to get rid of the non-propagating spurious divergence in the ${\bf E}$ fields. If we want to do this using high order finite differences (i.e. on high order Lagrange polynomial basis) our troubles begin.   

First, the exponential algorithm we proposed here is quite different from regular PIC routines since during propagation of Maxwell fields we enforce very high order precision both spatially and temporally - and this is not limited to group velocity curves of the one dimensional wave propagation. We thought that we might find more of common ground with the pseudospectral analytical time domain (PSATD) solvers \cite{godfrey2014stability_spectral} but those use the continuity Eq. (\ref{eq:maxwell_divJ}) in the spectral analytical solution itself, so they have to enforce the validity of the latter.

First mitigation technique was proposed by Boris \cite{boris1970relativistic}, which involved the correction of $\nabla \cdot {\bf E}$ with Poisson-equation. The other proposed way to handle this is to apply a \emph{current correction} step after the current has been deposited \cite{villasenor1992charge_conservation, esirkepov2001charge_conservation}. Let us denote the divergence error as $\delta \varrho = \partial _t \rho - \nabla \cdot {\bf J}$, then the solution of $\nabla^2 \phi = -\delta \varrho$ and $\tilde{\bf J} = {\bf J} - \nabla \phi$ provide the divergence corrected current. Now the problem reduces to the solution of Poisson's equation, the solution of which only can be efficiently done in spectral basis \cite{BOOK_PLASMA_SIMULATION, lehe2016fbpic} as we have discussed in \ref{subsubsec:poisson}. We are limited to iterative methods here such as Jacobi or conjugate gradient type iterations \cite{BOOK_NUMERICAL_RECIPIES, vorst1992bicgstab} which are totally unsuitable to provide this type of current correction due to their slow convergence properties. We state here that there is  a conceptual problem with this type of current correction: local errors in the divergence  $\delta \varrho$ may yield non-local corrections to the $\tilde{\bf J}$ current because the solution of Poisson's equation generally non-local. \revisee{If this is the case, the iterative methods converge very slowly.}

When we tested our code in typical laser wakefield acceleration scenarios we did not found any particular numerical error that we could link to the lack of current correction in Cartesian coordinates (even in sharp particle bunches or in hot plasma). We found some evidence that people could mitigate the self consistent errors in Yee codes using couple passes of Jacobi iteration \cite{pritchett2003pic_tutorial,minz2000divergence_corr}, which only affects the highest frequency waves near $\kk_{\max}$ (it converges with $\kk^{2}$). This suggest the presence of high frequency self-consistent error in usual Yee codes.
There is a possible explanation why this kind of error might not occur in our PIC method:  the underlying \emph{analytical deposition} formulas Eqs. (\ref{eq:particles_charge})-(\ref{eq:particles_chargeJ}) do satisfy the continuity equation. Therefore combining high order finite differences with smoother particle shapes (like 3rd order spline) automatically reduce divergence and high frequency interpolation errors by 1-2 magnitudes.

The other odd property that affects the high order finite difference representation is that the numerical form of the \emph{differential} and the \emph{integral continuity equations} are slightly different, therefore the meaning of a charge conserving deposition is ambiguous. The Maxwell solver itself only knows the differential form while the physical conservation laws are closer in nature to the integral form of the continuity equation. With increasing differential order (i.e. higher order Lagrange polynomial basis) the results of these two converge to each other (to the analytical limit). Interestingly, in the second order Yee-solution, the second order integral and differential continuity equations are identical. 
In practice, we implemented in our code a modified Esirkepov deposition scheme \cite{esirkepov2001charge_conservation} which we discuss in \ref{subsubsec:deposit_ES}.

\section{Modified Esirkepov's deposition scheme} \label{subsubsec:deposit_ES}

In our code we opted using \emph{Esirkepov's charge conserving current deposition scheme} (ES)  \cite{esirkepov2001charge_conservation} which is standard current deposition method for finite difference PIC codes \cite{arber2015pic_epoch, derouillat2018pic_smilei}. This method meets our preference that it provides only local corrections and it is separable according to the Cartesian directions $x,y,z$. During this scheme we deposit time differences $\delta \varrho_x$, $\delta \varrho_y$, $\delta \varrho_z$ of the charge density that contain only one dimensional motion of the particles in the respective dimensions.
These must satisfy
\begin{equation} \label{eq:depositES_rho}
\delta \varrho_x +\delta \varrho_y + \delta \varrho_z = \delta \varrho_{n+1/2}, 
\end{equation}
where $\delta \varrho_{n+1/2} = (\varrho_{n+1}-\varrho_{n})/\Delta t$ at $n$th time step.  For detailed formulas please refer to Section \ref{subsec:particles_deposit} or Esirkepov's paper \cite{esirkepov2001charge_conservation}. 
%These one dimensional density differentials contain particle motion only in $x,y$ or $z$ directions, respectively. 
The current components $J_x$,$J_y$,$J_z$ then can be calculated from the respective one dimensional continuity equations using $\delta \varrho_x$, $\delta \varrho_y$, $\delta \varrho_z$.

Using our discrete staggered differential matrices ${\rm D}^{(+)}_{x}$, ${\rm D}^{(+)}_{y}$ and ${\rm D}^{(+)}_{z}$, this can be achieved from the \emph{differential continuity equation} as: 
\begin{equation} \label{eq:depositES_Jderiv}
J_x = \left({\rm D}^{(+)}_{x}\right)^{-1} \delta \varrho_x, \ 
J_y = \left({\rm D}^{(+)}_{y}\right)^{-1} \delta \varrho_y, \ 
J_z = \left({\rm D}^{(+)}_{z}\right)^{-1} \delta \varrho_z.
\end{equation}
The advantage of this method is that computation in every time step is actually feasible with arbitrary order finite differences with lower and upper triangular matrix (LU) decomposition. The disadvantage is that this calculation is still non-local in nature along the particular dimensions, which limits decomposition and parallelization. These are also be susceptible to numerical errors and require the inverse matrices to be well behaved. This technique is reminiscent of the alternating-direction implicit (ADI) finite difference Poisson solvers \cite{BOOK_NUMERICAL_RECIPIES, sapagovas2023PoissonADI}.

We propose an alternative option that is we use \emph{integral continuity equation} to compute the current. For simplicity, we write it as the following in direction $x$ (the method is same in the  $y$, $z$ directions):
\begin{equation} \label{eq:depositES_continuity}
J_{x,k+1/2}-J_{x,k-1/2} =  \int_{-\Delta x /2}^{{\Delta x /2}} \delta \varrho_x {\rm d}x,
\end{equation}
on the left hand side of which we have also taken into account the proper staggering. On the highly accurate Lagrange polynomial basis the results of the integral and differential continuity equations converge to each other. %Curiously, in the second order Yee-solution, the second order integral and differential continuity equations are identical.
Let us substitute an $2n+2$ order central integral quadrature with coefficients $f_k$ in Eq. (\ref{eq:depositES_continuity}) to get:
\begin{equation}
J_{x,k+1/2}^{(n)}-J_{x,k-1/2}^{(n)} =  \sum_{k'=-n}^{n} f_{k'} \delta \varrho_{x,k'+k} \Delta x_{k'+k},
\end{equation}
where the sampling $\Delta x_{k} = g_k \Delta x$ may have spatial dependence if we use coordinate transformations (otherwise $\Delta x_{k} = \Delta x$).
Next, we unroll the recursive definition of $J_{x,k+1/2}^{(n)}$ as follows:
\begin{equation}
J_{x,k+1/2}^{(n)} =  \sum_{k' = 0}^{k} \sum_{k''=-n}^{n} f_{k''} \delta \varrho_{x,k''+k'} \Delta x_{k''+k'} 
\end{equation}
We change the order of the summation here, and we evaluate the one with $k'$ first. This introduces the following second order current formula in its recursive form:
\begin{equation} \label{eq:depositES_continuity2}
J_{x,k+1/2}^{(0)} = J_{x,k-1/2}^{(0)} + \delta \varrho_{x,k} \Delta x_{k}.
\end{equation}
This is actually the same Esirkepov current formula as the one that can be found in any other second order Yee PIC codes.

Finally, to acquire the current on high order Lagrange interpolation basis we need to compute:
\begin{equation} \label{eq:depositES_continuityN}
J_{x,k+1/2}^{(n)} =   \sum_{k'=-n}^{n} f_{k'} J_{x,k'+k+1/2}^{(0)} \  \rightarrow \  J^{(n)}_x = {\rm F}^{({\text {div}}, A)} J_{x}^{(0)},
\end{equation}
which entails the matrix multiplication of the current with the banded diagonal matrix  ${\rm F}^{({\text {div}}, A)}$. This latter contains the integral quadrature coefficients $f_k$ in Eq. (\ref{eq:depositES_continuity}) with anti symmetric boundary conditions. We show the form of this is using 10th order approximation in Table \ref{tab:spatial_filter} labeled as div-10.

During the current deposition step itself we deposit the current ${\bf J}_\pp = {\bf J}^{(0)}_\pp$ at particle $\pp$ same as any standard Yee code. Notice that this algorithm is completely local and well behaved, while the one from Eq. (\ref{eq:depositES_Jderiv}) is generally not. This modified Esirkepov scheme is very convenient because it also staggers the deposited current properly in a single pass. We incorporate Eq. (\ref{eq:depositES_continuityN}) in the current filtering step to improve the validity of the continuity equation in Cartesian coordinates.

\bibliographystyle{elsarticle-num}
\bibliography{0BIB_pic_theory.bib, 0BIB_laser_theory.bib, 0BIB_laser_plasma.bib, 0Introduction.bib}
\end{document}